\documentclass[a4paper, 11pt]{article}
\pdfoutput=1

\usepackage[british]{babel}
\usepackage[T1]{fontenc}
\usepackage{jheppub}

\usepackage{microtype}
\usepackage{capt-of}
\usepackage{xfrac}
\usepackage{mleftright}
\usepackage{siunitx}
\usepackage{csquotes}
\usepackage[noabbrev]{cleveref}

\usepackage{amsfonts}
\usepackage{amsthm}
\usepackage{float}
\usepackage{array}
\usepackage{booktabs}
\usepackage{multirow}
\usepackage{subfigure}
\usepackage{theorem}
\usepackage{textcomp}
\usepackage[utf8]{inputenc}

\allowdisplaybreaks[1]

\numberwithin{equation}{section}

\hypersetup{
	pdftitle={PDF Profiling Using the Forward-Backward Asymmetry in Neutral Current Drell-Yan Production},
	pdfauthor={Elena Accomando, Valerio Bertone, Juri Fiaschi, Francesco Giuli, Alexandre Glazov, Francesco Hautmann, Stefano Moretti, Oleksandr Zenaiev},
	pdfkeywords={Electroweak interaction, Lepton production, Particle and resonance production, proton-proton scattering, Nonperturbative Effects}
}

\preprint{DESY-19-127, MS-TP-19-19, CERN-TH-2019-110}

\arxivnumber{1907.07727}

\title{PDF Profiling Using the Forward-Backward Asymmetry in Neutral Current Drell-Yan Production}

\author[a,b]{Elena Accomando,}
\author[c,a,b]{Juri Fiaschi,}
\author[b,d,e,f,g]{Francesco Hautmann,}
\author[a,b]{Stefano Moretti}
\author[h]{and the xFitter Developers' team: Hamed~Abdolmaleki,}
\author[i]{Valerio Bertone,}
\author[j]{Francesco Giuli,}
\author[k]{Alexander Glazov,}
\author[l]{Agnieszka Luszczak,}
\author[m]{Ivan Novikov,}
\author[n]{Fred Olness,}
\author[o]{Oleksandr Zenaiev}

\affiliation[a]{School of Physics \& Astronomy, University of Southampton, Highfield, Southampton SO17 1BJ, UK}
\affiliation[b]{Particle Physics Department, Rutherford Appleton Laboratory, Chilton, Didcot, Oxon OX11 0QX, UK}
\affiliation[c]{Institut f{\" u}r Theoretische Physik, Universit{\" a}t M{\" u}nster, D 48149 M{\" u}nster, Germany}
\affiliation[d]{Elementaire Deeltjes Fysica, Universiteit Antwerpen, B 2020 Antwerpen, Belgium}
\affiliation[e]{Theoretical Physics Department, University of Oxford, Oxford OX1 3NP, UK}
\affiliation[f]{UPV/EHU University of the Basque Country, Bilbao 48080}
\affiliation[g]{CERN, Theoretical Physics Department, CH 1211 Geneva}
\affiliation[h]{Faculty of Physics, Semnan University, 35131-19111 Semnan, Iran}
\affiliation[i]{Dipartimento di Fisica, University of Pavia, Pavia, Italy}
\affiliation[j]{University of Rome Tor Vergata and INFN, Sezione di Roma 2, Via della Ricerca Scientifica 1, 00133 Roma, Italy}
\affiliation[k]{Deutsches Elektronen-Synchrotron (DESY), Notkestrasse 85, D-22607 Hamburg, Germany}
\affiliation[l]{T. Kosciuszko Cracow University of Technology, PL-30-084, Cracow, Poland}
\affiliation[m]{Joint Institute for Nuclear Research, Joliot-Curie 6, Dubna, Moscow region, Russia, 141980}
\affiliation[n]{SMU Physics, Box 0175 Dallas, TX 75275-0175, United States of America}
\affiliation[o]{Hamburg University, II. Institute for Theoretical Physics, Luruper Chaussee 149, D-22761 Hamburg, Germany}

\emailAdd{e.accomando@soton.ac.uk}
\emailAdd{fiaschi@uni-muenster.de}
\emailAdd{hautmann@thphys.ox.ac.uk}
\emailAdd{s.moretti@soton.ac.uk}
\emailAdd{h.abdolmalki@gmail.com}
\emailAdd{valerio.bertone@cern.ch}
\emailAdd{francesco.giuli@roma2.infn.it}
\emailAdd{alexander.glazov@desy.de}
\emailAdd{agnieszka.luszczak@desy.de}
\emailAdd{ivan.novikov@desy.de}
\emailAdd{olness@smu.edu}
\emailAdd{oleksandr.zenaiev@desy.de}

\abstract{
Non-perturbative QCD effects from Parton Distribution Functions (PDFs) may be constrained by using high-statistics Large Hadron Collider (LHC) data.
Drell-Yan (DY) measurements in the Charged Current (CC) case provide one of the primary means to do this, in the form of the lepton charge asymmetry.
We investigate here the impact of measurements in Neutral Current (NC) DY data mapped onto the Forward-Backward Asymmetry ($A_{\rm FB}$) on PDF determinations, by using the open source fit platform {\tt{xFitter}}.
We demonstrate the potential impact of $A_{\rm FB}$ data on PDF determinations and perform a thorough analysis of related uncertainties.
}


\begin{document}
\maketitle
\flushbottom

\section{Introduction}
\label{sec:Intro}

The high-statistics data collected at the LHC during the Runs 2, 3, and the forthcoming High Luminosity LHC (HL-LHC) phase open the door to precision measurements at the TeV scale, which will be used in both studies of the Standard Model (SM) and searches for Beyond the SM (BSM) physics.
In order to keep up with the increasing statistical precision of experimental measurements, an impressive effort is being made on the theoretical side to provide calculations of perturbative Quantum Chromo-Dynamics (QCD) radiative corrections at Next-to-Leading Order (NLO) and Next-to-NLO (NNLO) as well as perturbative resummations of enhanced QCD corrections with Leading-Logarithmic (LL), Next-to-LL (NLL), and Next-to-NLL (NNLL) accuracy~\cite{Wackeroth:2019xib}. 
 
With improving perturbative accuracy, non-perturbative QCD contributions, such as those present in PDFs, increasingly become a crucial limiting factor in the theoretical systematics affecting both precision SM studies and (in)direct BSM searches.
Therefore, identification of measurements at the LHC, both current and upcoming, that can place strong constraints on the PDFs, thus reducing their uncertainties, is an important part of the LHC physics programme.

In the DY-induced lepton pair production channel, measurements of the cross section, differential with respect to the di-lepton mass (transverse or invariant) and rapidity have long been used to constrain PDFs (see e.g.,\cite{Aaboud:2016btc,Dulat:2015mca,Ball:2017nwa,Alekhin:2017kpj,Harland-Lang:2014zoa,Abramowicz:2015mha} for recent results).
In fact, also the lepton charge asymmetry in CC processes has been an effective way to constrain PDFs~\cite{Ball:2011gg,Ball:2014uwa,Harland-Lang:2014zoa,Dulat:2015mca,Alekhin:2017kpj,Ball:2017nwa}. 
However, it was observed in~\cite{Accomando:2018nig,Accomando:2017scx} that measurements in the NC process of the $A_{\rm{FB}}$, which are traditionally used in the context of precision determinations of the weak mixing angle $\theta_W$ (see e.g.,~\cite{Aad:2015uau,Chatrchyan:2011ya,Aaij:2015lka,Bodek:2016olg,ATLAS:2018gqq,Bodek:2018sin,Sirunyan:2018swq} and references therein), can usefully be employed for PDF determinations as well.
Refs.~\cite{Accomando:2018nig,Accomando:2017scx} studied the role of the angular information encoded in the $A_{\rm FB}$, which is related to the single-lepton pseudorapidity and, once combined with di-lepton mass and rapidity, would qualitatively correspond to triple-differential cross sections.
Precision measurements of triple-differential observables have been presented in~\cite{Aaboud:2017ffb}, while a recent study of DY differential cross sections in the context of PDFs has been presented in~\cite{Willis:2018yln}.
Furthermore, recently the ATLAS and CMS Collaborations determined the weak mixing angle in Refs.~\cite{ATLAS:2018gqq,Sirunyan:2018swq} through their DY measurements using methods which constrain PDF uncertainties. The CMS paper~\cite{Sirunyan:2018swq} uses the Bayesian $\chi^2$ reweighting technique~\cite{Giele:1998gw,Sato:2013ika,Bodek:2016olg} to constrain PDF uncertainties, while profiling of PDF error eigenvectors is used as a cross check. In the ATLAS note~\cite{ATLAS:2018gqq} the PDF uncertainties are included in the likelihood fit and thus constrained.

The DY triple-differential cross section for di-lepton production at LO is given by:

\begin{equation}
 \label{eq:triple_diff}
 \frac{d^3 \sigma}{dM_{\ell\ell}dy_{\ell\ell}d\cos\theta^*} = \frac{\pi\alpha^2}{3M_{\ell\ell}s} \sum_q P_q \left[f_q (x_1, Q^2) f_{\bar{q}} (x_2, Q^2) + f_{\bar{q}} (x_1, Q^2) f_q (x_2, Q^2)\right],
\end{equation}

\noindent
where $s$ is the square of the centre-of-mass energy of the colliding protons and $x_{1,2} = M_{\ell\ell} e^{\pm y_{\ell\ell}}/\sqrt{s}$ are the parton momentum fractions, $f_{q,\bar{q}} (x_i, Q^2)$ are the PDFs of the involved partons (either quark or anti-quark), $Q^2$ is the squared factorization scale (in our analysis always set equal to the di-lepton centre of mass energy), and $M_{\ell\ell}$ and $y_{\ell\ell}$ are the invariant mass and rapidity of the final state di-lepton system.
The function $P_q$ contains the propagators of the neutral SM gauge bosons and their couplings to the involved fermions:

\begin{align}
 \label{eq:propagators}
 P_q &= e^2_\ell e^2_q (1 + \cos^2\theta^*)\\ \nonumber
 &+ \frac{2M^2_{\ell\ell}(M^2_{\ell\ell} - M^2_Z)}{\sin^2\theta_W \cos^2\theta_W\left[(M^2_{\ell\ell} - M^2_Z)^2 + \Gamma^2_Z M^2_Z\right]} (e_\ell e_q) \left[v_\ell v_q (1 + \cos^2\theta^*) + 2 a_\ell a_q \cos\theta^*\right]\\ \nonumber
 &+ \frac{M^4_{\ell\ell}}{\sin^4\theta_W \cos^4\theta_W\left[(M^2_{\ell\ell} - M^2_Z)^2 + \Gamma^2_Z M^2_Z\right]} [(a^2_\ell + v^2_\ell) (a^2_q + v^2_q) (1+\cos^2\theta^*)\\
 &+ 8 a_\ell v_\ell a_q v_q \cos\theta^*], \nonumber
\end{align}

\noindent
where $\theta_W$ is the Weinberg angle, $M_Z$ and $\Gamma_Z$ are the mass and the width of the $Z$ boson, $e_\ell$ and $e_q$ are the lepton and quark electric charges, $v_\ell = -\frac{1}{4} + \sin^2\theta_W$, $a_\ell = -\frac{1}{4}$, $v_q = -\frac{1}{2}I^3_q - e_q \sin^2\theta_W$, $a_q = \frac{1}{2}I^3_q$ are the vector and axial couplings of leptons and quarks respectively, with $I^3_q$ the third component of the weak isospin; the angle $\theta^*$ is the lepton decay angle in the partonic centre-of-mass frame.
The first and third terms in Eq.~(\ref{eq:propagators}) are the square of the $s$-channel diagram with photon and $Z$ boson mediators respectively, while the second term is the interference between the two.

The $A_{\rm{FB}}^*$ is defined as:

\begin{equation}
  A_{\rm{FB}}^* = \frac { d \sigma / d M(\ell^+\ell^-)[\cos\theta^*>0] - d \sigma / d M(\ell^+\ell^-)[\cos\theta^*<0] }      { d \sigma / d M(\ell^+\ell^-)[\cos\theta^*>0] + d \sigma / d M(\ell^+\ell^-)[\cos\theta^*<0] }.
  \label{eq:afb}
\end{equation}

\noindent
From this expression it follows that the dominant contribution is given by the interference term, and in particular by the linear term in $\cos\theta^*$~\cite{Accomando:2015cfa}, which does not cancel in the numerator of Eq.~(\ref{eq:afb}).
The contribution of up-type and down-type quarks varies with the invariant mass and with the rapidity of the system as shown in Ref.~\cite{Accomando:2017scx}.
The $A_{\rm{FB}}^*$ is sensitive to the chiral couplings combination $v_\ell a_\ell v_q a_q$ and is proportional to valence quark PDFs.
In particular we expect the $A_{\rm{FB}}^*$ to be sensitive to the linear combination:

\begin{equation}
 e_\ell a_\ell [e_u a_u u_V(x, Q^2) + e_d a_d d_V(x, Q^2)] \propto \frac{2}{3} u_V(x, Q^2) + \frac{1}{3} d_V(x, Q^2).
\end{equation}

\noindent
However, when constraining valence quark PDFs we get constraints on sea PDFs too, since other data are sensitive to the sum of the valence and sea quark PDFs.
In particular we note a strong complementarity of the constraints coming from DY CC asymmetry, which is sensitive to the combination $u_V - d_V$ at LO~\cite{Harland-Lang:2014zoa}.

This paper is devoted to investigating the impact of the $A_{\rm FB}$ data on PDF extractions by using the open-source QCD fit platform {\tt{xFitter}}~\cite{Alekhin:2014irh}. 
We consider three different scenarios for luminosities, ranging from Runs 2, 3 to the HL-LHC stage~\cite{Azzi:2019yne}.
We perform PDF profiling~\cite{Paukkunen:2014zia} with {\tt{xFitter}} and present results for several PDFs, i.e., we quantitatively estimate the impact of the $A_{\rm FB}$ data on the uncertainties of these PDF sets, including different scenarios corresponding to different selection cuts for the di-lepton rapidity. 

The paper is organised as follows.
In Sec.~\ref{sec:xFitter} we describe technical aspects of the {\tt{xFitter}} implementation and treatment of $A_{\rm FB}$ pseudodata, while in Sec.~\ref{sec:NLO} we describe the inclusion of NLO QCD corrections in the analysis.
In Sec.~\ref{sec:Profiling} we present result of the PDF profiling.
In Sec.~\ref{sec:Sys} we discuss theoretical and systematic uncertainties affecting the $A_{\rm FB}$ observable.
We give our conclusions in Sec.~\ref{sec:Conclusions}. 

\section{$A_{\rm{FB}}$ in {\tt xFitter} and pseudodata generation} 
\label{sec:xFitter} 

In this section we describe the implementation of the $A_{\rm{FB}}$ observable in {\tt{xFitter}}~\cite{Alekhin:2014irh}, the generation of the pseudodata and the fitting procedure.

A suitable {\tt{C++}} code has been developed and integrated in the {\tt{xFitter}} environment for the analysis of the reconstructed forward-backward asymmetry ($A_{\rm{FB}}^*$) of two leptons with opposite charges in the final state from DY production in the NC channel.

Initially we implemented the observable at LO, where the initial state interaction occurs between a quark and an anti-quark of the same flavour ($q\bar{q}$) and the angle $\theta^*$ is defined with respect to the direction of the incoming quark. The latter is reconstructed accordingly to the direction of the boost of the di-lepton system, as discussed in Refs.~\cite{Dittmar:1996my,Rizzo:2009pu,Accomando:2015cfa,Accomando:2016tah,Accomando:2016ehi}.

Using the analytical expression in Eq. (\ref{eq:triple_diff}) for the hadronic triple differential cross section, numerical integrations for the calculation of the $A_{\rm{FB}}^*$ in different invariant mass bins and rapidity regions are performed using the \verb|GSL| public library, adopting the ``Adaptive Gauss-Kronrod'' rule with 61 points within each integration interval~\cite{piessens1983quadpack,DBLP:journals/corr/abs-1006-3962}.
This choice provides a sufficient precision in all integration intervals, including the more problematic high rapidity regions and $Z$-peak resonance neighbourhood.
Adaptive methods in principle could be problematic for fits using numerical estimation of derivatives, however there are no issues for profiling purposes.
Adjustable parameters of the analysis, such as collider energy, acceptance and rapidity cuts, have been implemented in the associated parameter card.
The mass effects of charm and bottom quarks in the matrix element are neglected, as appropriate for a high-scale process, and the calculation is performed in the $n_{f} = 5$ flavour scheme~\cite{Thorne:2008xf}.
Acceptance cuts reflect the usual ATLAS and CMS detector fiducial region, defined by $|\eta_\ell| < 2.5$ and $p^{\ell}_{{T}} >$ 20 GeV.
The input theoretical parameters have been chosen to be the ones from the Electro-Weak (EW) $G_{\mu}$ scheme~\cite{Hollik:1988ii}.
The explicit values for the relevant parameters in our analysis are the following: $M_Z = 91.188$ GeV, $\Gamma_Z = 2.441$ GeV, $M_W = 80.149$ GeV, $\alpha_{em} = 1/132.507$ and $\sin^2\theta_W = 0.222246$ (the last one does not matter for this specific profiling exercise).

Suitable datafiles with pseudodata have been generated for the analysis.
An important component contained in the datafiles is the statistical precision associated to the $A_{\rm{FB}}^*$ experimental measurements in each invariant mass bin.
The statistical error on the observable is given by:

\begin{equation}
  \Delta A_{\rm{FB}}^* = \sqrt{\frac{1-{A_{\rm{FB}}^*}^2}{N}}, 
  \label{eq:afb_error}
\end{equation}

\noindent
where $N$ is the total number of events in a specific invariant mass interval.
In order to obtain estimates as close as possible to the projected experimental accuracy, we have computed the number of events by convoluting the LO cross section without any acceptance cut with an acceptance times efficiency factor with typical value $\sim$20\% corresponding to realistic detector response~\cite{Khachatryan:2014fba}, and with a mass dependent $k$-factor reproducing the NNLO QCD corrections~\cite{Hamberg:1990np, Harlander:2002wh}.
We stress that the latter is used in the evaluation of the number of events in Eq.~\ref{eq:afb_error}, not in the evaluation of the observable itself.

The pseudodata have been generated according to this procedure fixing the collider centre-of-mass energy to 13 TeV for the three projected integrated luminosities of:

\begin{enumerate}
  \item 30 fb$^{-1}$, a subset of the currently available LHC data after the end of Run 2;
  \item 300 fb$^{-1}$, the designed integrated luminosity at the end of the LHC Run 3;
  \item 3000 fb$^{-1}$, the designed integrated luminosity at the end of the HL-LHC stage~\cite{Gianotti:2002xx}.
\end{enumerate}

In order to study the effects of data in the high di-electron rapidity region, the pseudodata have also been generated imposing various low rapidity cuts as $|y_{\ell\ell}| > 0$ (no rapidity cut), $|y_{\ell\ell}| > 1.5$ and $|y_{\ell\ell}| > 4.0$ (the last one required the extension of the detector acceptance region up to pseudorapidities $|\eta_\ell| < 5$).
Despite the possibility of exploring the impact of the $A_{\rm{FB}}^*$ in rapidity bins, instead of rapidity cuts, we opted for the latter choice in order to have data with larger statistic, which benefits the profiling of the PDFs.

With the goal of an exhaustive analysis over several PDF sets, multiple datafiles have been generated adopting the recent CT14nnlo~\cite{Dulat:2015mca}, NNPDF3.1nnlo (Hessian set)~\cite{Ball:2017nwa}, ABMP16nnlo~\cite{Alekhin:2017kpj}, HERAPDF2.0nnlo (EIG)~\cite{Abramowicz:2015mha} and MMHT2014nnlo~\cite{Harland-Lang:2014zoa} PDF sets.

\section{NLO study} 
\label{sec:NLO} 
For the calculation of the NLO $A^*_{\rm FB}$, the {\tt{MadGraph5{\_}aMC@NLO}}~\cite{Alwall:2014hca} program was used, interfaced to {\tt{APPLgrid}}~\cite{Carli:2010rw} through {\tt{aMCfast}}~\cite{Bertone:2014zva}. These NLO theoretical predictions correspond to the analysis cuts of the ATLAS data from Ref.~\cite{Aaboud:2017ffb}.
These NLO calculations are not supplemented by any $k$-factors to match higher-order accuracy. 

The asymmetry distribution is provided in 62 bins 2.5 GeV wide between $M_{\ell\ell}=$ 45 and 200 GeV\footnote{In this paper we work in the region near the $Z$ boson mass and assume this region to be free of BSM effects. See~\cite{Carrazza:2019sec} for a recent study of cross-contamination effects between BSM and PDF analyses.} (the pseudodata are prepared for the same invariant mass interval and bin size) for 5 different di-lepton rapidity $|y_{\ell\ell}|$ regions: 0.0 $< |y_{\ell\ell}| <$ 0.5, 0.5 $< |y_{\ell\ell}| <$ 1.0, 1.0 $< |y_{\ell\ell}| <$ 1.5, 1.5 $< |y_{\ell\ell}| <$ 2.0, and 2.0 $< |y_{\ell\ell}| <$ 2.5.
The asymmetry distribution is defined as a function of the angular variable $\cos\theta^{*}$ between the outgoing lepton and the incoming quark in the Collins–Soper (CS) frame~\cite{Collins:1977iv}, in which the decay angle is measured from an axis symmetric with respect to the two incoming partons. The decay angle $\theta^{*}$ in the CS frame is given by:

\begin{equation}
\cos\theta^{*} = \dfrac{p_{{Z},\ell\ell}}{M_{\ell\ell}|p_{{Z},\ell\ell}|} \dfrac{p_{1}^{+}p_{2}^{-}-p_{1}^{-}p_{2}^{+}}{\sqrt{M^{2}_{\ell\ell}+p^{2}_{{T},\ell\ell}}},
\end{equation}

\noindent
where $p_{i}^{\pm}=E_{i}\pm p_{{Z},i}$ and the index $i = 1,2$ corresponds to the positive and negative charged lepton respectively. Here, $E$ and $p_{{Z}}$ are the energy and the $z$-components of the leptonic four-momentum, respectively; $p_{{Z},\ell\ell}$ is the di-lepton $z$-component of the momentum and $p_{{T},\ell\ell}$ is the di-lepton transverse momentum. 
Then, the experimental measurement of the $A_{\rm{FB}}^*$ is obtained differentially in $M_{\ell\ell}$ according to Eq.~(\ref{eq:afb}) for the five aforementioned di-lepton rapidity regions.

\begin{figure}[h]
\begin{center}
\includegraphics[width=0.48\textwidth]{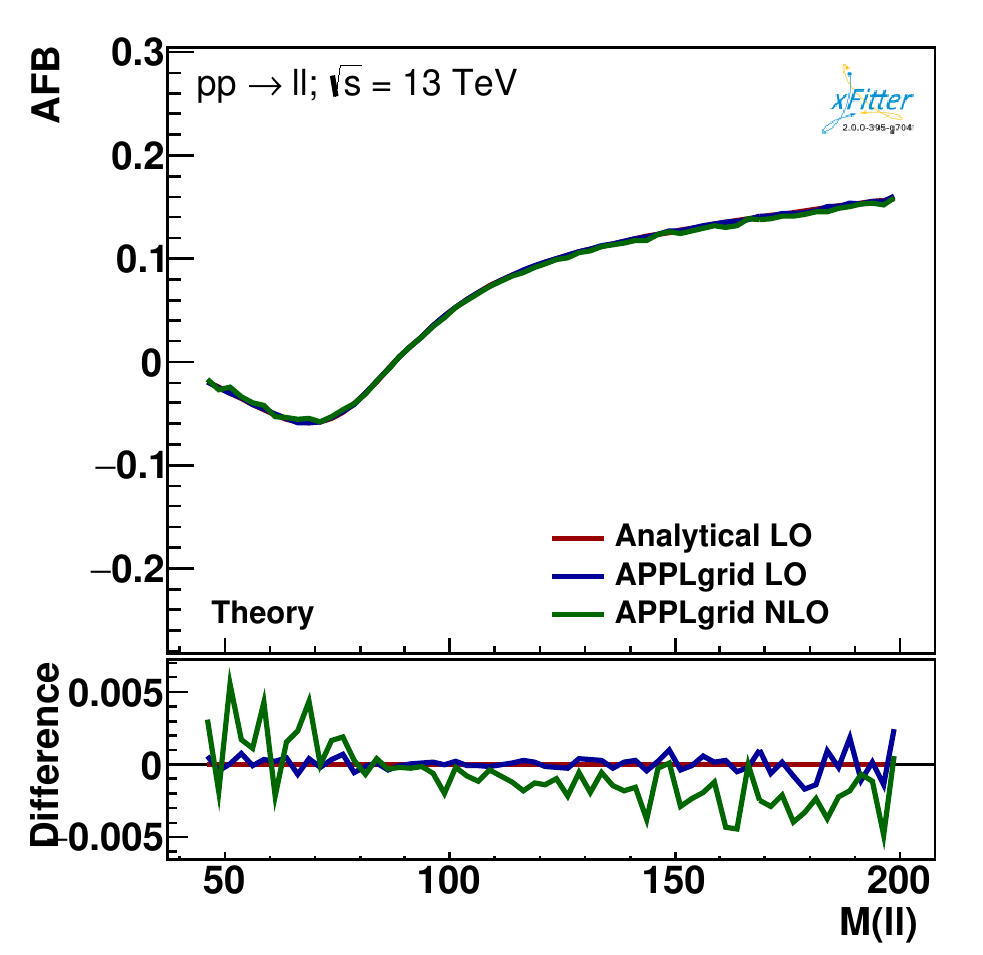}
\caption{The $A_{\rm{FB}}^*$ invariant mass distribution output of {\tt{xFitter}} obtained for the HERAPDF2.0nnlo PDF set with the independent analytical code as well as with the grids computed with {\tt{MadGraph5{\_}aMC@NLO}} at LO and NLO.}
\label{fig:AFB_NLO}
\end{center}
\end{figure}

Because of the definition of the $A_{\rm{FB}}^*$ observable, NLO corrections largely cancel in the ratio of cross sections, thus there is no significant difference between the observable calculated at LO or NLO.
In Fig.~\ref{fig:AFB_NLO} we show the $A_{\rm{FB}}^*$ curves from {\tt{xFitter}} obtained with the LO analytical code and when employing the LO and NLO grids.
As visible in the lower panel, the differences between the results obtained with the LO analytical code and with LO grids match very well up to purely statistical fluctuations, while NLO corrections slightly dilute the $A_{\rm{FB}}^*$ shape, being positive (negative) in the region below (above) the $Z$ peak where the $A_{\rm{FB}}^*$ is negative (positive).

We have verified that no differences are visible when comparing the profiled curves obtained using either LO or NLO calculations.
The results that follow have been obtained by means of the described NLO grids, unless stated differently.

\section{PDF profiling and numerical results} 
\label{sec:Profiling}
In this section we present the results of the profiling on the aforementioned PDF sets, using various combinations of $A_{\rm{FB}}^*$ pseudodata, varying the integrated luminosity and the rapidity cut.
The qualitative behaviour of the profiled distributions does not change when varying the $Q^2$ scale, thus, unless otherwise stated, in the following, results will be shown for a reference scale $Q^2 = M_Z^2$.
A more extensive discussion on the effects of the choice of the scales (both factorisation and renormalisation) is presented in Sect.~\ref{sec:Sys}.

\subsection{PDF profiling}

The profiling technique~\cite{Paukkunen:2014zia} is based on minimizing $\chi^2$ between data and theoretical predictions.
The PDF uncertainties are included in the $\chi^2$ using nuisance parameters.
The values of the PDF nuisance parameters at the minimum are interpreted as optimised, or profiled, PDFs, while their uncertainties determined using the tolerance criterion of $\Delta\chi^2 = 1$ correspond to the new PDF uncertainties. 
The profiling approach assumes that the new data are compatible with the theoretical predictions using the existing PDF set and, under this assumption, the central values of the data points are set to the central values of the theoretical predictions.
No theoretical uncertainties except the PDF uncertainties are considered when calculating the $\chi^2$.

\begin{figure}[h]
\begin{center}
\includegraphics[width=0.33\textwidth]{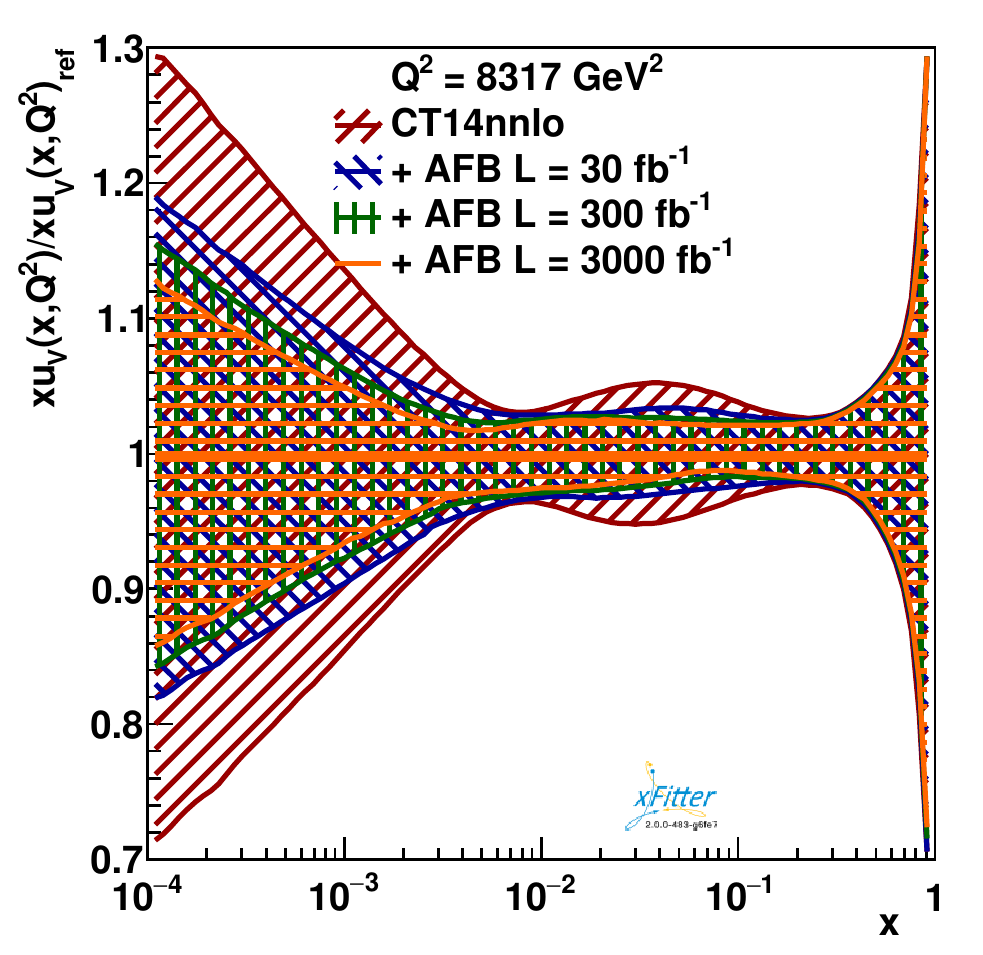}%
\includegraphics[width=0.33\textwidth]{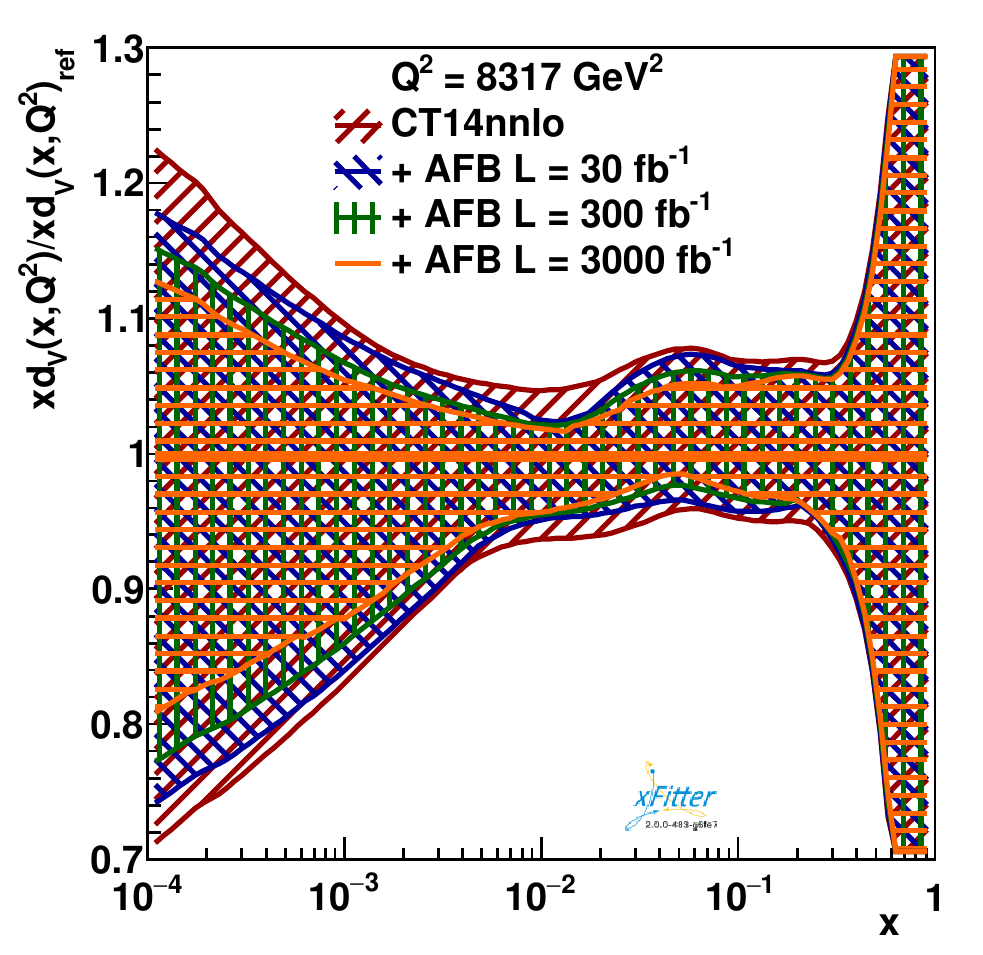}%
\includegraphics[width=0.33\textwidth]{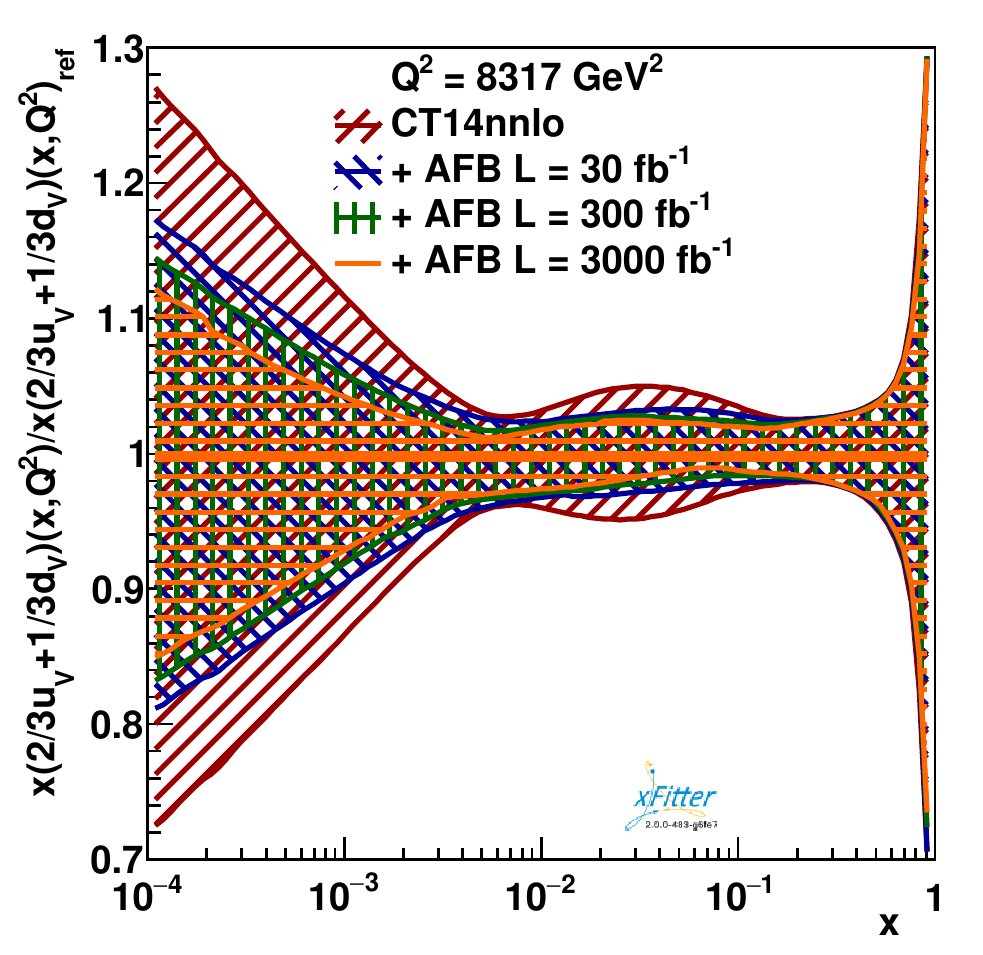}\\
\includegraphics[width=0.33\textwidth]{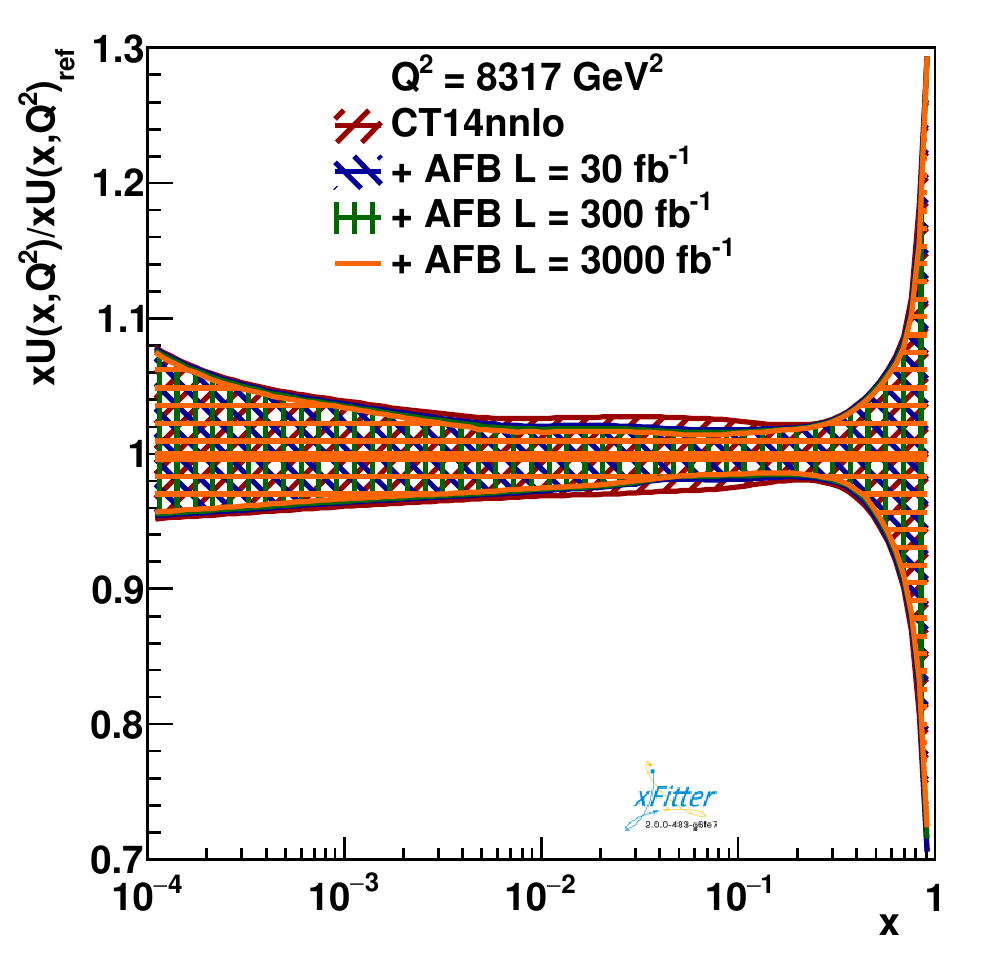}%
\includegraphics[width=0.33\textwidth]{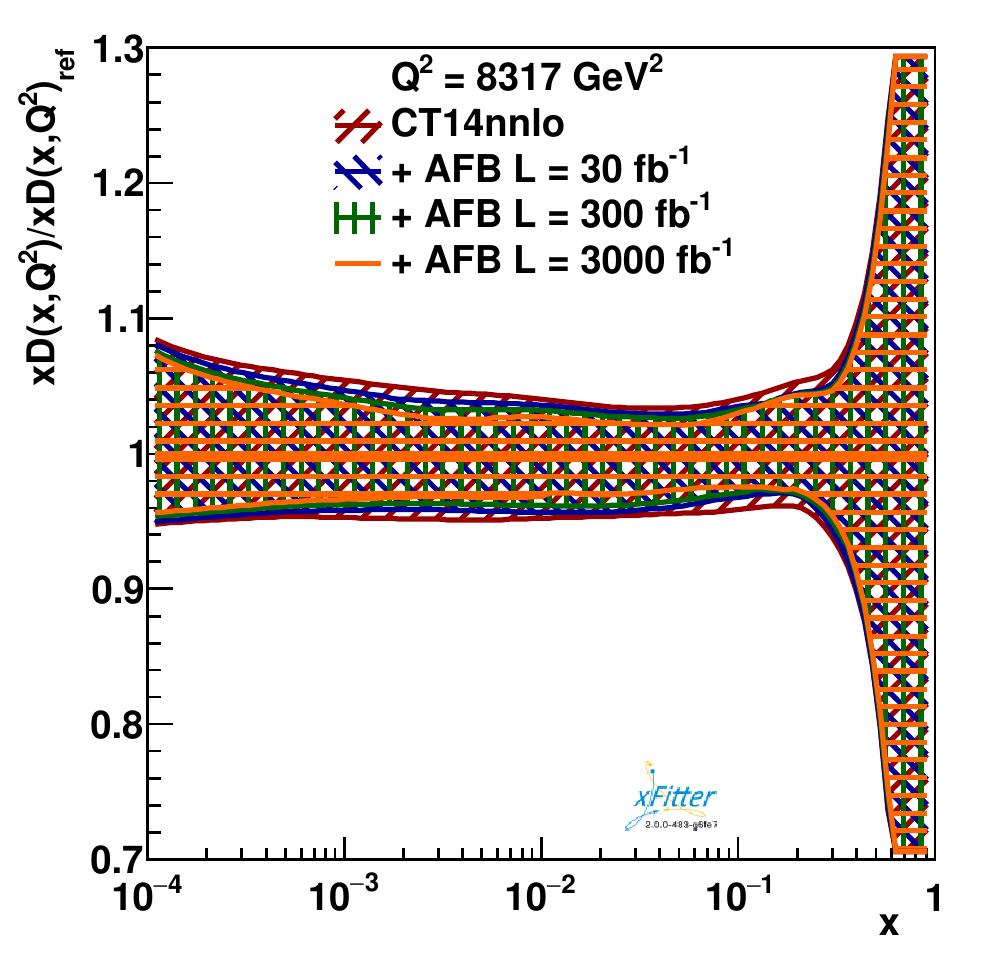}%
\includegraphics[width=0.33\textwidth]{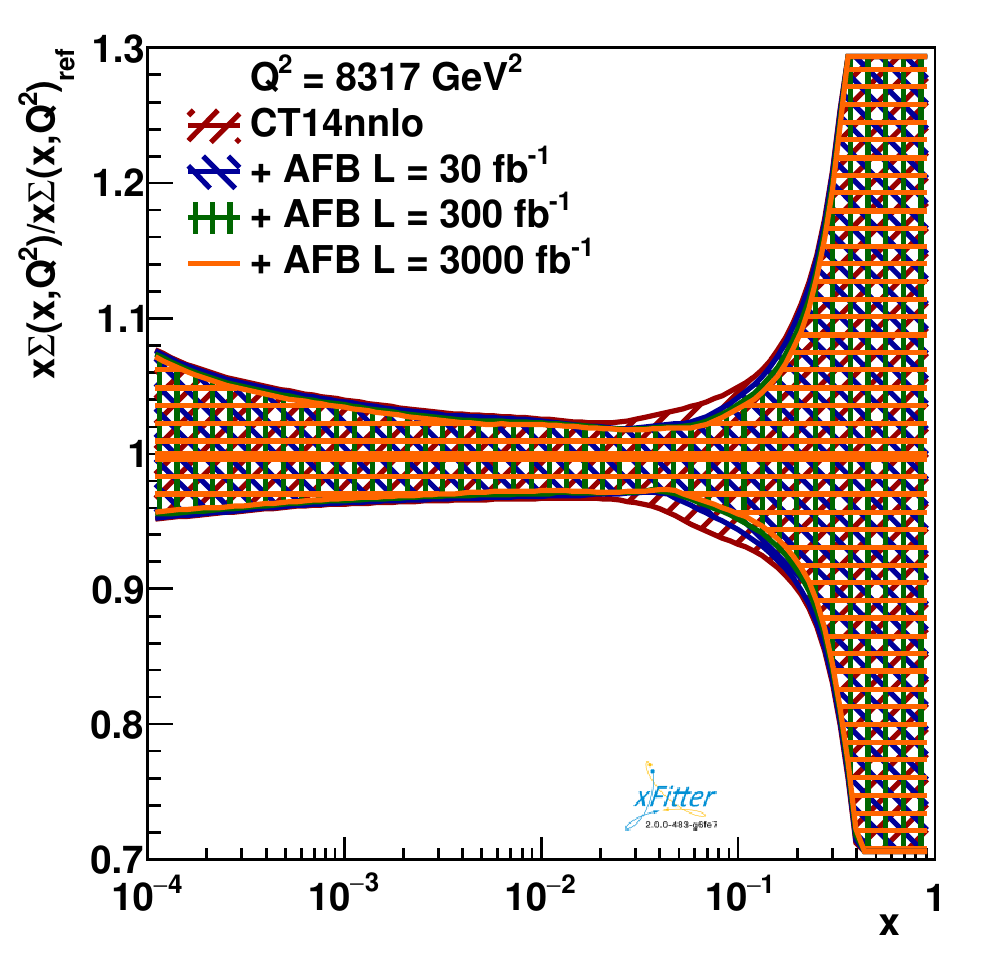}
\caption{Original (red) and profiled curves distributions for the normalised distribution of the ratios of (top row, left to right) $u$-valence, $d$-valence and $((2/3)u+(1/3)d)$-valence and (bottom row, left to right) $u$-sea, $d$-sea quarks and $(u+d)$-sea quarks of the CT14nnlo PDF set using $A_{\rm{FB}}^*$ pseudodata corresponding to an integrated luminosity of 30 fb$^{-1}$ (blue), 300 fb$^{-1}$ (green) and 3000 fb$^{-1}$ (orange).}
\label{fig:prof_CT14nnlo_lum}
\end{center}
\end{figure}

Fig.~\ref{fig:prof_CT14nnlo_lum} shows the impact of the profiling on the CT14nnlo PDF set.
For this specific PDF set we also rescale the error bands to 68\% Confidence Level (CL), for a better comparison with the results obtained with the other PDF sets.
As visible, the largest reduction of the uncertainty band is obtained for the $u$-valence quark distribution. As the luminosity grows, also the distribution for the $d$-valence quark displays a visible improvement.
The main effects are concentrated in the region of intermediate and small momentum fraction $x$.
The sea quark distributions show a moderate improvement, progressively increasing with the integrated luminosity. While the contraction of the error band in the $u$-sea distribution seems to saturate above 300 fb$^{-1}$, the $d$-sea quark distribution appears to show continued improvement with an integrated luminosity of 3000 fb$^{-1}$.
For the sea quark distributions, these effects are concentrated in the region of intermediate $x$.

\begin{figure}[h]
\begin{center}
\includegraphics[width=0.25\textwidth]{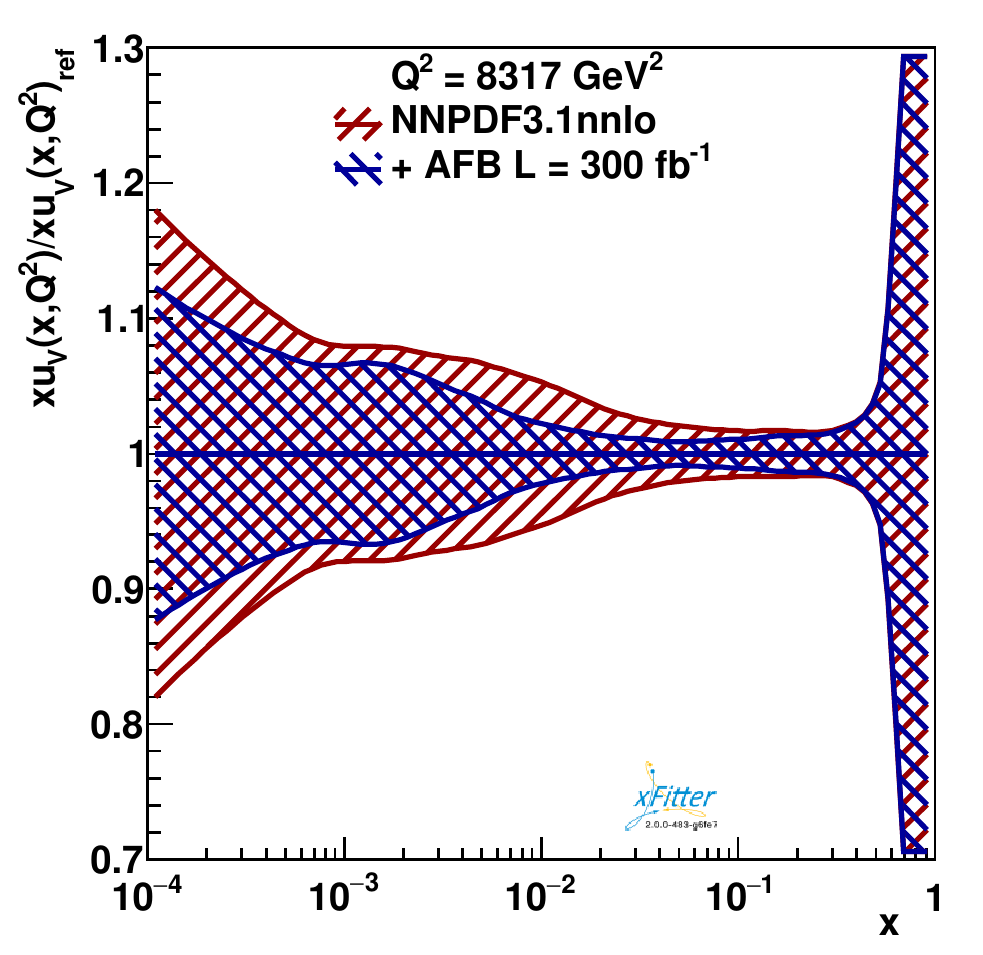}%
\includegraphics[width=0.25\textwidth]{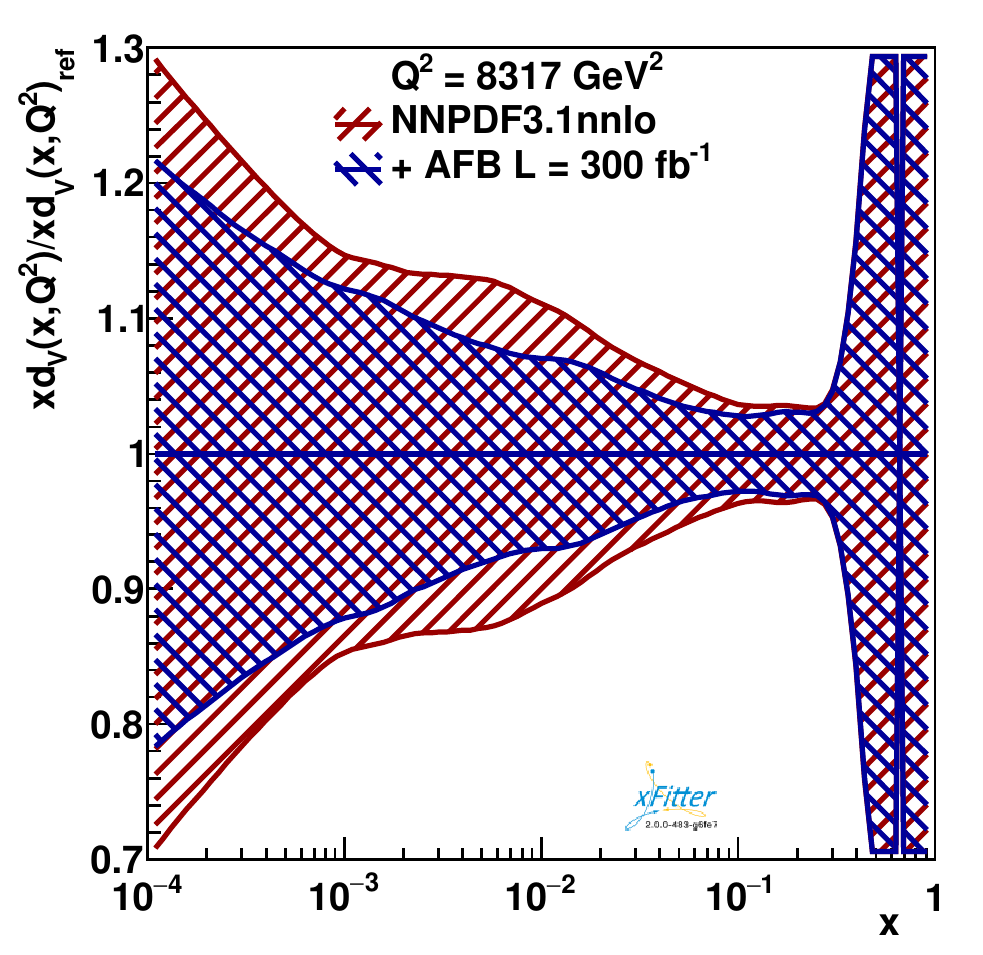}%
\includegraphics[width=0.25\textwidth]{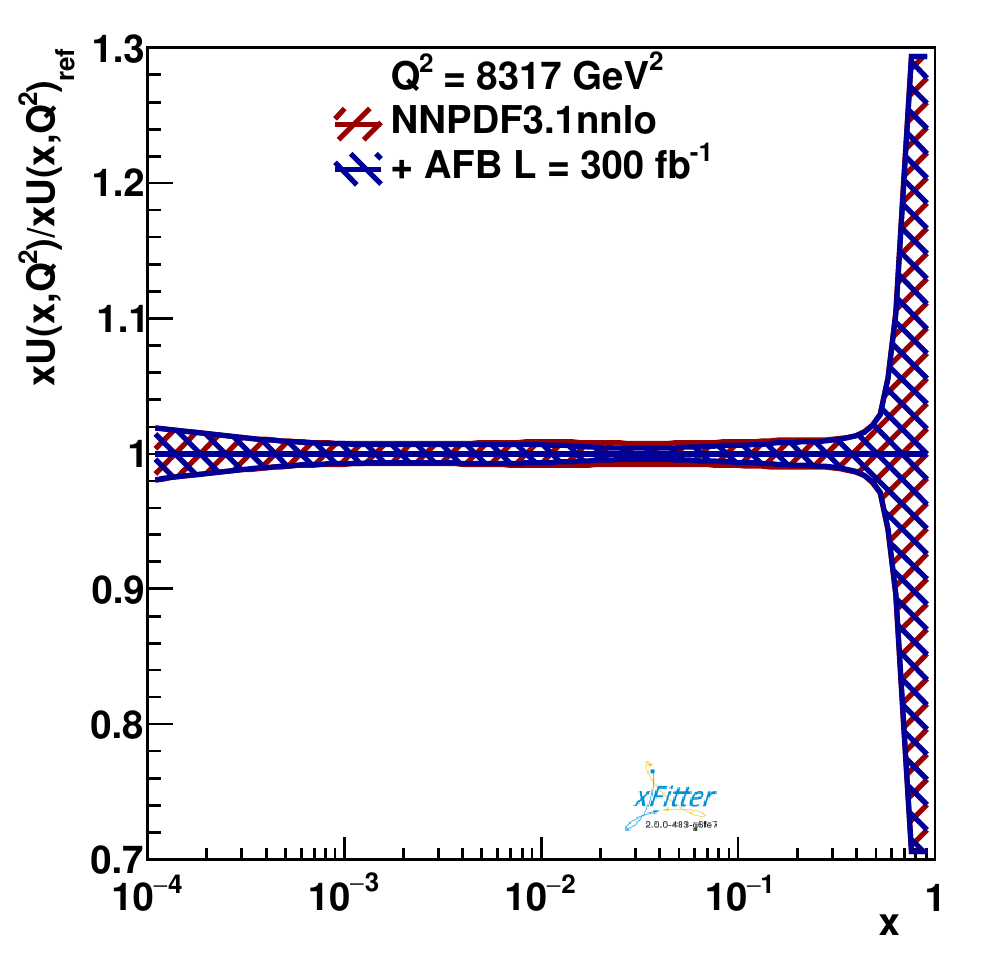}%
\includegraphics[width=0.25\textwidth]{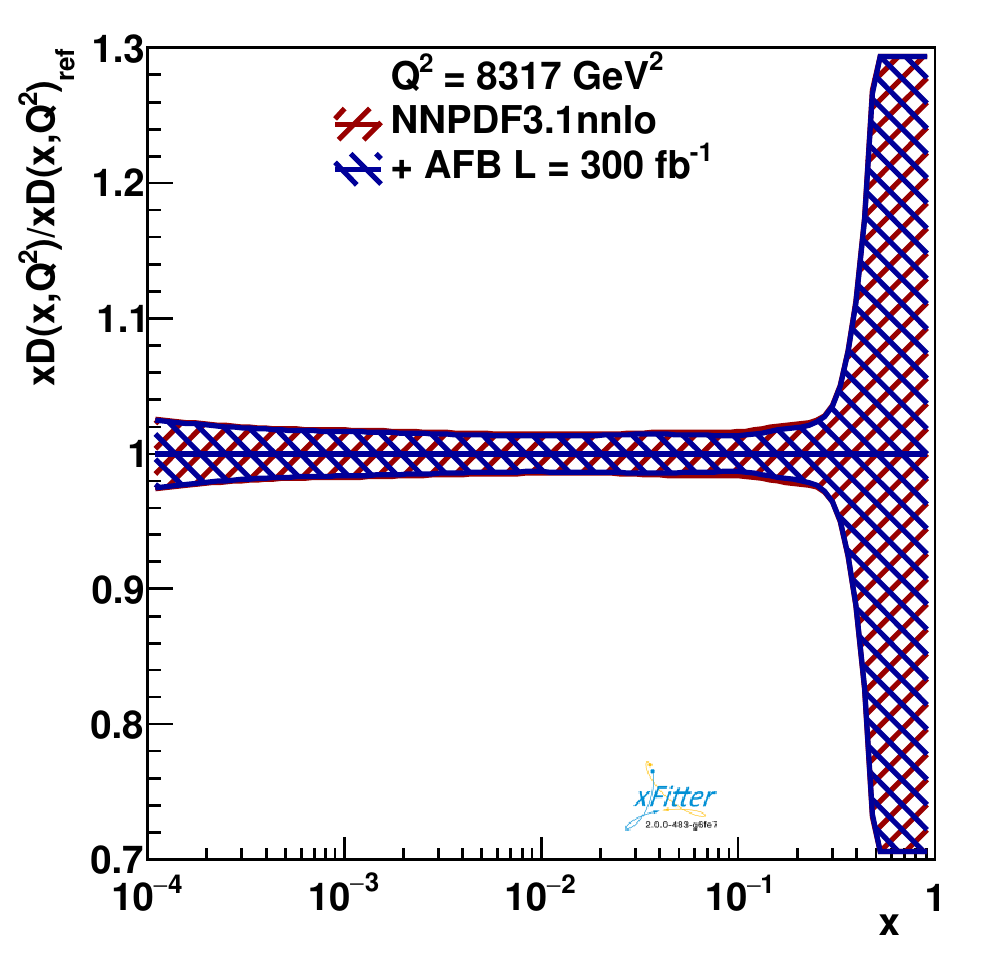}\\
\includegraphics[width=0.25\textwidth]{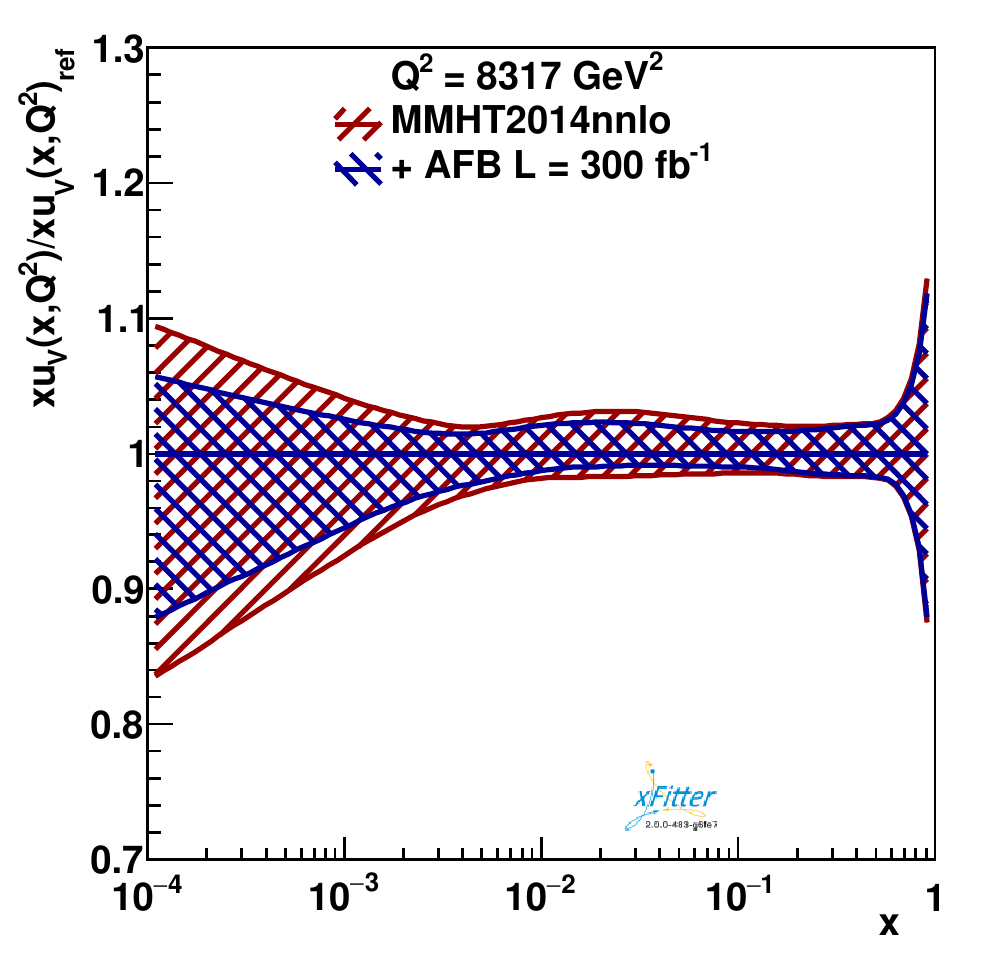}%
\includegraphics[width=0.25\textwidth]{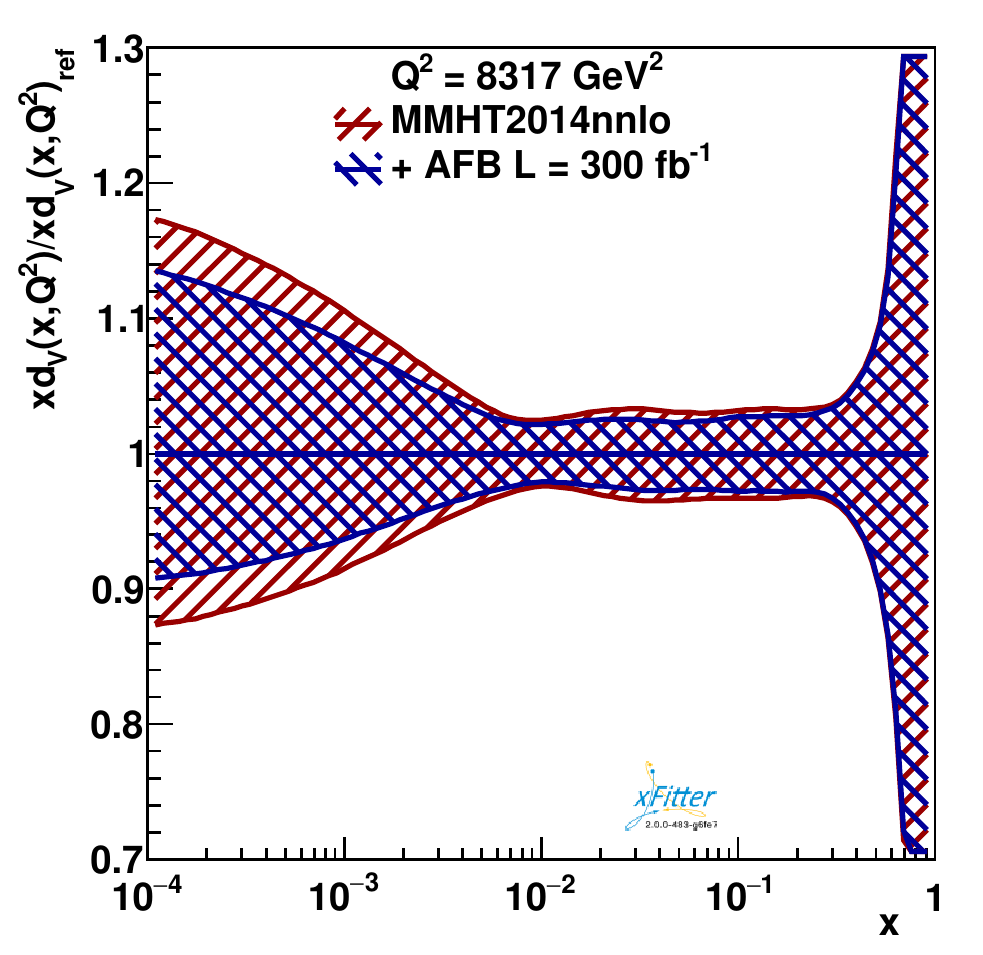}%
\includegraphics[width=0.25\textwidth]{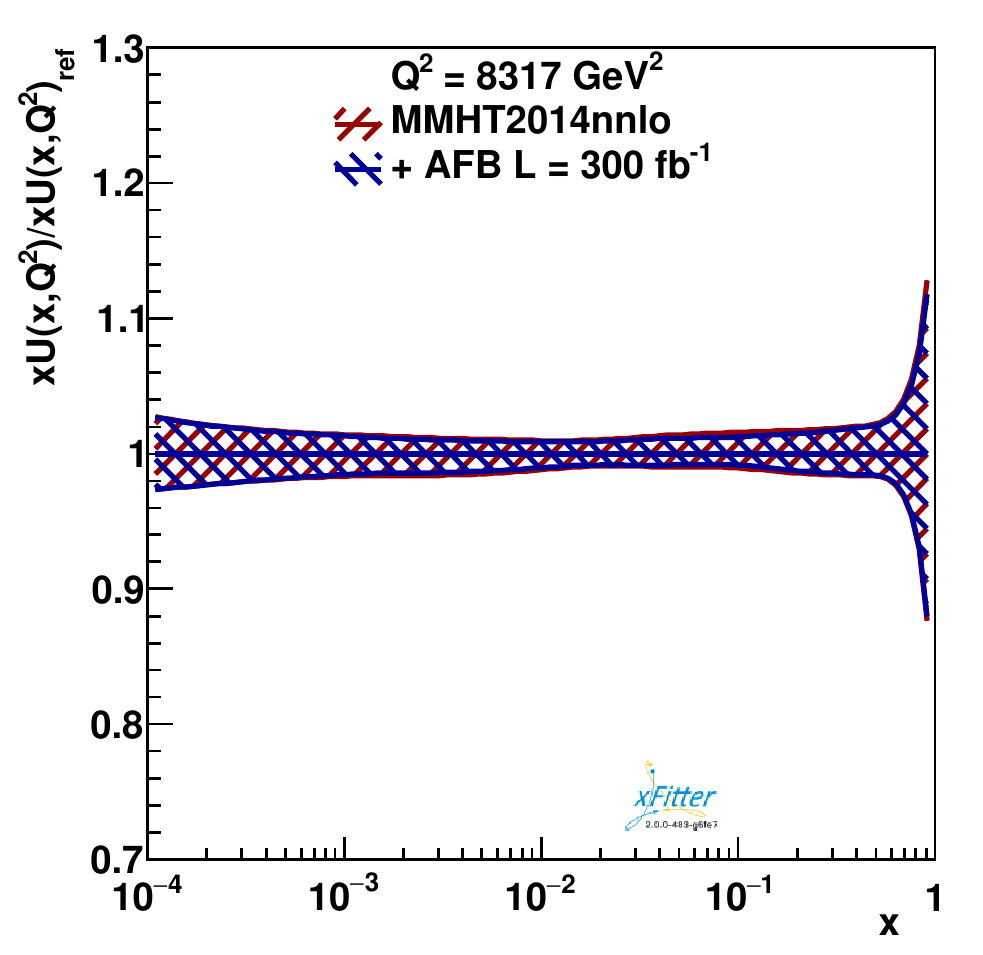}%
\includegraphics[width=0.25\textwidth]{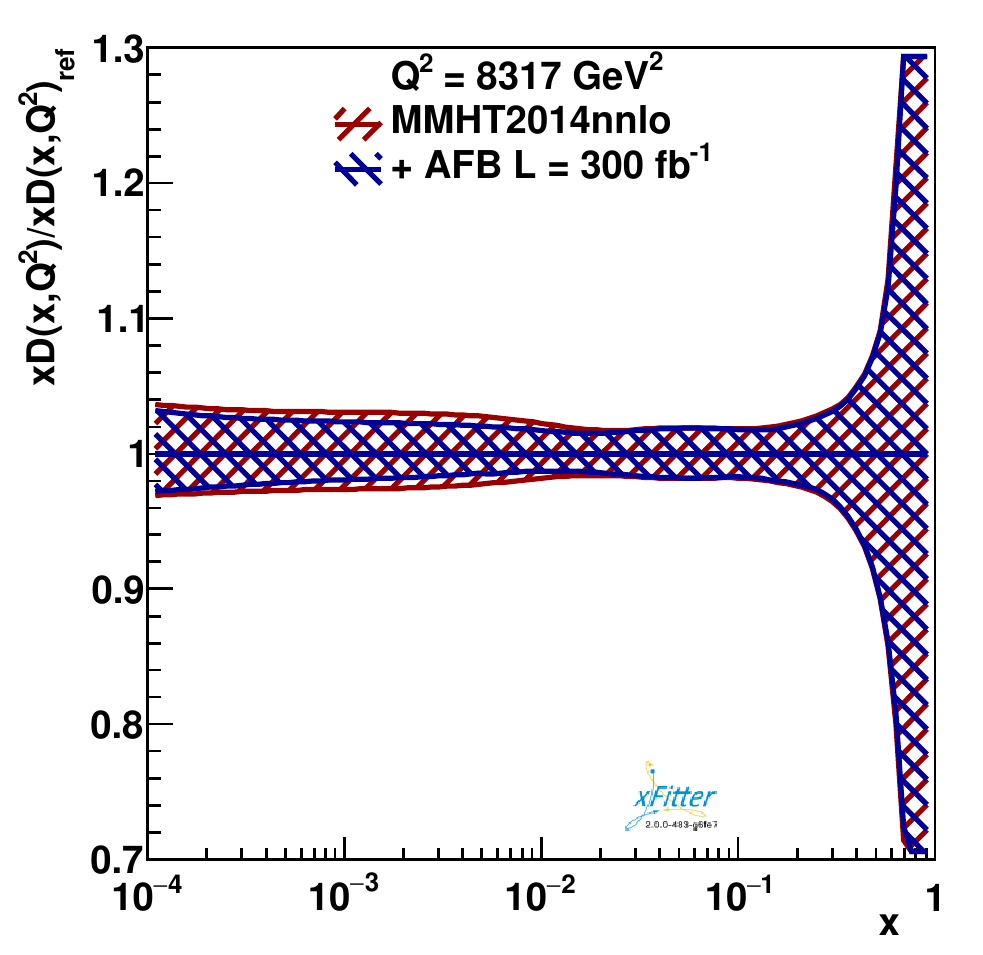}\\
\includegraphics[width=0.25\textwidth]{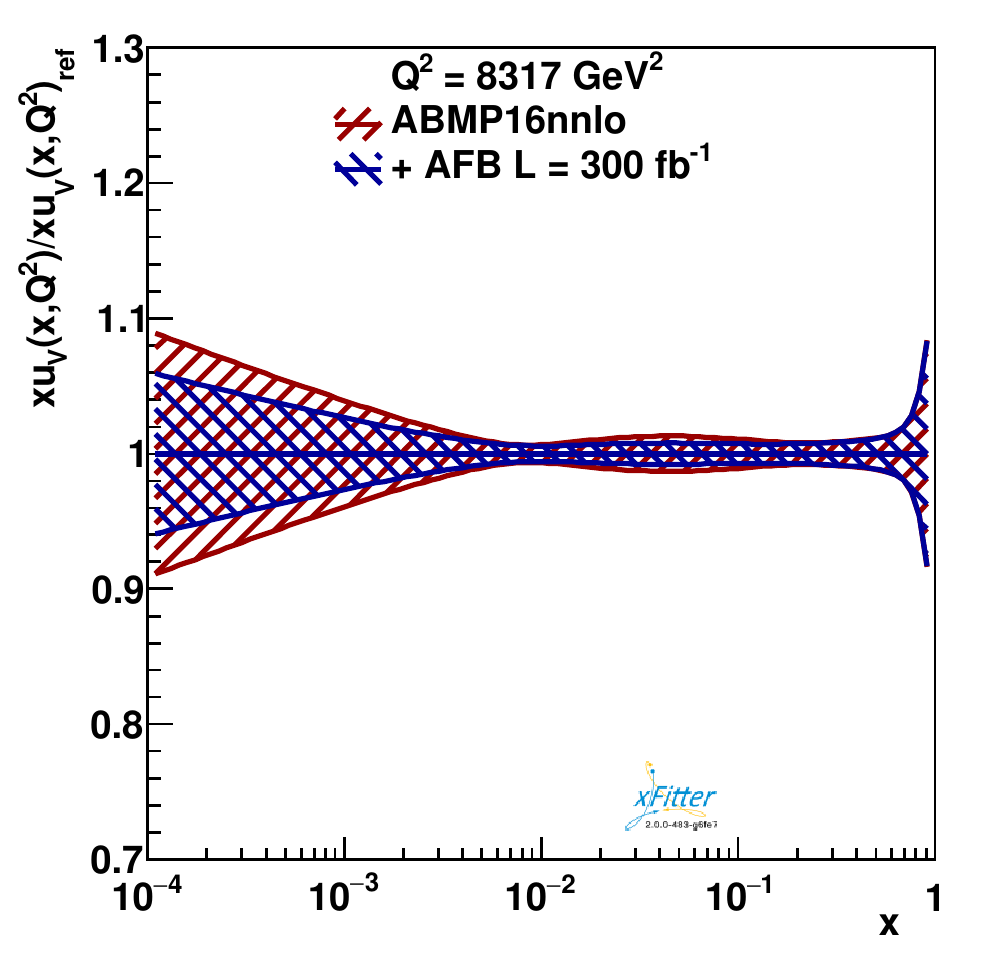}%
\includegraphics[width=0.25\textwidth]{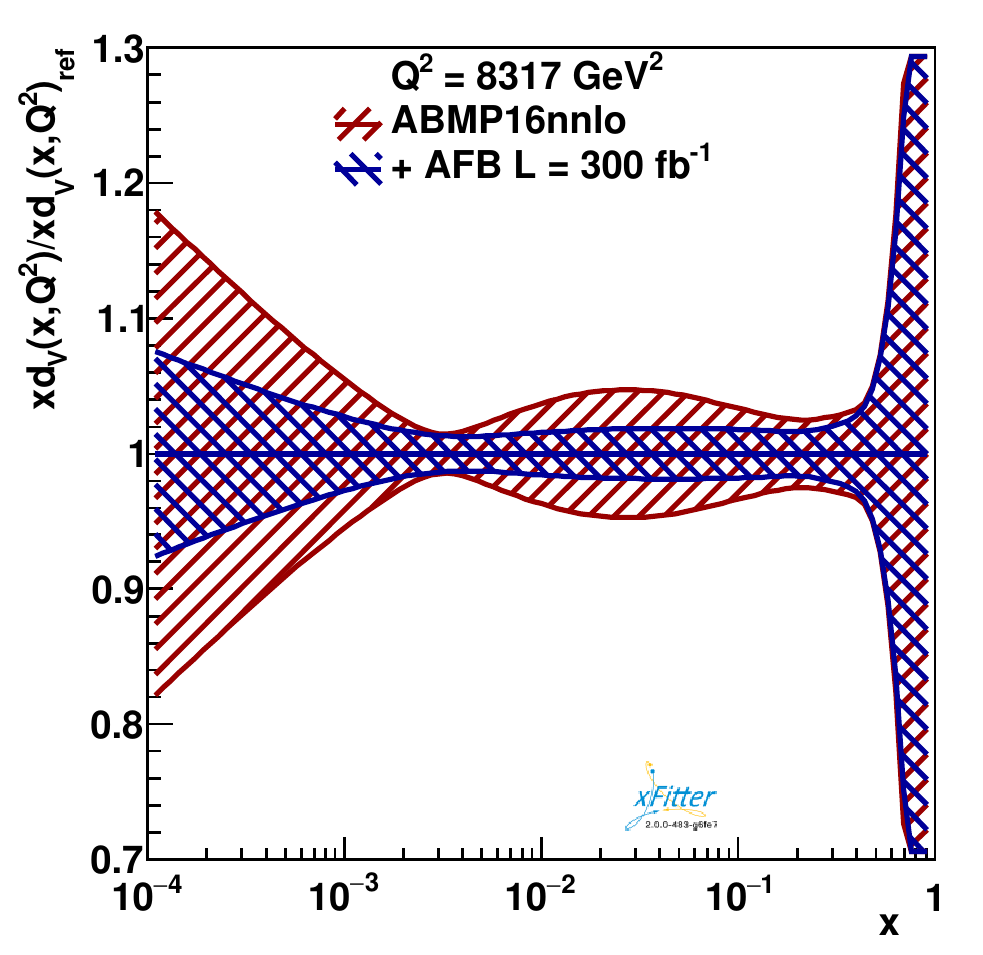}%
\includegraphics[width=0.25\textwidth]{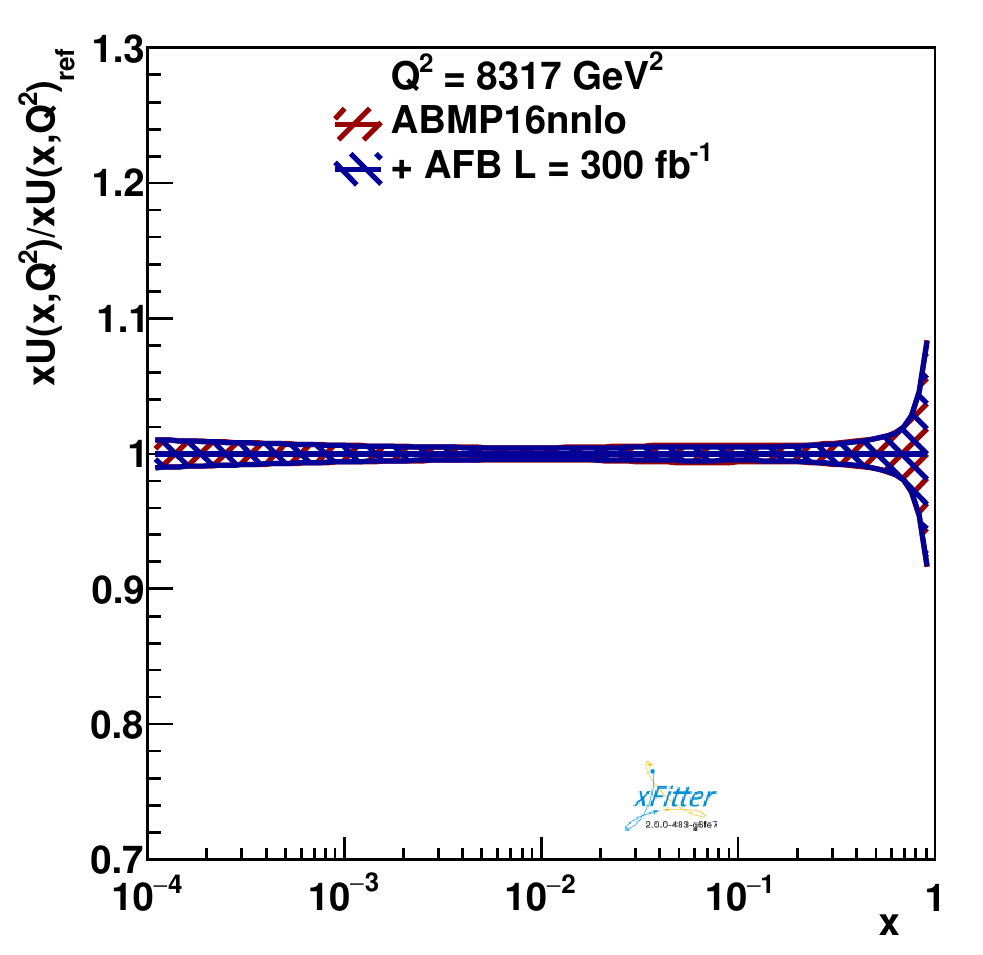}%
\includegraphics[width=0.25\textwidth]{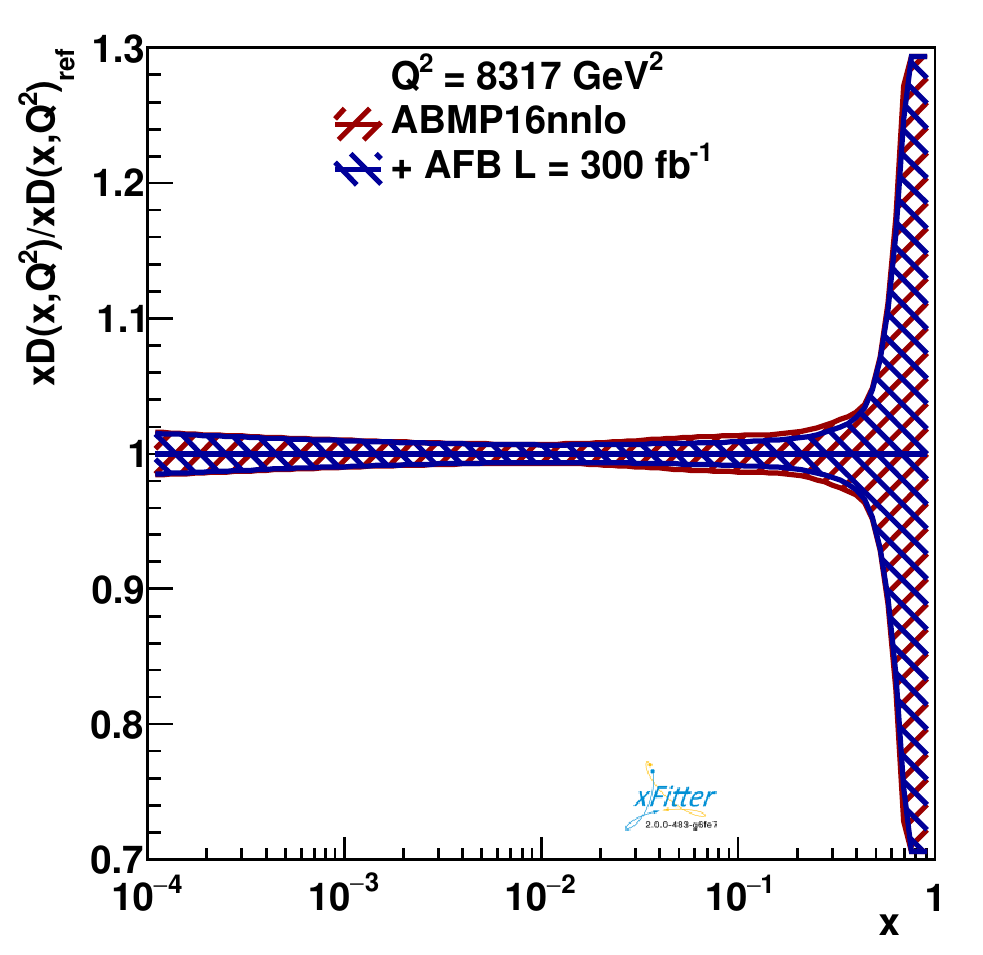}\\
\includegraphics[width=0.25\textwidth]{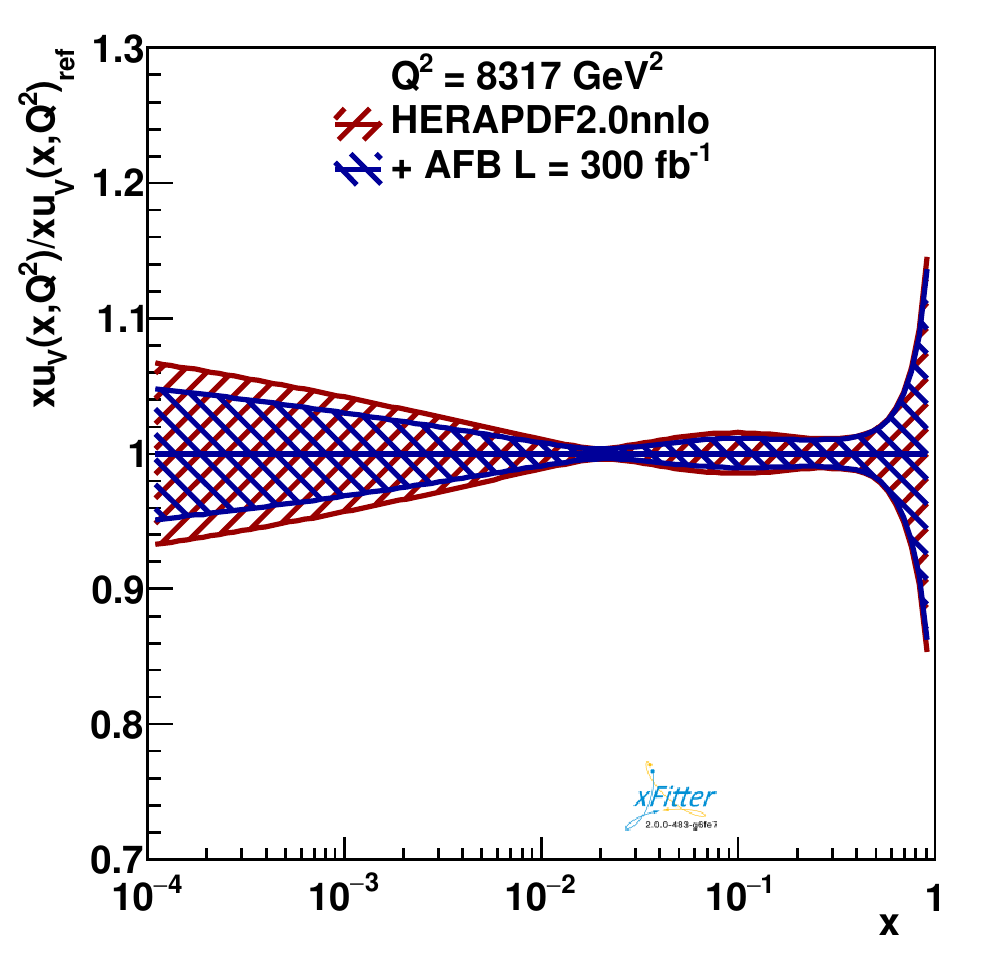}%
\includegraphics[width=0.25\textwidth]{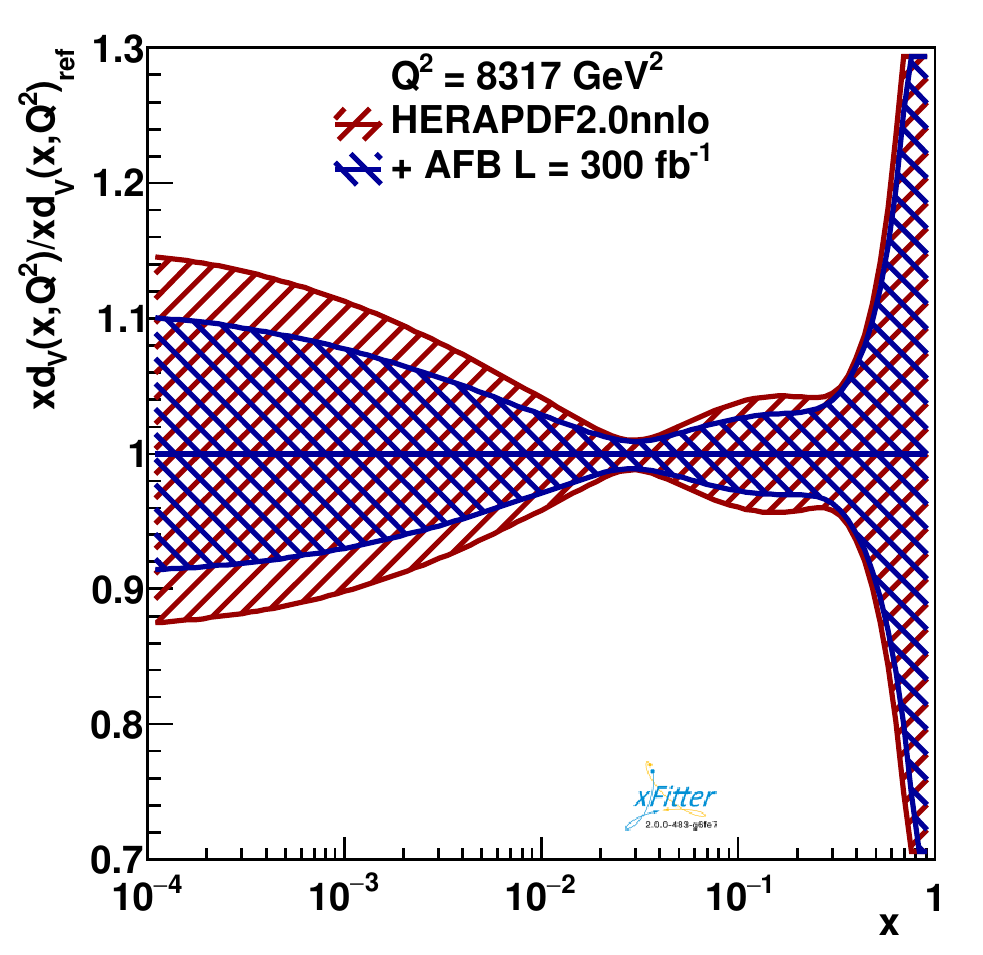}%
\includegraphics[width=0.25\textwidth]{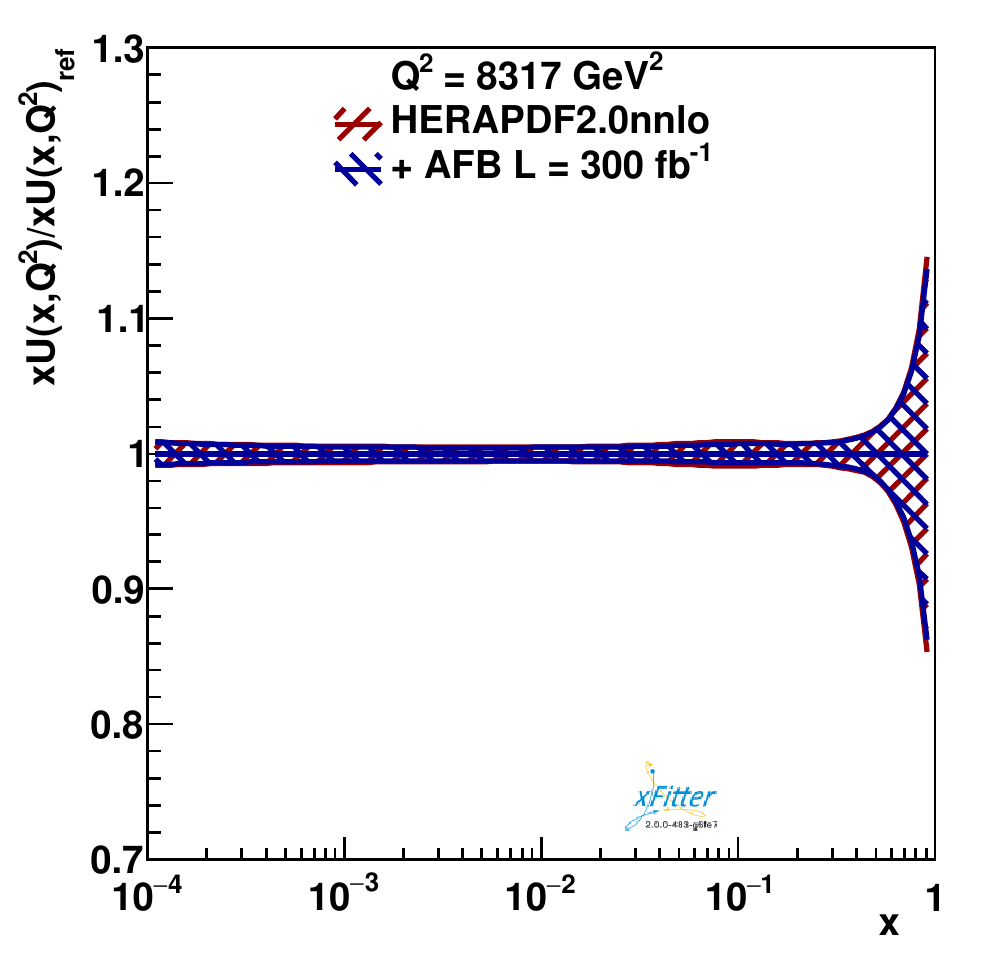}%
\includegraphics[width=0.25\textwidth]{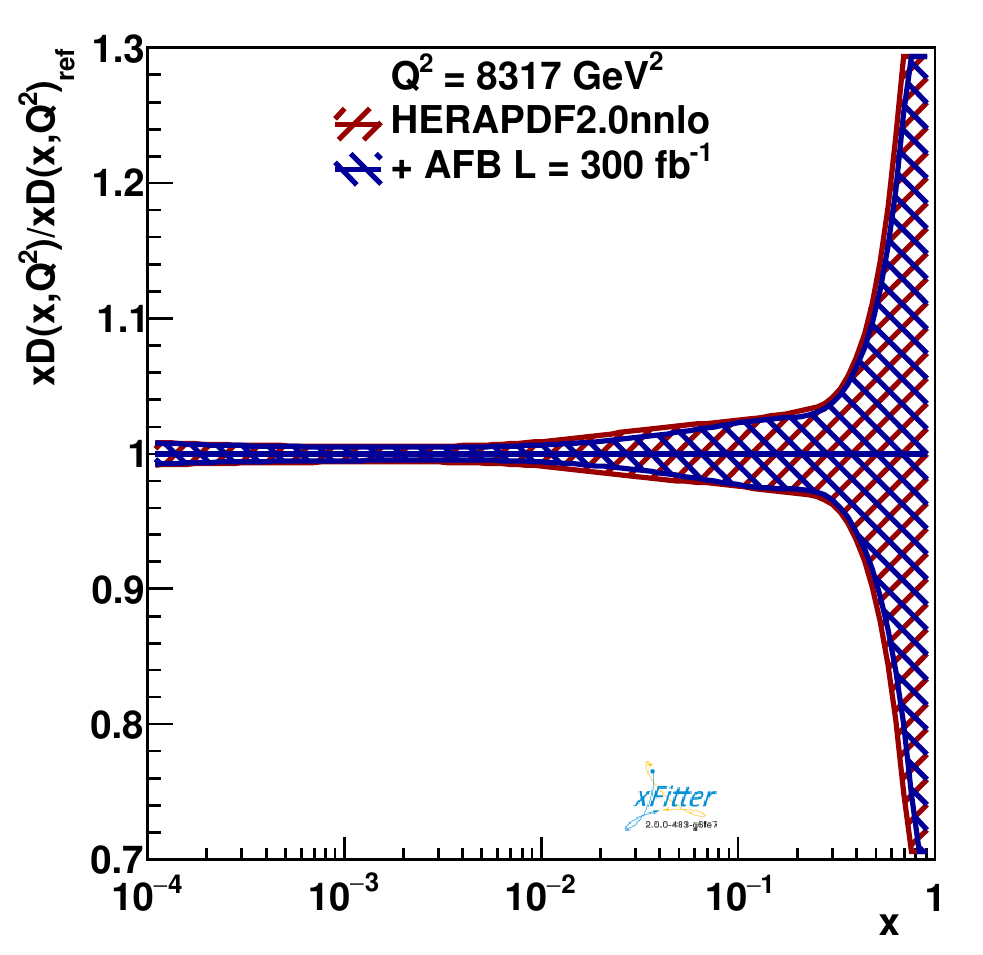}\\
\caption{Original (red) and profiled (blue) distributions for the normalised distribution of the ratios of (left to right) $u$-valence, $d$-valence, $u$-sea and $d$-sea quarks.
The profiled curves are obtained using $A_{\rm{FB}}^*$ pseudodata corresponding to an integrated luminosity of 300 fb$^{-1}$. Distributions are shown for the PDF sets (rows top to bottom) NNPDF3.1nnlo, MMHT2014nnlo, ABMP16nnlo and HERAPDF2.0nnlo.}
\label{fig:prof_PDFs}
\end{center}
\end{figure}

Fig.~\ref{fig:prof_PDFs} presents the results for the other PDF sets under analysis.
From top to bottom there are the profiling for the NNPDF3.1nnlo, MMHT2014nnlo, ABMP16nnlo, and HERAPDF2.0nnlo sets, obtained using pseudodata with an integrated luminosity of 300 fb$^{-1}$.

The NNPDF3.1nnlo set shows an intermediate sensitivity to the $A_{\rm{FB}}^*$ data.
The distributions that are more affected are those of the $u$-valence and $d$-valence quarks in the intermediate and small $x$ regions.
Also the $u$-sea distribution displays some sensitivity in the region of intermediate $x$.
The MMHT2014nnlo set appears as the least sensitive to the new data.
A mild improvement on the error bands is visible in the distribution of the $u$-valence, $d$-valence and $u$-sea quark distributions in the small $x$ region.
The ABMP16nnlo set is the most sensitive to $A_{\rm{FB}}^*$ data.
A remarkable improvement is visible especially in the distribution of the $d$-valence quark in the region of small to intermediate $x$.
A visible improvement is also obtained in the distribution of the $u$-valence quark, while the sea quarks are less affected.
In the HERAPDF2.0nnlo set, a noticeable reduction of the error bands is obtained for the valence quarks in the small and intermediate $x$ regions, while the sea quarks appear not as sensitive to the new data.

In the following we study the effects on the profiling from the application of low rapidity cuts on the data.
Since this procedure in general reduces the amount of data, thus increasing the statistical uncertainty of the measurements, we carry out the following analysis adopting an integrated luminosity of 3000 fb$^{-1}$, and we select a PDF set which showed an intermediate sensitivity to the $A_{\rm{FB}}^*$ data, such as the HERAPDF2.0nnlo set.
For an exhaustive discussion on the differences between the various PDF sets, we refer to the PDF reviews in Ref.~\cite{Butterworth:2015oua,Accardi:2016ndt}.

\begin{figure}[h]
\begin{center}
\includegraphics[width=0.33\textwidth]{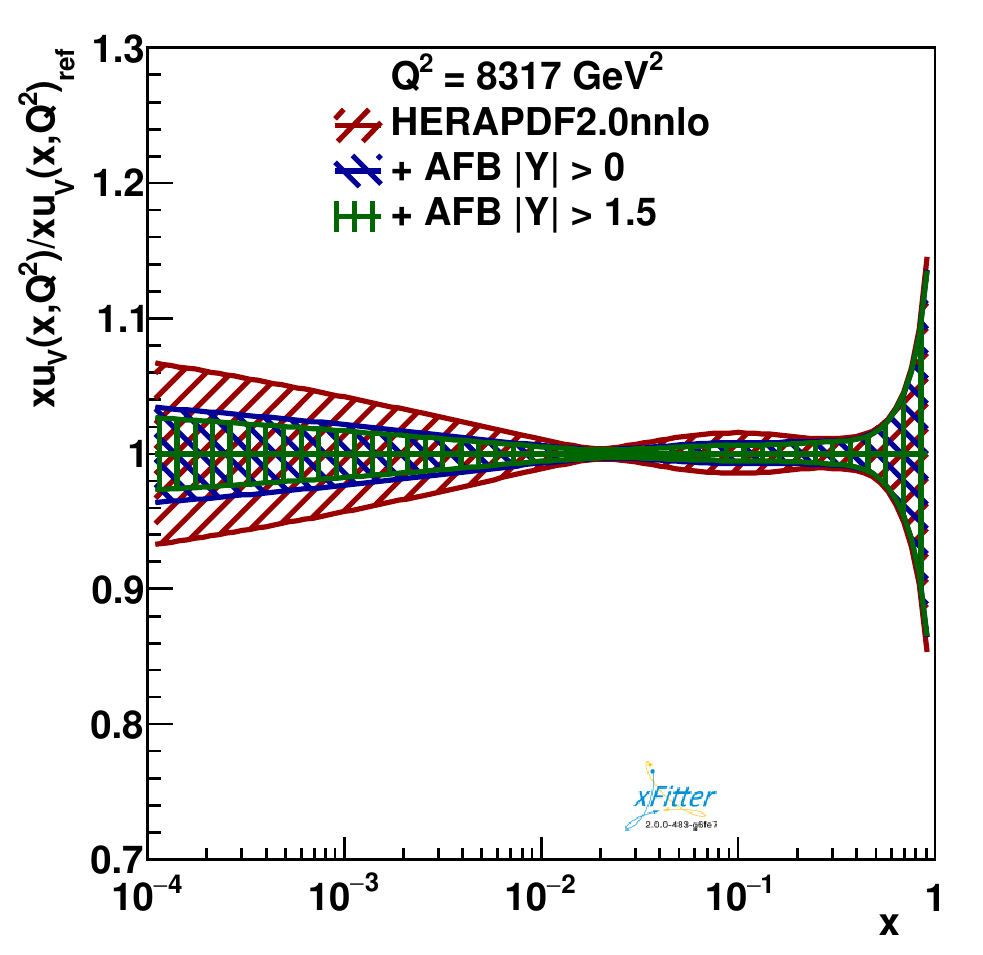}%
\includegraphics[width=0.33\textwidth]{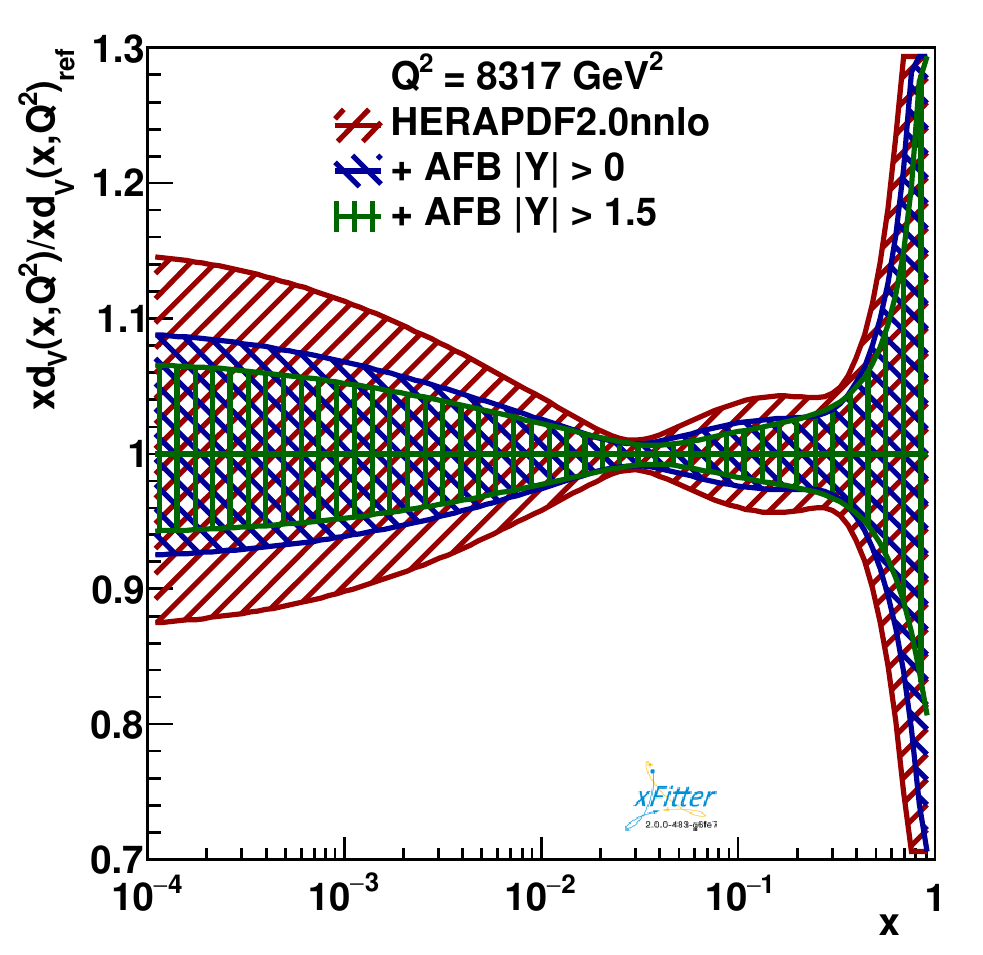}%
\includegraphics[width=0.33\textwidth]{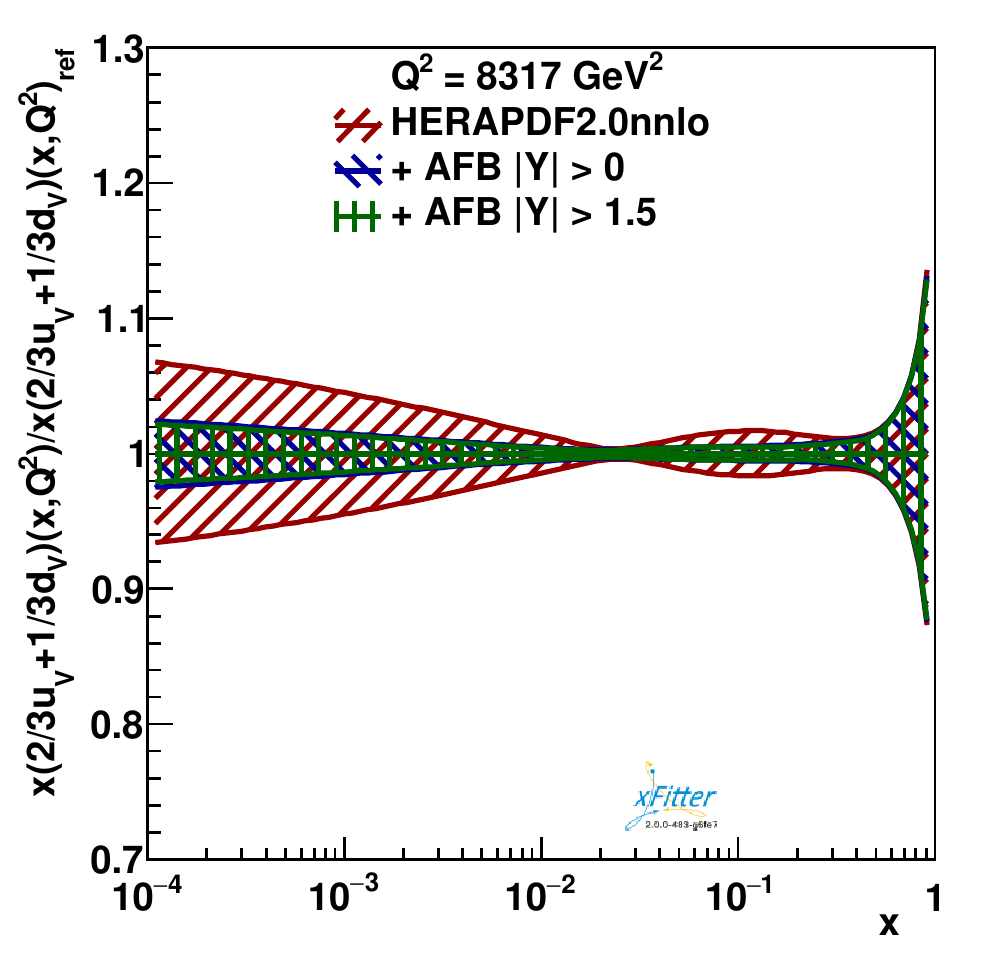}\\
\includegraphics[width=0.33\textwidth]{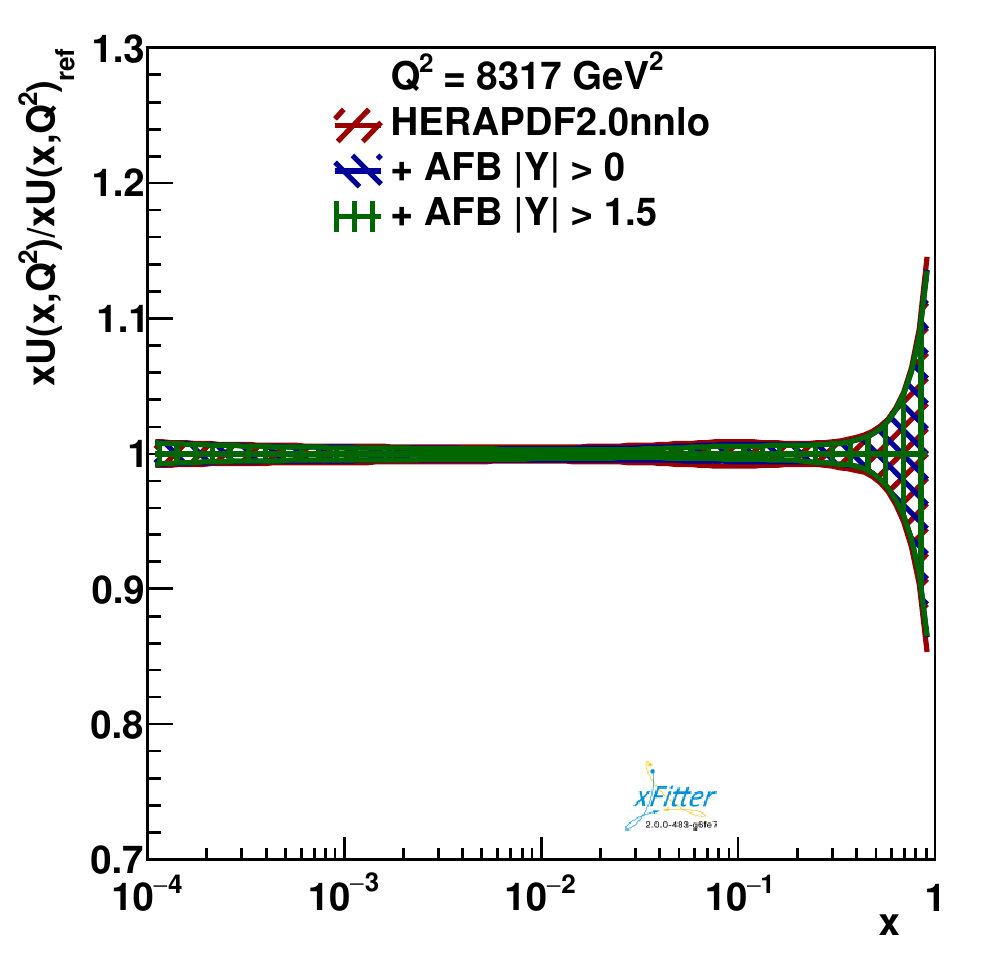}%
\includegraphics[width=0.33\textwidth]{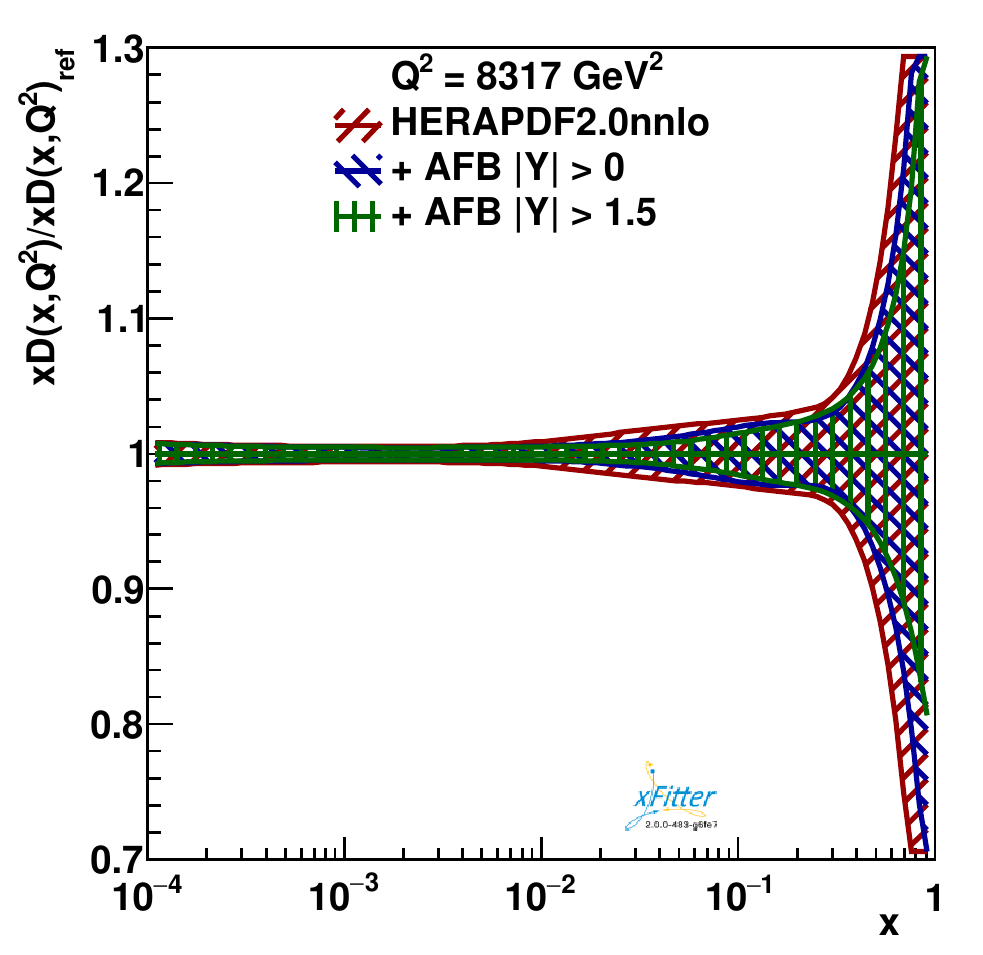}%
\includegraphics[width=0.33\textwidth]{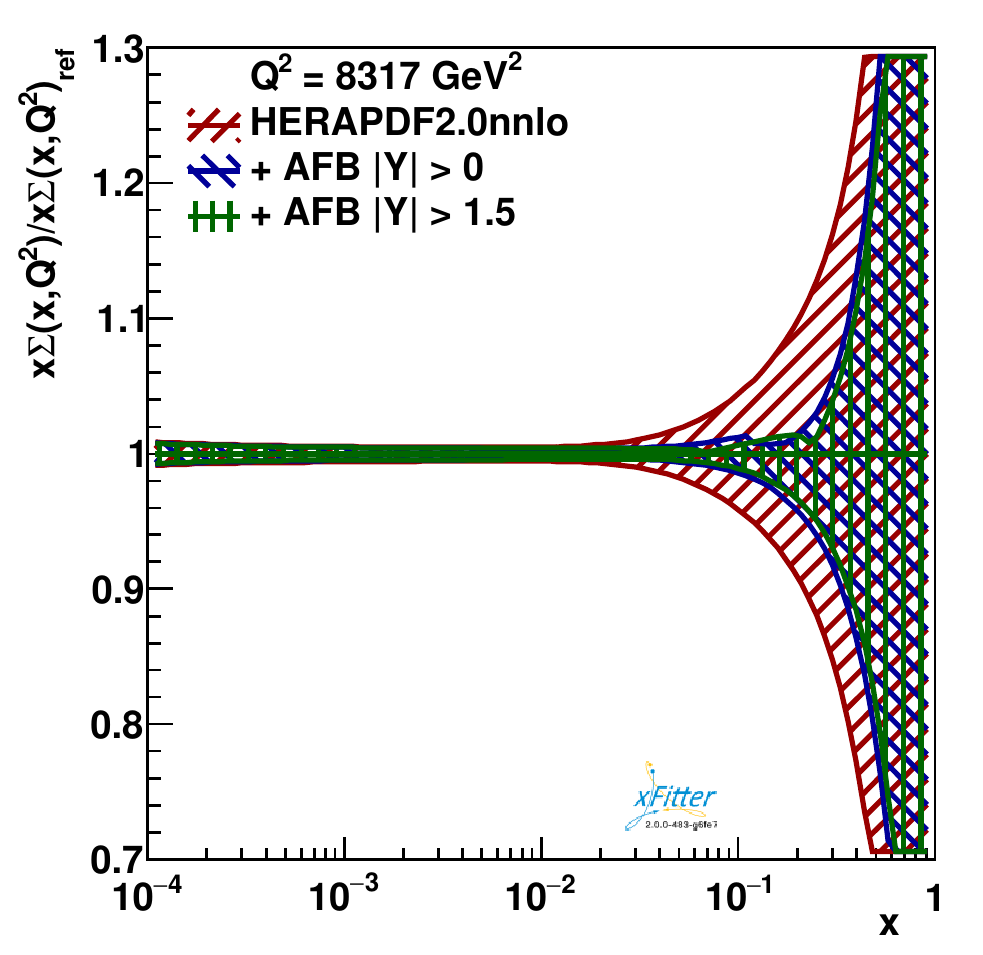}\\
\caption{Original (red) and profiled distributions for the normalised distribution of the ratios of (top row, left to right) $u$-valence, $d$-valence and $((2/3)u+(1/3)d)$-valence and (bottom row, left to right) $u$-sea, $d$-sea quarks and $(u+d)$-sea quarks of the HERAPDF2.0nnlo PDF set obtained using $A_{\rm{FB}}^*$ pseudodata corresponding to an integrated luminosity of 3000 fb$^{-1}$, applying rapidity cuts of $|y_{\ell\ell}| > 0$ (blue) and $|y_{\ell\ell}| > 1.5$ (green).}
\label{fig:prof_HERA_Y_15}
\end{center}
\end{figure}

\begin{figure}[h]
\begin{center}
\includegraphics[width=0.33\textwidth]{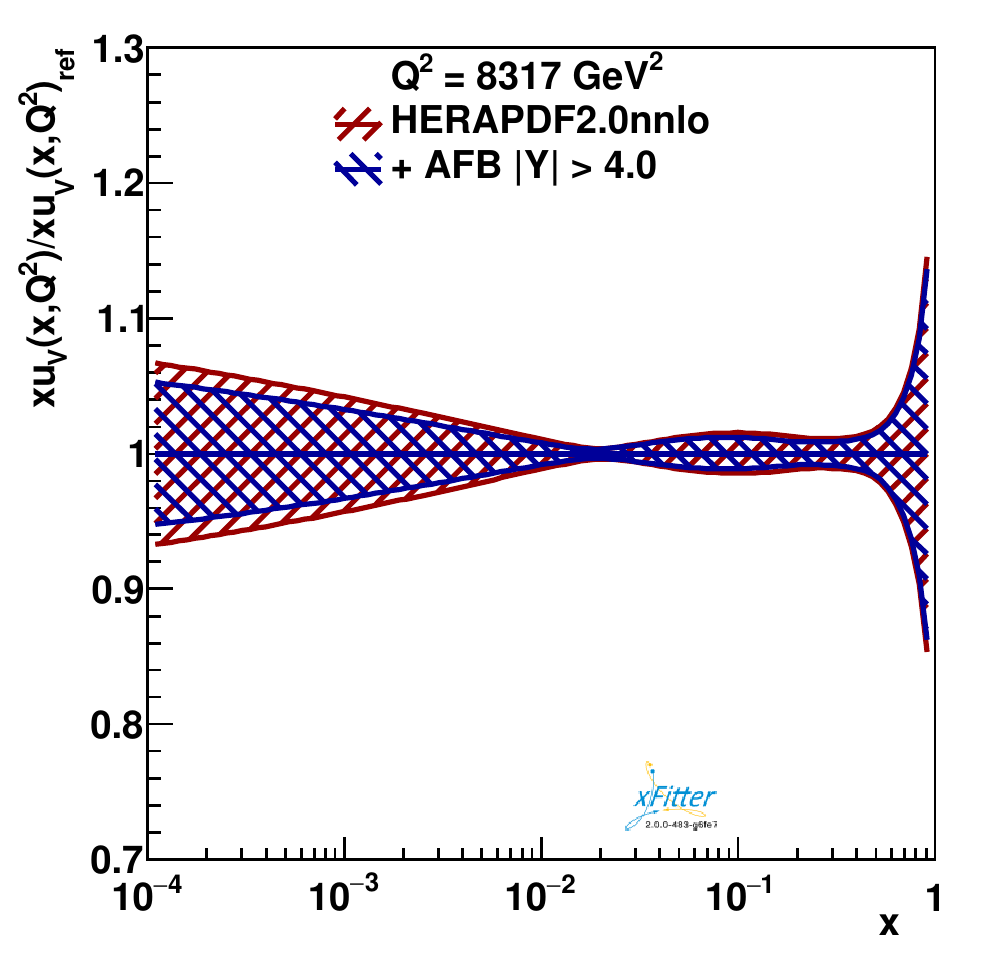}%
\includegraphics[width=0.33\textwidth]{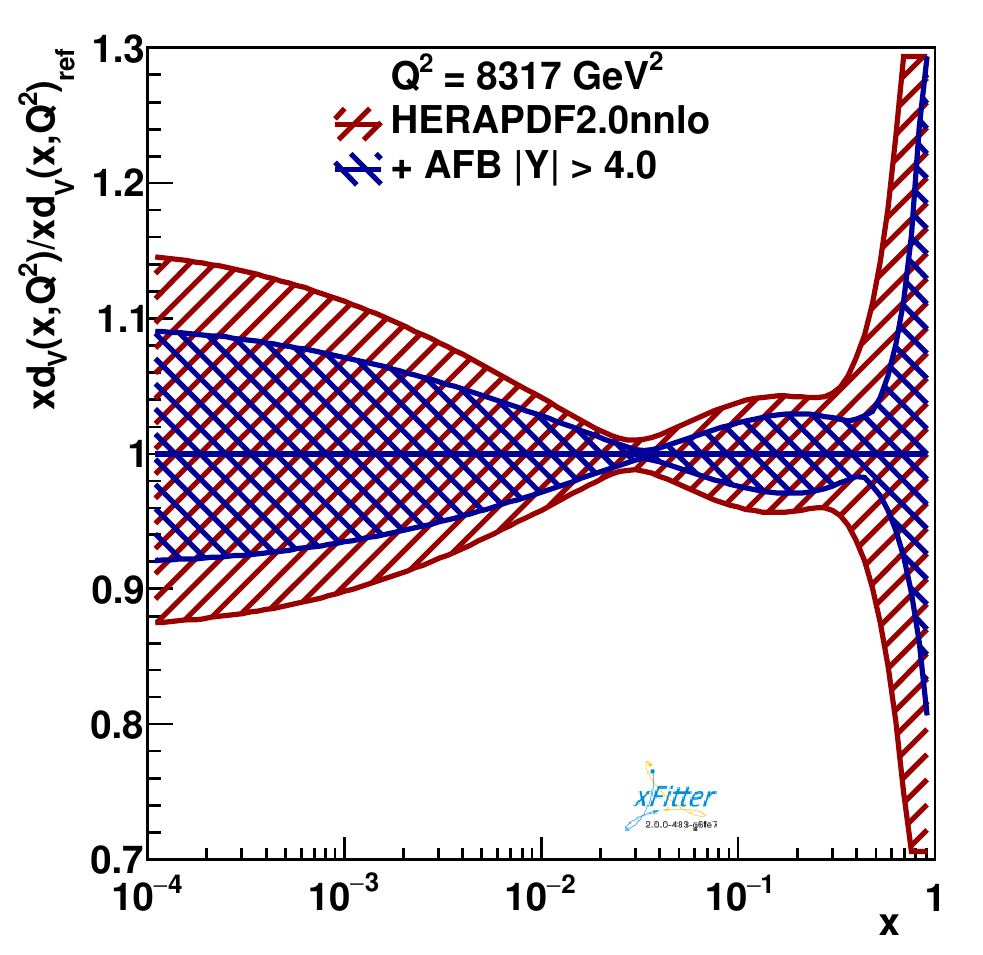}%
\includegraphics[width=0.33\textwidth]{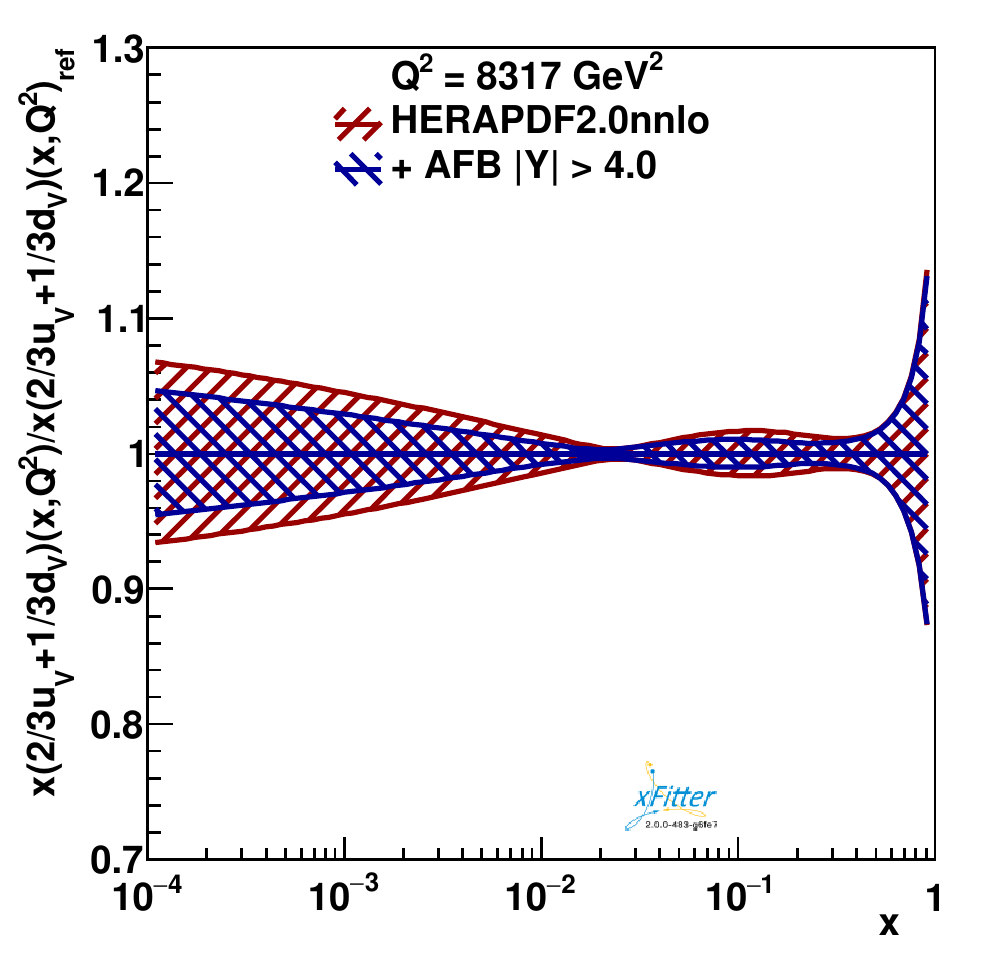}\\
\includegraphics[width=0.33\textwidth]{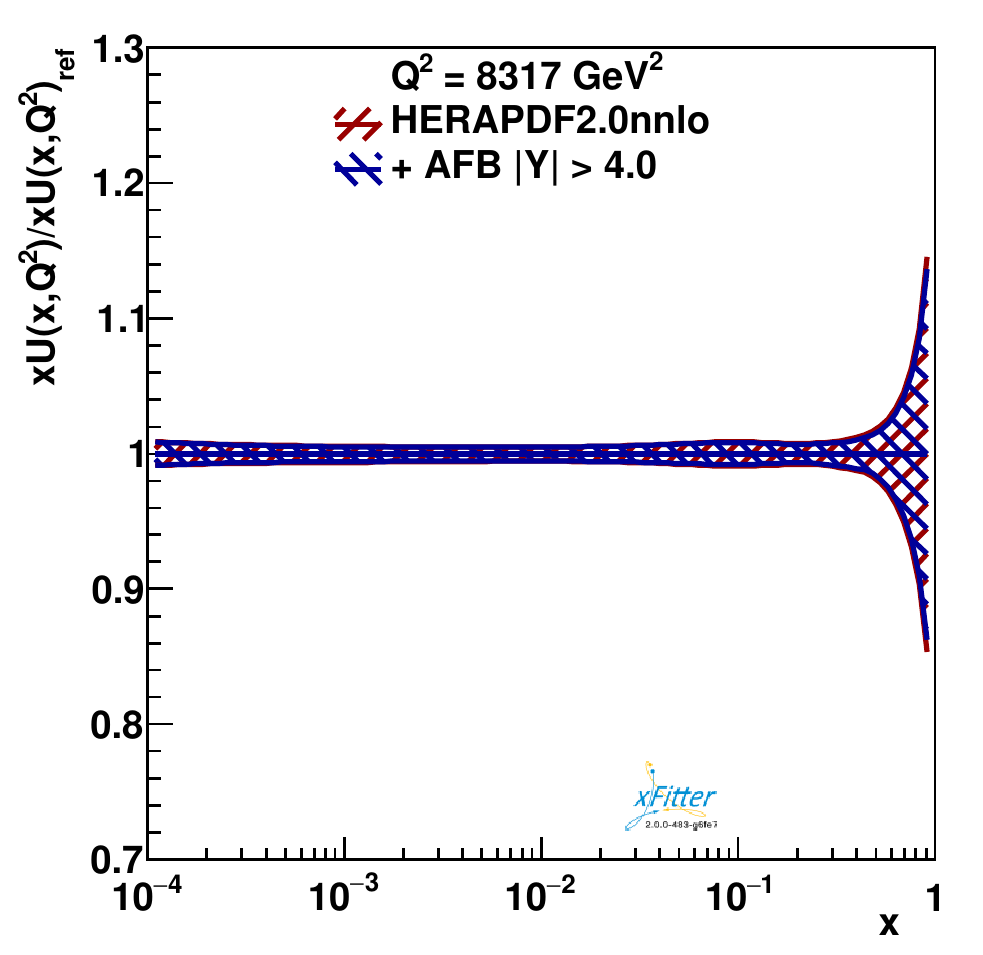}%
\includegraphics[width=0.33\textwidth]{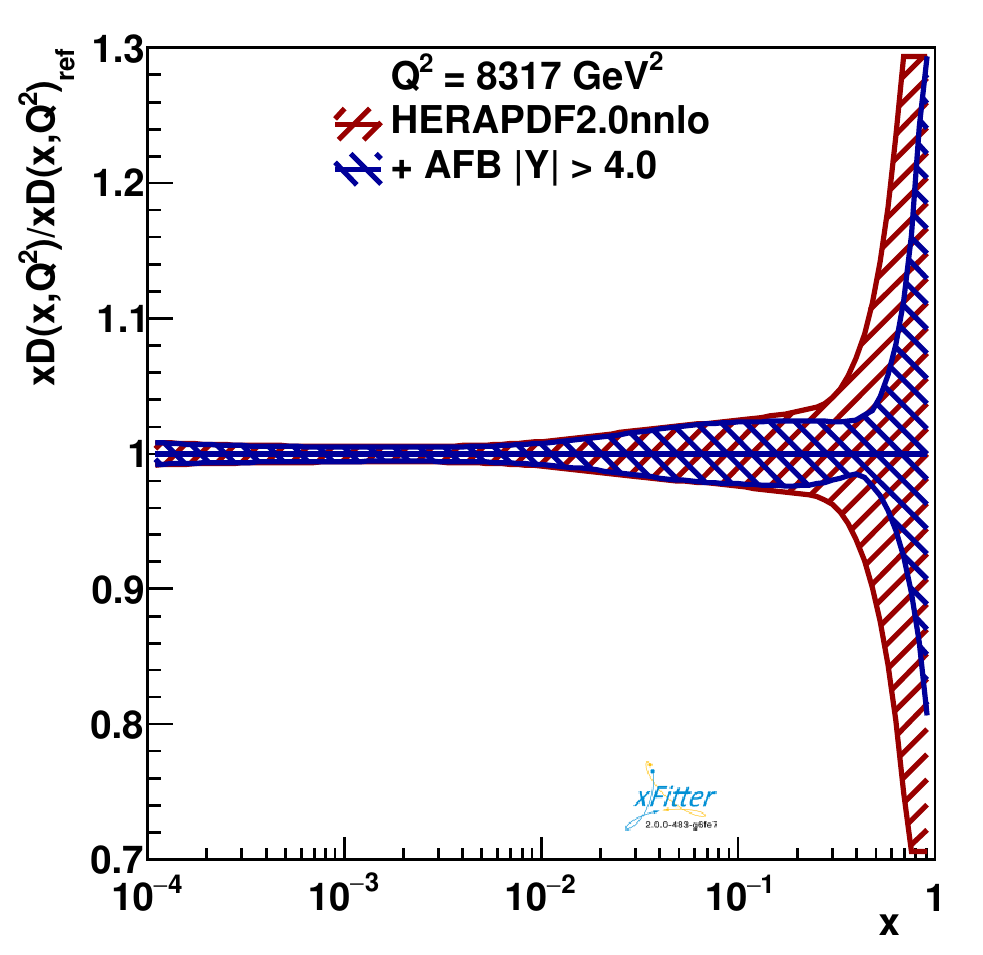}%
\includegraphics[width=0.33\textwidth]{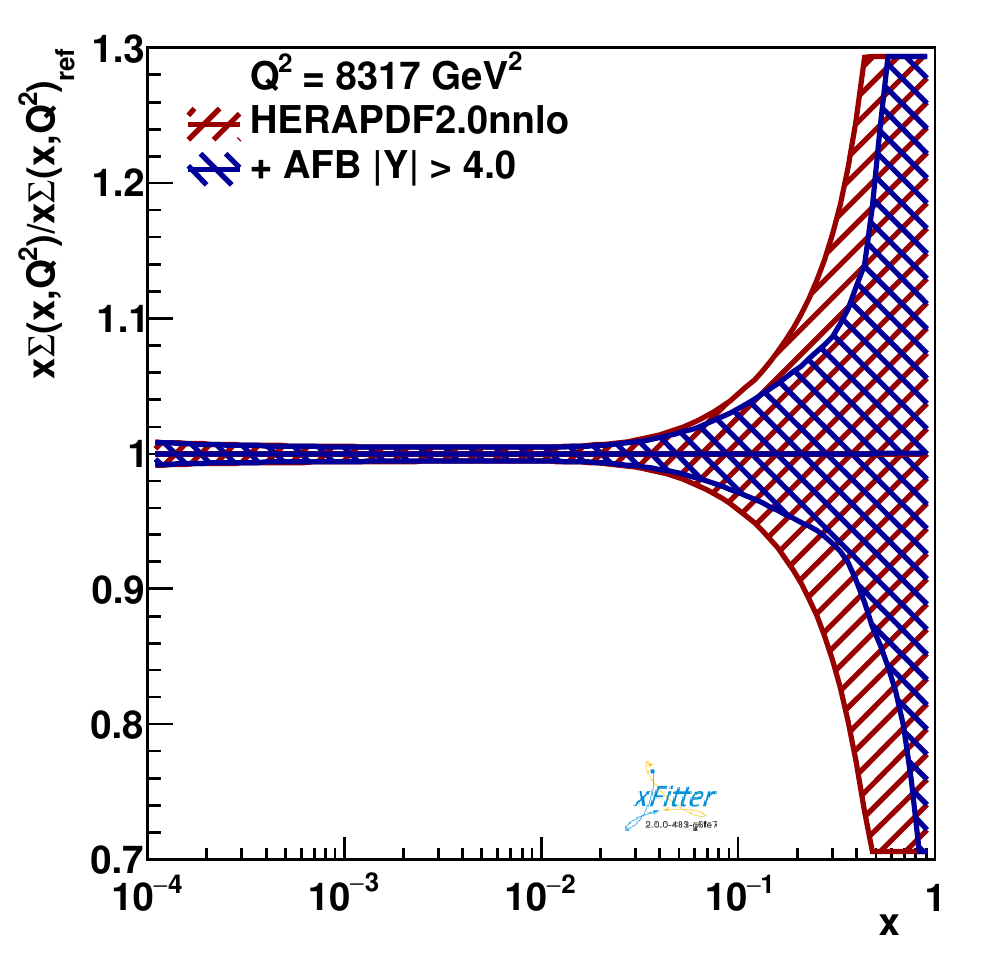}\\
\caption{Original (red) and profiled (blue) distributions for the normalised distribution of the ratios of (top row, left to right) $u$-valence, $d$-valence and $((2/3)u+(1/3)d)$-valence and (bottom row, left to right) $u$-sea, $d$-sea quarks and $(u+d)$-sea quarks of the HERAPDF2.0nnlo PDF set obtained using $A_{\rm{FB}}^*$ pseudodata corresponding to an integrated luminosity of 3000 fb$^{-1}$, applying a rapidity cut of $|y_{\ell\ell}| > 4.0$.
The acceptance region of the detector has been enlarged up to $|\eta_\ell| < 5$, and the profiling is performed through the LO code.}
\label{fig:prof_HERA_Y_4}
\end{center}
\end{figure}

In Figs.~\ref{fig:prof_HERA_Y_15} and~\ref{fig:prof_HERA_Y_4} are presented the effects on the profiling when imposing rapidity cuts on the pseudodata.
Comparing those profiled error bands, we note some improvement in the distribution of the $d$-valence quark, especially in the region of small $x$.
A visible reduction of the error bands can also be appreciated in the distribution of both $u$-sea and $d$-sea quarks in the region of intermediate $x$.

In Fig.~\ref{fig:prof_HERA_Y_4} we instead consider the profiling obtained when imposing a rapidity cut $|y_{\ell\ell}| > 4.0$ on the data.
In order to analyse this scenario, which probes the very high rapidity region, we need to enlarge the acceptance region of the detector.
Experimentally it is possible to explore pseudorapidity regions up to $|\eta_\ell| < 5$ in the di-electron channel.
However, in this case the experimental analysis requires that at least one lepton falls in the usual acceptance region $|\eta_\ell| < 2.5$~\cite{Aaboud:2017ffb}.
We drop this requirement and we impose instead a symmetric acceptance cut $|\eta_\ell| < 5$ on both leptons.
The profiled curves in this case have been obtained by means of the LO code implemented into {\tt{xFitter}}, while the pseudodata contains 120 bins of 1 GeV covering the invariant mass region $80~{\rm GeV} < M_{\ell\ell} < 200~{\rm GeV}$.
In the curve obtained in this scenario we notice how the reduced statistics due to the phase space cut leads to an overall poorer profiling compared to the previous cases.
Conversely, in this setup the reduction of uncertainty is concentrated in the region of high $x$, which was not accessible before.
The high rapidity cut indeed forces more asymmetric combination of $x_1$ and $x_2$ of the incoming interacting partons, such that one parton has to lie in the high $x$ region while the other in the small $x$ region, as it was already pointed out in Ref.~\cite{Accomando:2018nig,Accomando:2017scx}.
In particular, we observe a remarkable improvement on the distribution of $d$-valence and $d$-sea quarks in the high $x$ region.

\subsection{Eigenvectors rotation}

In this section we want to determine the PDFs (and their combinations) which are more sensitive to the $A_{\rm{FB}}^*$ data.
We perform a reparameterisation of the eigenvectors of selected PDF sets~\cite{Pumplin:2009nm}. 
The new set of eigenvectors will be the result of a rotation of the original set, and they will be sorted according to their sensitivity to the new data.
We have performed this exercise on two sets with Hessian eigenvectors: the CT14nnlo and HERAPDF2.0nnlo PDFs. 

In Tab.~\ref{tab:rot_CT14_HERA} are shown the $\chi^2$/d.o.f. (degrees of freedom) values for the rotated eigenvectors of the two PDF sets. 
The larger the number, the stronger the effect of the new data on the eigenvector. Clearly, the first two eigenvectors of the CT14nnlo and HERAPDF2.0nnlo PDF sets, which correspond to one pair of asymmetric Hessian uncertainties, are the most sensitive to the $A_{\rm{FB}}$ data.

\begin{figure}[h]
\begin{center}
\includegraphics[width=0.33\textwidth]{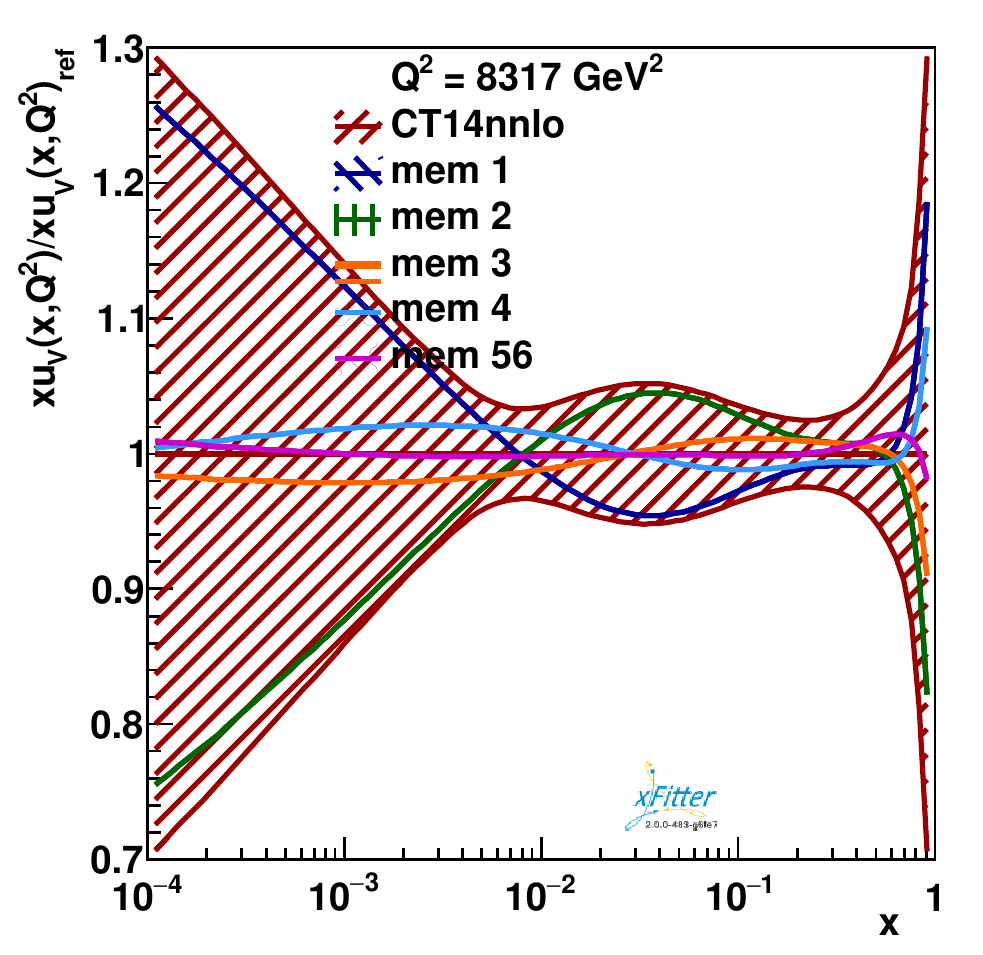}%
\includegraphics[width=0.33\textwidth]{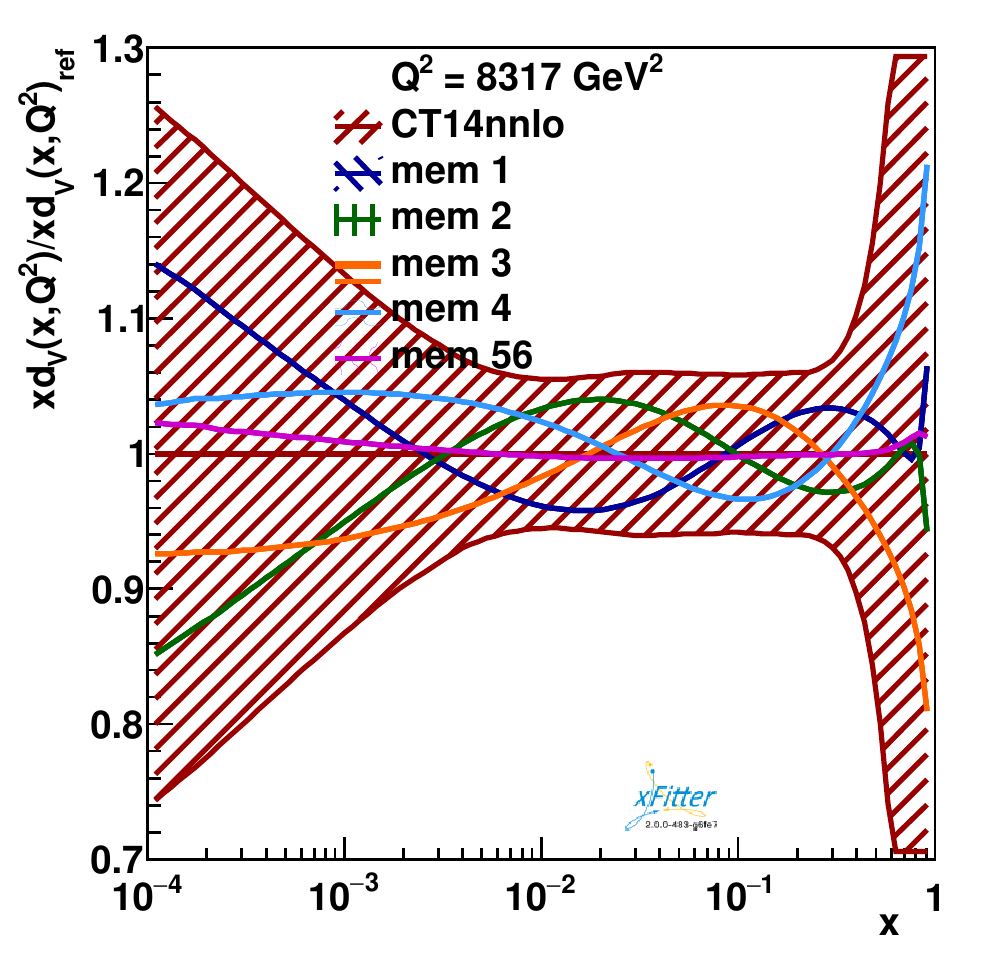}%
\includegraphics[width=0.33\textwidth]{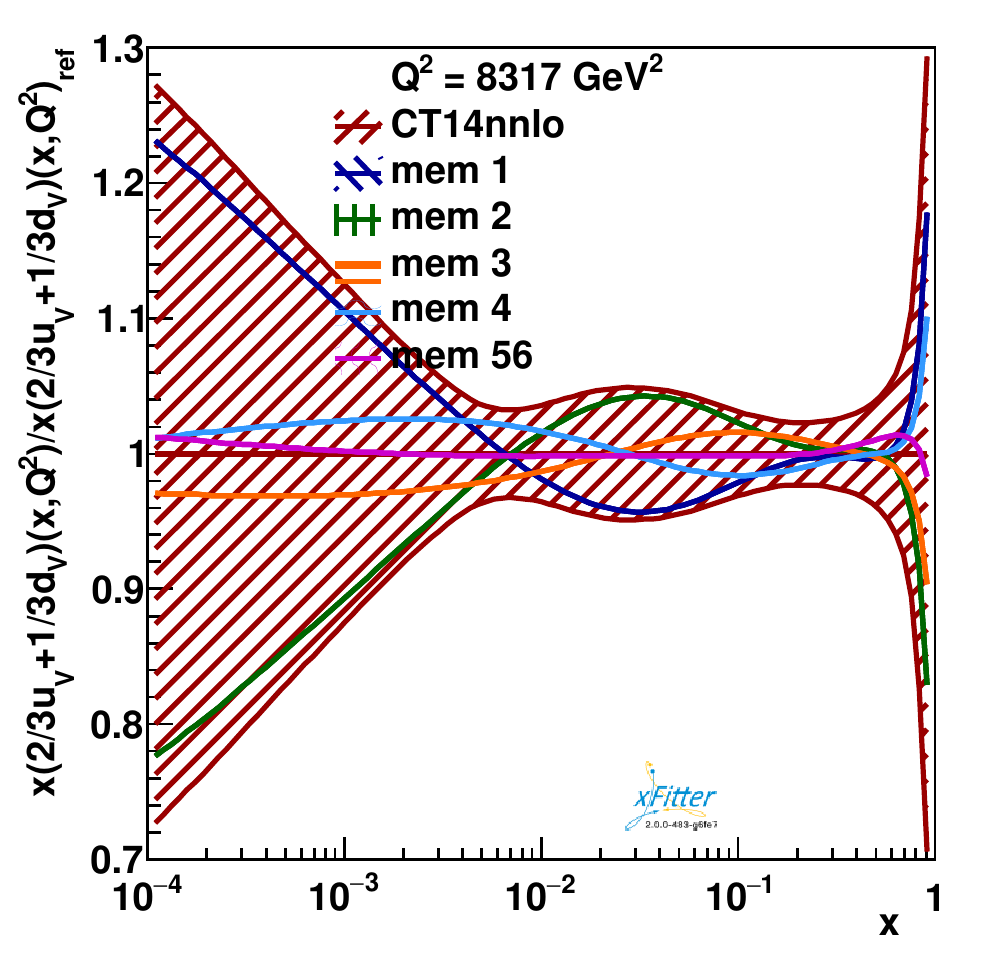}\\%
\includegraphics[width=0.33\textwidth]{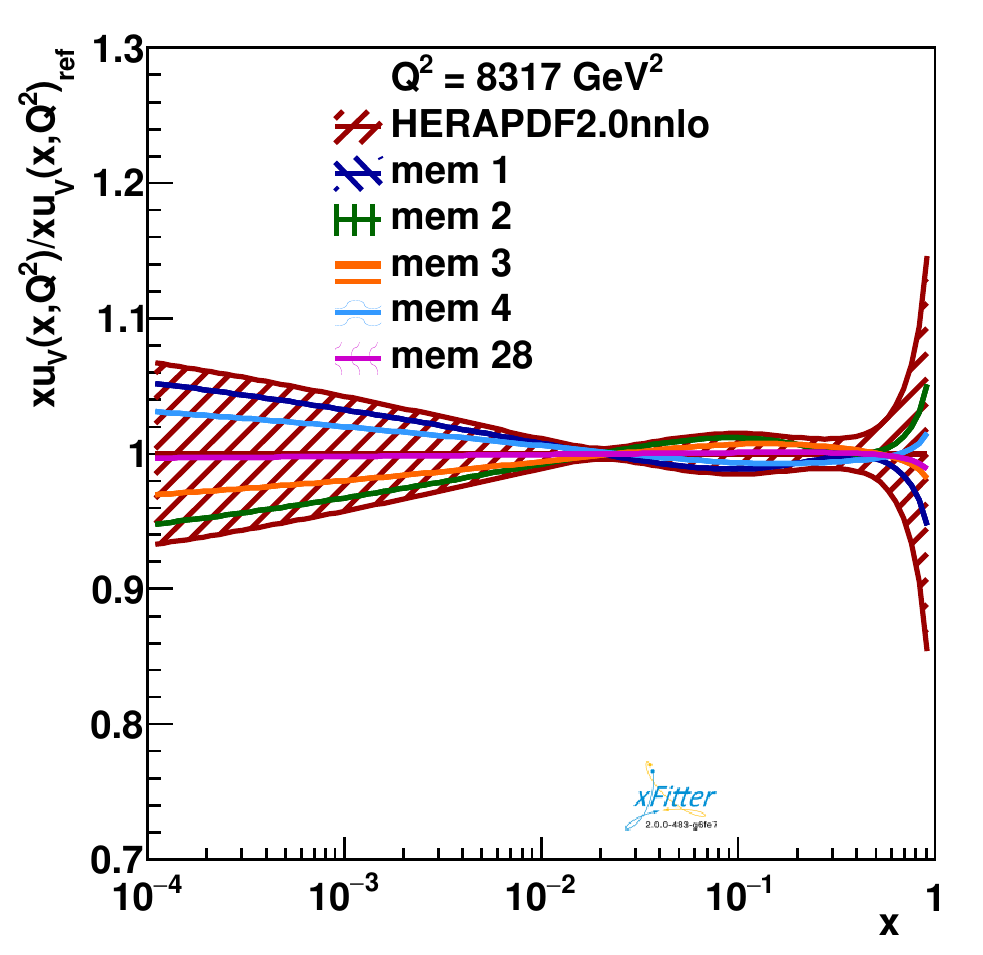}%
\includegraphics[width=0.33\textwidth]{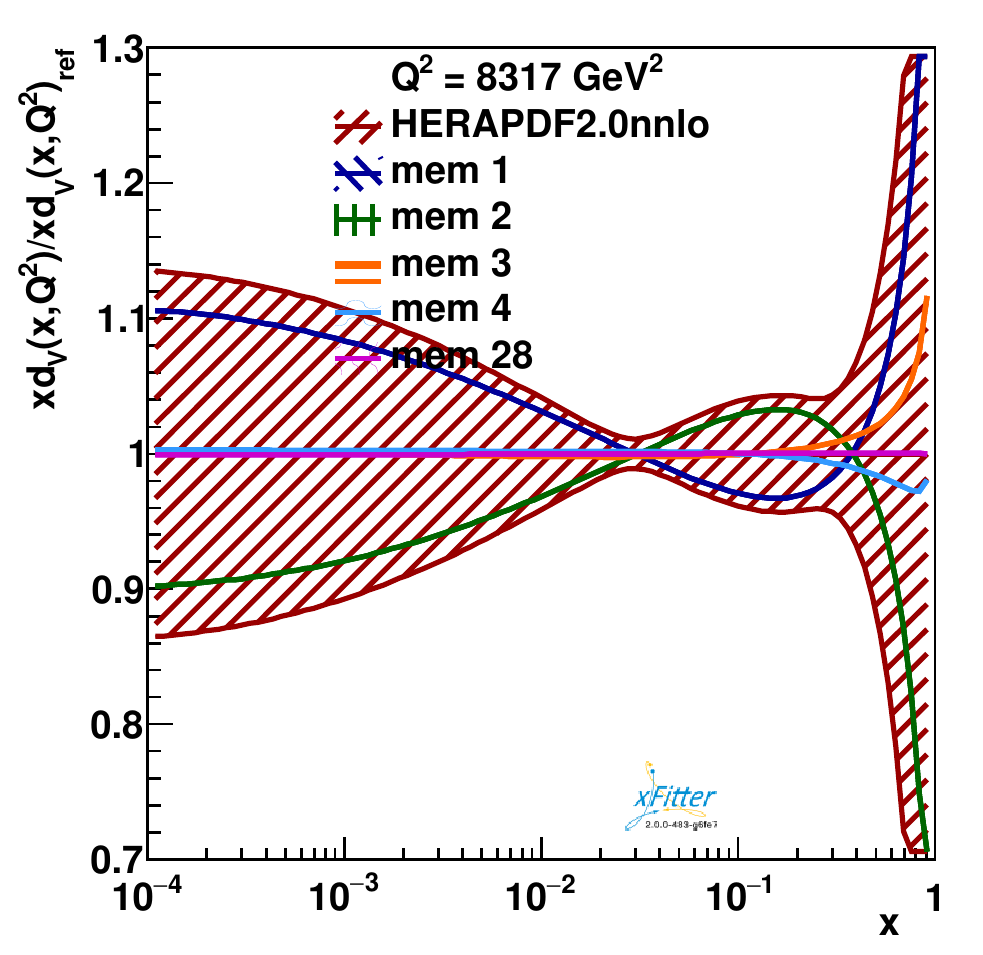}%
\includegraphics[width=0.33\textwidth]{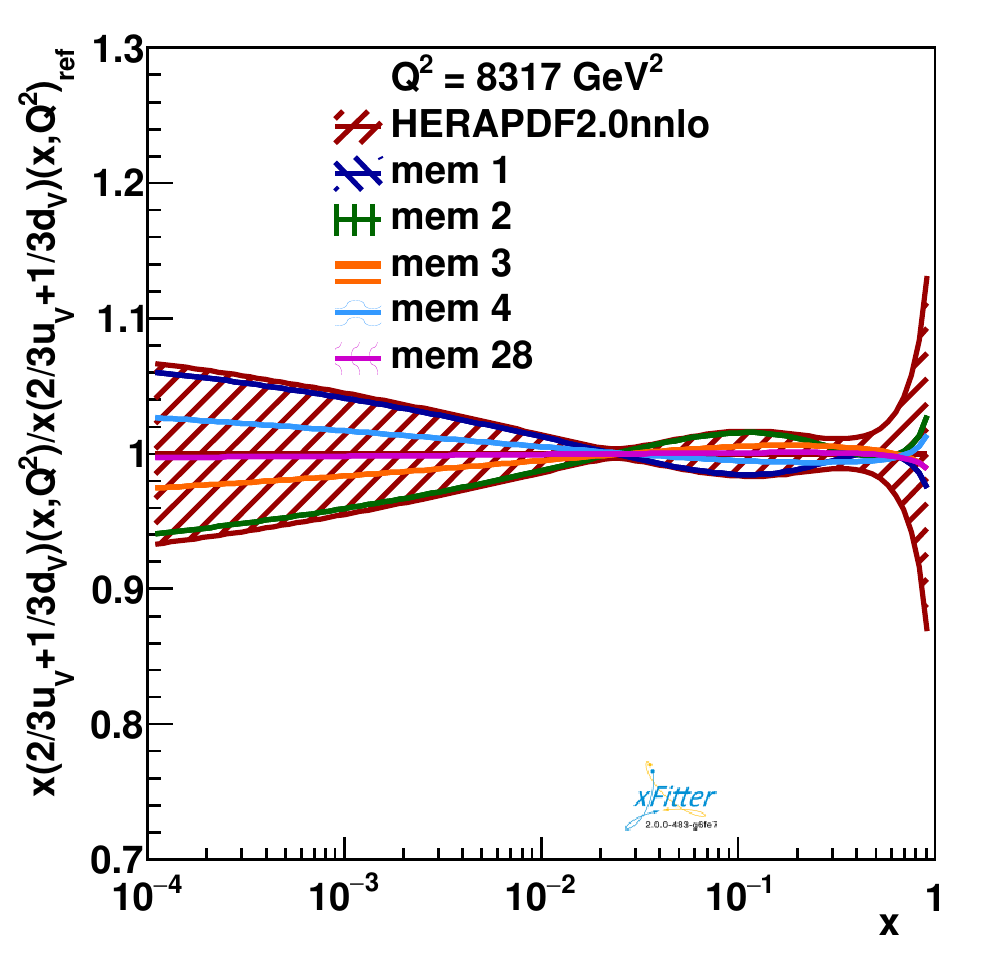}%
\caption{Contribution of the first four and last rotated eigenvectors to the uncertainty error bands of the normalised distribution of the ratios of (left to right) $u$-valence, $d$-valence and $((2/3)u+(1/3)d)$-valence of the CT14nnlo (top row) and HERAPDF2.0nnlo (bottom row) PDF sets.
The eigenvectors are rotated and sorted according to their sensitivity to $A_{\rm{FB}}^*$ pseudodata corresponding to an integrated luminosity of 300 fb$^{-1}$.}
\label{fig:rot_eigen}
\end{center}
\end{figure}

\begin{table}
\begin{center}
\begin{tabular}{|c|c|c|c|c|c|}
  \hline
  \textbf{CT14nnlo} & mem1 & mem2 & mem3 & mem4 & mem56\\
  \hline
  Total $\chi^2$/d.o.f. & 164/106 & 169/106 & 10/106 & 14/106 & 0.98/106\\
  \hline
  \hline
  \textbf{HERAPDF2.0nnlo} & mem1 & mem2 & mem3 & mem4 & mem28\\
  \hline
  Total $\chi^2$/d.o.f. & 4.8/106 & 8.0/106 & 0.48/106 & 0.74/106 & 0.01/106\\
  \hline
\end{tabular}
\caption{The $\chi^2$ table for the CT14nnlo and HERAPDF2.0nnlo sets with rotated eigenvectors.}
\label{tab:rot_CT14_HERA}
\end{center}
\end{table}

Now we plot the contribution of the rotated first four and last eigenvectors to the error bands of the valence quark distributions and their sum.
The results are displayed in Figs.~\ref{fig:rot_eigen} for the two PDF sets.
We observe that the first two eigenvectors almost completely determine the error bands for the distribution of the $u$-valence and $d$-valence quarks and their sum.
In particular we observe that $u$-valence and $d$-valence eigenvectors are very correlated and the $A_{\rm{FB}}^*$ data will constrain their charge weighted sum $\frac{2}{3}u_V(x,Q^2) + \frac{1}{3}d_V(x,Q^2)$.
This is in contrast to CC lepton asymmetry data which at LO constrain instead the combination $u_V - d_V$~\cite{Harland-Lang:2014zoa}.
In the light of these results, we conclude that $A_{\rm{FB}}^*$ data will mostly constrain the distribution of the valence quarks and this outcome is in agreement with the results presented in the previous section.

We conclude this section by noting that, following the observation made in  Refs.~\cite{Accomando:2018nig,Accomando:2017scx}, in which detailed comparisons were made between statistical errors and PDF errors for various scenarios with different selection cuts and luminosities, in this paper we have obtained for the first time quantitative results for the reduction of PDF uncertainties from using the $A_{\rm FB}$ asymmetry, and we have identified the charge-weighted sum of $u$-valence and $d$-valence PDFs as the distribution which is most sensitive to $A_{\rm FB}$. We arrived at this result by analyzing the structure of the axial and vector couplings in the part of the differential DY cross section which contributes to the asymmetry, and this has been confirmed by the explicit numerical exercise of eigenvectors rotation carried out in this section.  

We present further new results analyzing theoretical and systematic uncertainties in the next section: in particular, the correlation between different choices of factorization and renormalization scales  in the forward and backward regions, and the impact on PDFs from the most accurate LEP/SLD $\theta_W$ measurement and global fit of EW parameters.

\section{Theoretical and systematic uncertainties on the $A^*_{\rm FB}$ predictions}
\label{sec:Sys}
In this section we discuss the dependence of the $A^*_{\rm{FB}}$ observable on the most important sources of theoretical uncertainty.
We first check the theoretical uncertainty from the choice of factorisation ($\mu_F$) and renormalisation ($\mu_R$) scales.
For this purpose we employed the ``seven points" method, which considers the predictions obtained for the combinations obtained with a relative factor no larger than two between the two scales, from $\mu_{F,R}/M_{\ell\ell} = 0.5$ to $\mu_{F,R}/M_{\ell\ell} = 2.0$.

\begin{figure}[h]
\begin{center}
\includegraphics[width=0.46\textwidth]{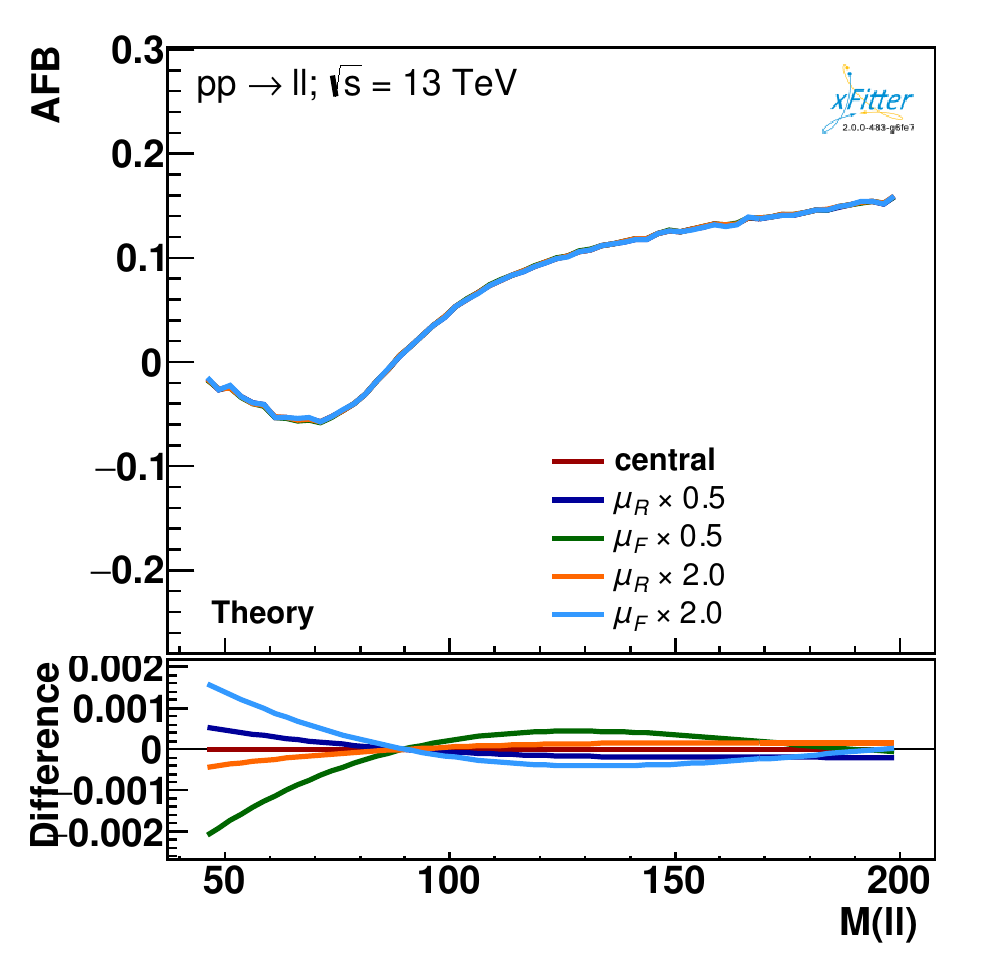}{(a)}
\includegraphics[width=0.46\textwidth]{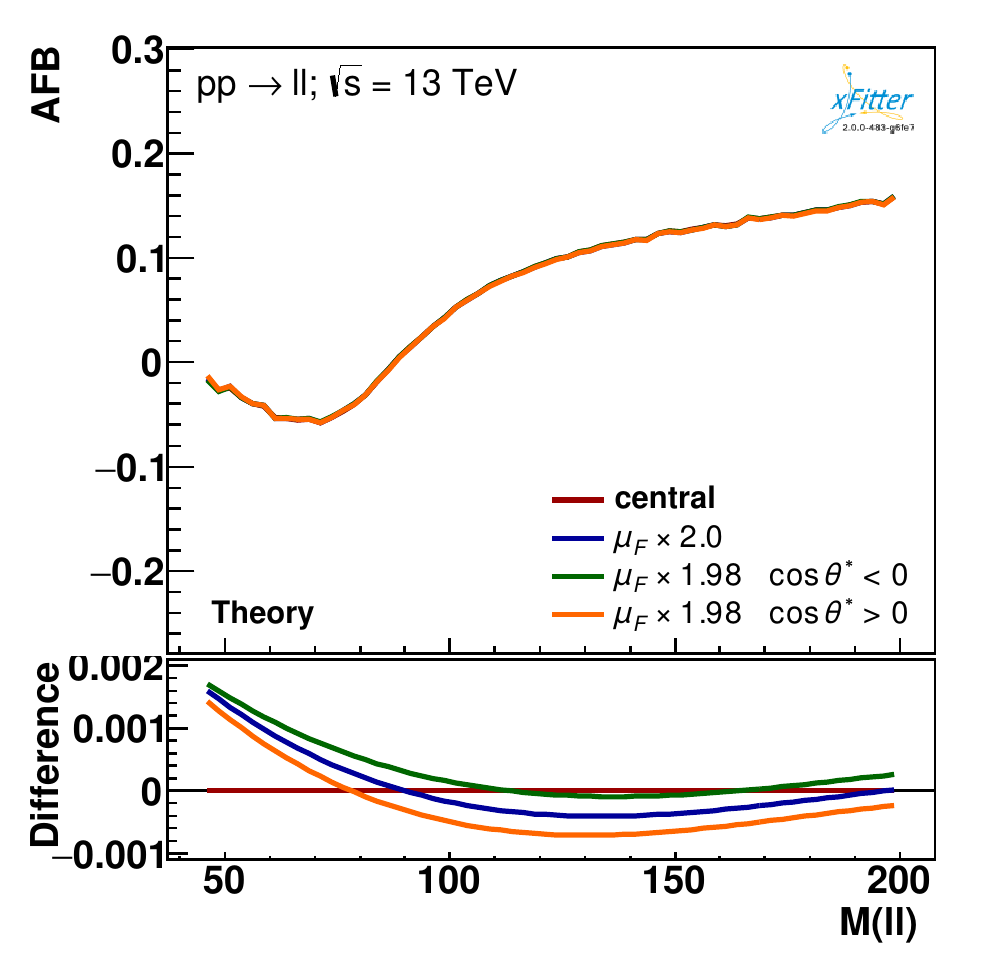}{(b)}
\caption{(a) $A_{\rm{FB}}$ predictions obtained with the HERAPDF2.0nnlo PDF set for some combinations of factorisation and renormalisation scales (top panel) and their deviations with respect to the central curve with $\mu_{F,R}/M_{\ell\ell} = 1.0$ (bottom panel). (b) $A_{\rm{FB}}$ predictions for the curve with $\mu_{F}/M_{\ell\ell} = 2.0, \mu_{R}/M_{\ell\ell} = 1.0$ in both APPLgrids with $\cos\theta^* > 0$ and $\cos\theta^* < 0$ (blue line) and with $\mu_{F}/M_{\ell\ell} = 2.0, \mu_{R}/M_{\ell\ell} = 1,0$ for $\cos\theta^* > 0$ and $\mu_{F}/M_{\ell\ell} = 1.98, \mu_{R}/M_{\ell\ell} = 1.0$ for $\cos\theta^* < 0$ (green line) and with $\mu_{F}/M_{\ell\ell} = 1.98, \mu_{R}/M_{\ell\ell} = 1.0$ for $\cos\theta^* > 0$ and $\mu_{F}/M_{\ell\ell} = 2.0, \mu_{R}/M_{\ell\ell} = 1.0$ for $\cos\theta^* < 0$ (orange line) (top panel) and their deviations with respect to the central curve with $\mu_{F,R}/M_{\ell\ell} = 1.0$ (bottom panel).
For presentation purposes the curves in the bottom panels are smoothed using a cubic polynomial function.}
\label{fig:AFB_scale}
\end{center}
\end{figure}

For this exercise, we have employed the HERAPDF2.0nnlo PDF set.
The predictions for the $A^*_{\rm{FB}}$ and their deviation with respect the baseline represented by the ``central" ($\mu_{F,R}/M_{\ell\ell} = 1.0$) are visible in Fig.~\ref{fig:AFB_scale}(a).
Here we have omitted the curves with $\mu_{F,R}/M_{\ell\ell} = 0.5$ and $\mu_{F,R}/M_{\ell\ell} = 2.0$ since they produced the smallest variations with respect to the baseline.

In Fig.~\ref{fig:AFB_scale}(b) are shown instead the predictions for the $A_{\rm{FB}}$ with factorisation scale $\mu_{F}/M_{\ell\ell} = 2.0$ and and renormalisation scale $\mu_{R}/M_{\ell\ell} = 1.0$, and the corresponding curves when the factorisation scale is chosen differently in the phase space regions with $\cos\theta^* > 0$ and $\cos\theta^* < 0$.
In the bottom panel are also shown their differences.

It is worth to mention that recently a dedicated study on the errors in the PDFs propagating from the missing higher order uncertainty, and an extensive discussion on the analysis of scale variations to quantify their weight has been proposed in Ref.~\cite{AbdulKhalek:2019ihb}.

Another source of uncertainty lies in the employed value of the Weinberg mixing angle.
The most accurate measurement comes from LEP and SLD data~\cite{ALEPH:2005ab} and gives an absolute error $\Delta\sin^2\theta_W = 16\times 10^{-5}$, while the most precise estimate is obtained from a global fit of EW parameters~\cite{Haller:2018nnx} resulting in the uncertainty $\Delta\sin^2\theta_W = 6\times 10^{-5}$.
The deviations of the $A^*_{\rm{FB}}$ observable due to the variations of $\sin^2\theta_W$ are generally small when compared to statistical or the other systematical uncertainties, however, they lead to visible differences in the PDF central values.
Again, we use the HERAPDF2.0nnlo PDF set to estimate this effect.
In the invariant mass region under analysis, using predictions obtained at LO, we obtain $|\Delta A_{\rm{FB}}| < 10^{-4}$ when including the error from LEP and SLD measurements or $|\Delta A_{\rm{FB}}| < 4 \times10^{-5}$ when employing the uncertainty from the global EW fit.

\begin{figure}[h]
\begin{center}
\includegraphics[width=0.33\textwidth]{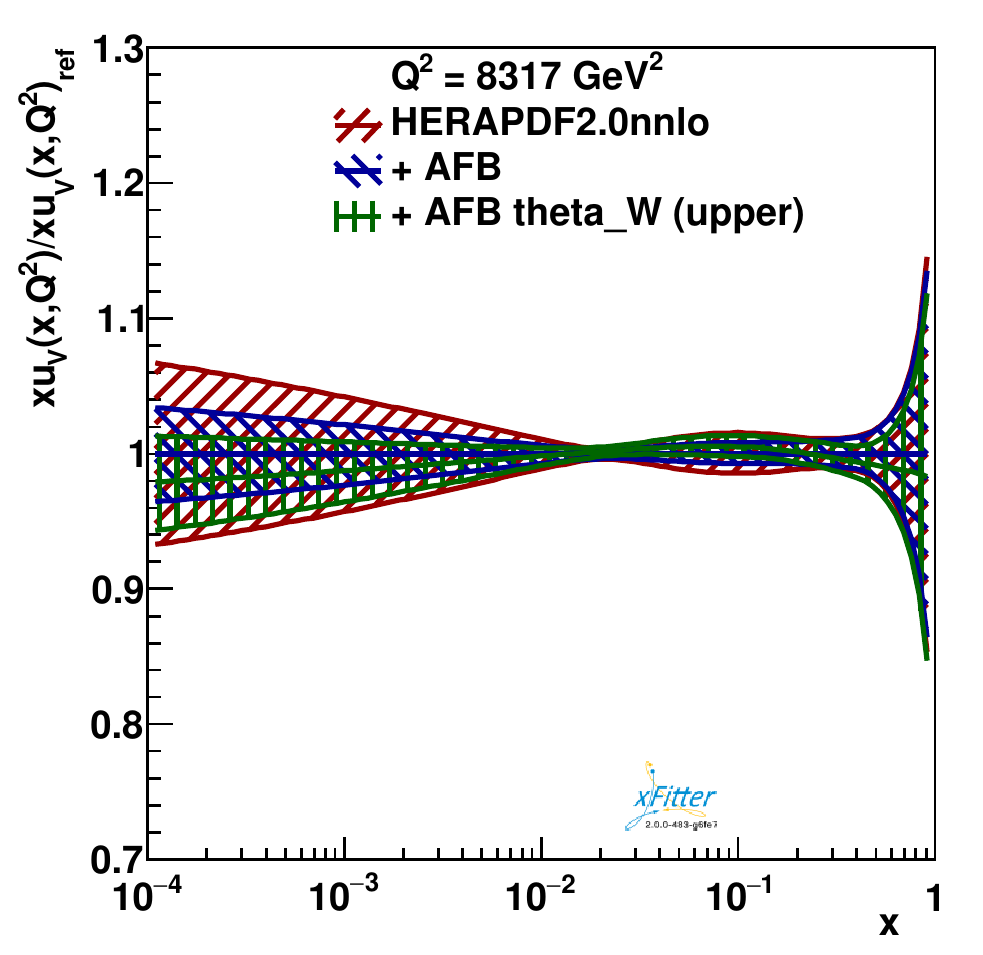}%
\includegraphics[width=0.33\textwidth]{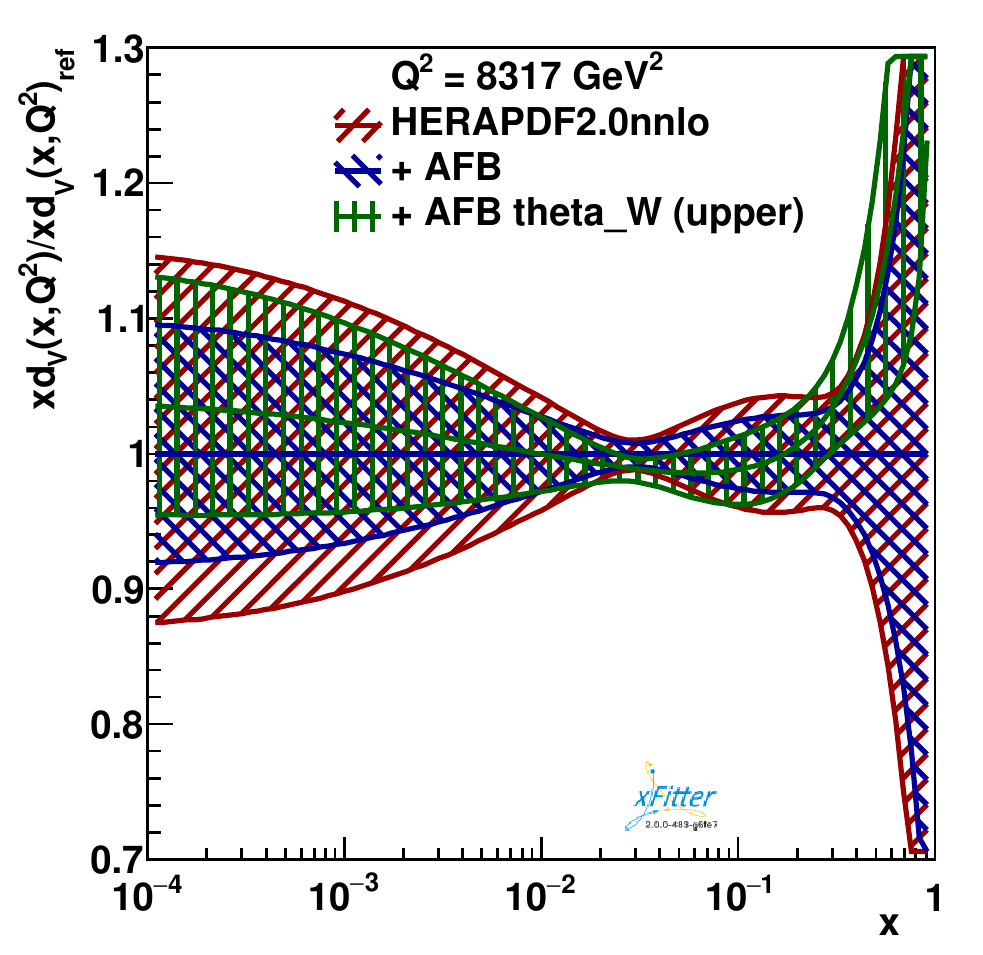}%
\includegraphics[width=0.33\textwidth]{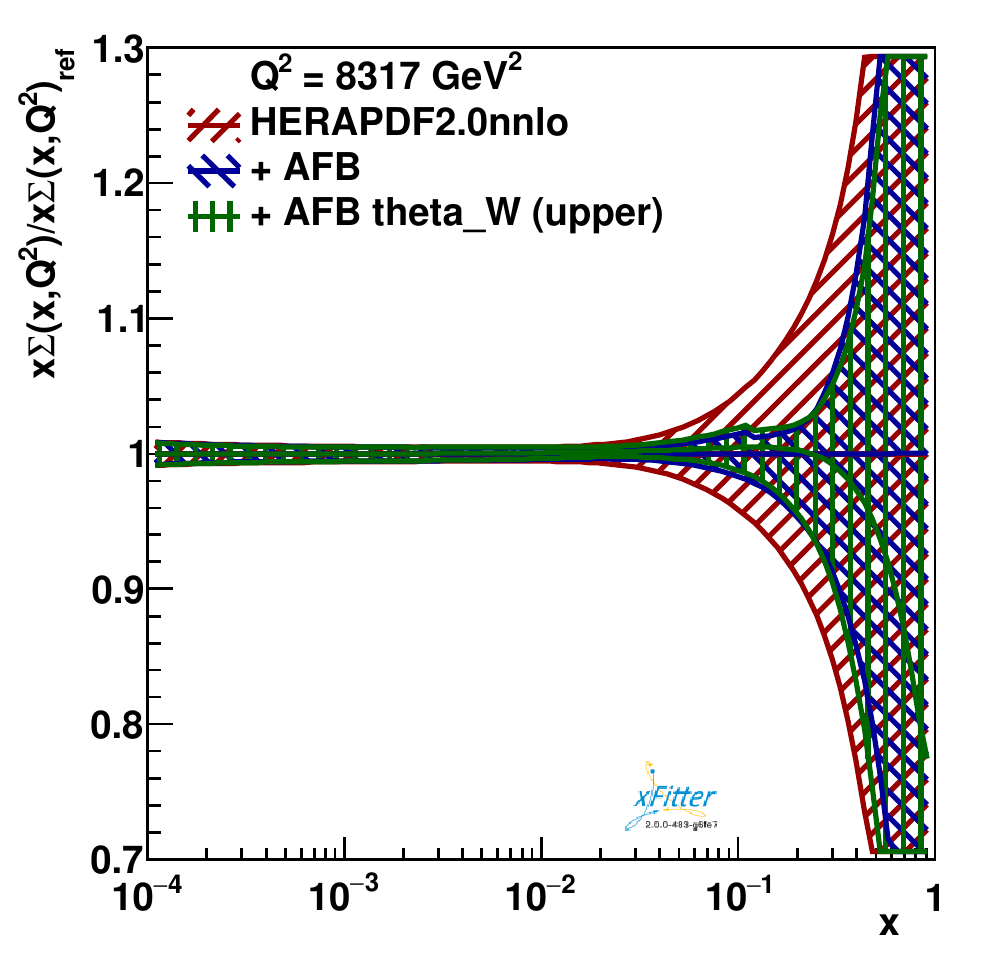}\\%
\includegraphics[width=0.33\textwidth]{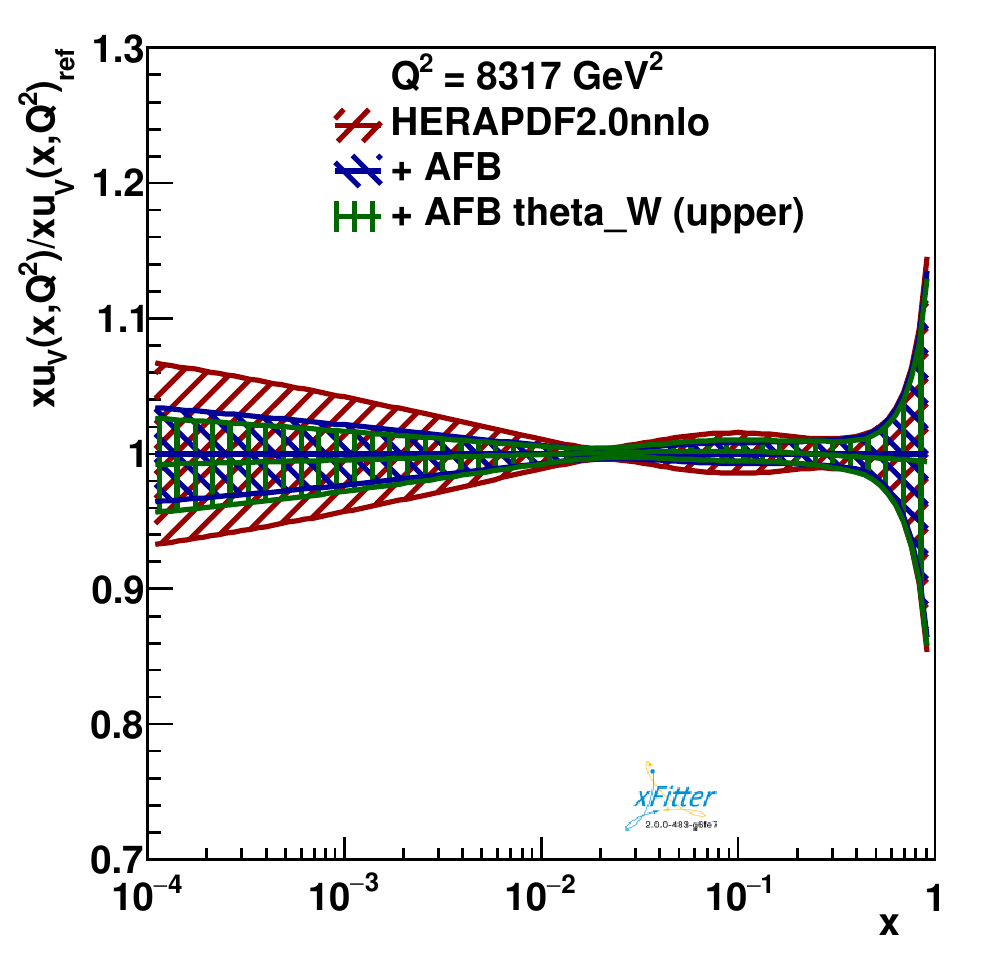}%
\includegraphics[width=0.33\textwidth]{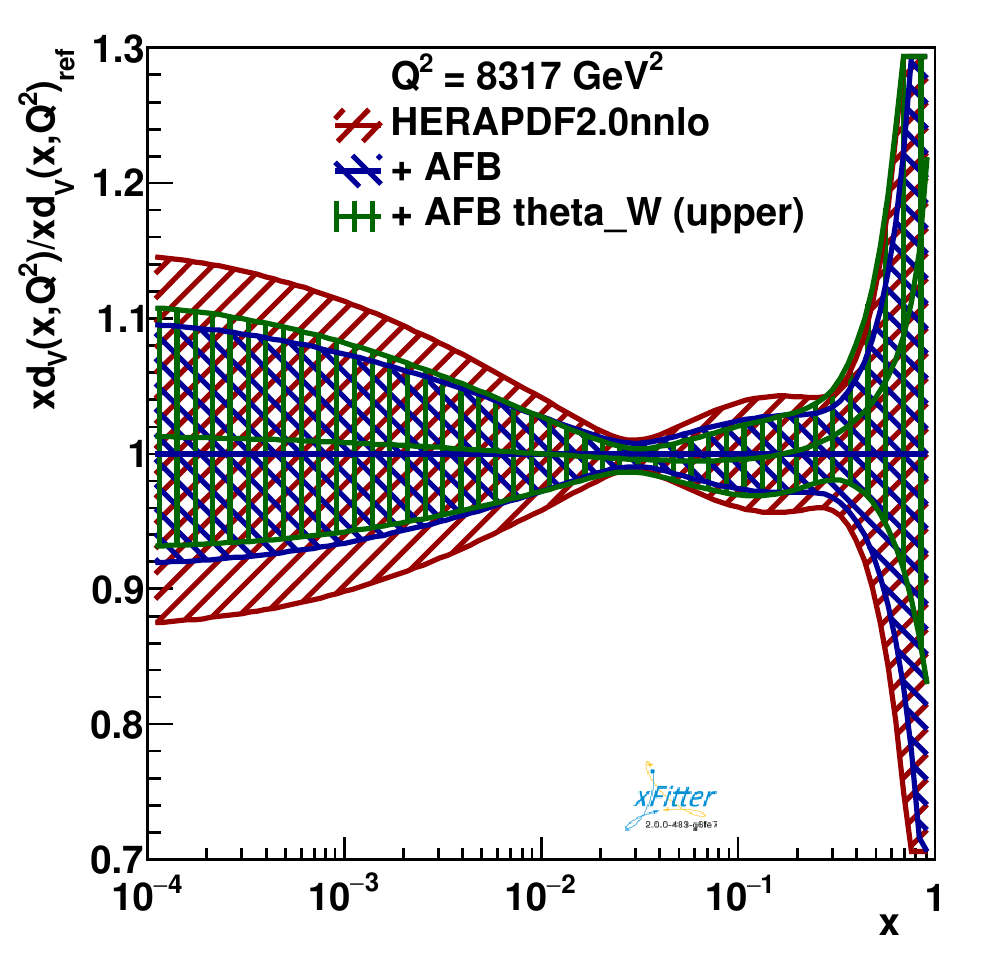}%
\includegraphics[width=0.33\textwidth]{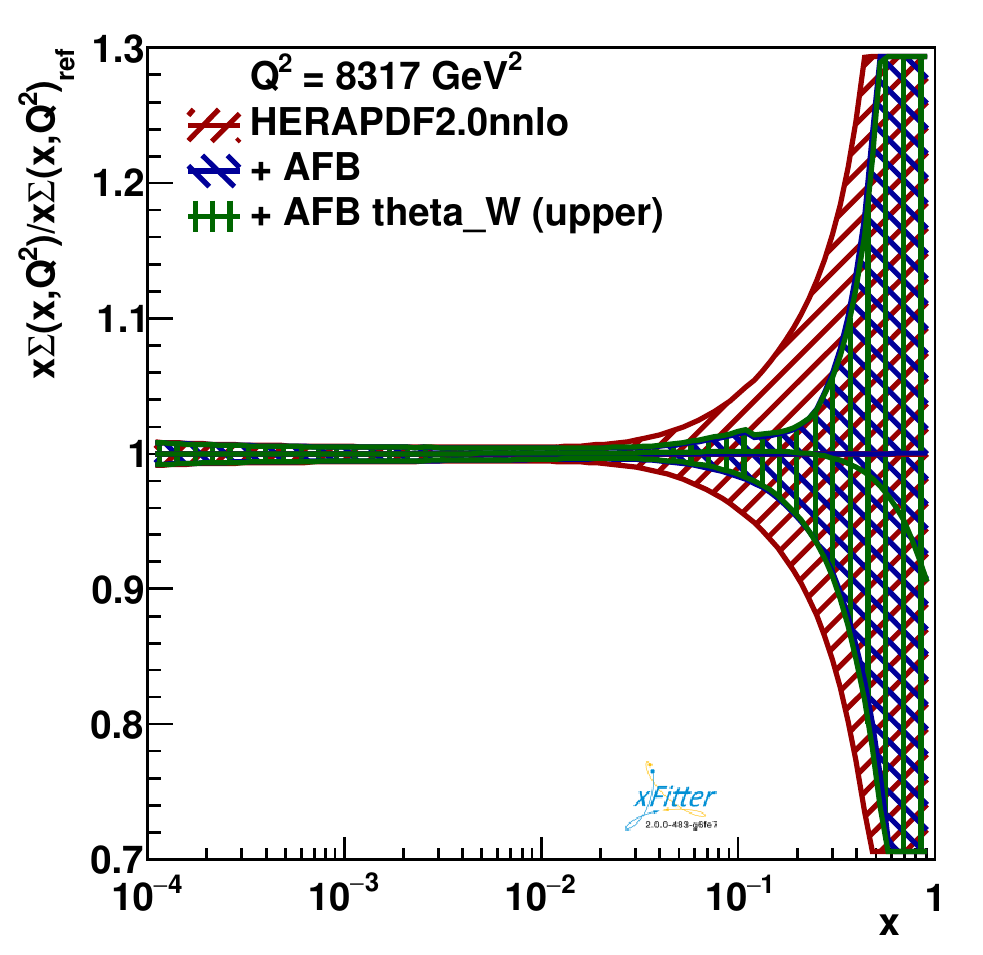}%
\caption{Profiled curves obtained using the upper limit of $\sin^2\theta_W$ allowed by LEP-SLD measurements (top row) and by a global fit of electroweak parameters (bottom row).
The pseudodata correspond to an integrated luminosity of 3000 fb$^{-1}$.}
\label{fig:AFB_theta_W}
\end{center}
\end{figure}

When adopting values for $\sin^2\theta_W$ at the extremes of the LEP-SLD confidence interval, we obtain some differences in the profiled curves, due to the shift of the central value predictions.
We show the results of the profiling in the two cases in Fig.~\ref{fig:AFB_theta_W}, adopting the upper limit of the value of $\sin^2\theta_W$.
Using instead the lower limit of the value of $\sin^2\theta_W$ one obtains profiled curves mirrored to those with respect to the longitudinal axis.
The deviations are clearly more visible in the first case with LEP and SLD accuracy, while we observe smaller differences when employing EW global fit estimates.
It is important to mention that historically measurements of the $A_{\rm{FB}}$ have been used to set constrains on the $\theta_W$ angle~\cite{ATLAS:2018gqq, Bodek:2018sin, Sirunyan:2018swq}.
One very interesting proposal, to which the results of this work provide strong support, is the implementation of a simultaneous fit of both PDFs and $\sin^2\theta_W$.

In the analysis carried out so far, we have neglected any EW radiative corrections to the considered process.
Terms of $\mathcal{O}(\alpha)$ have nowadays been included in the Dokshitzer-Gribov-Lipatov-Altarelli-Parisi (DGLAP) evolution equations, and Quantum Electro-Dynamics (QED) PDF sets, which consistently account for a photon component within the proton, are well established.
In this work we do not include QED or EW corrections, and we limit ourselves to estimating the impact on our analysis when going from a PDF set which includes QED PDFs to a set which does not.

More precisely, we want to check whether in these sets we would obtain substantial differences when importing $A^*_{\rm{FB}}$ data in the profiling (while no QED corrections are taken into account in the matrix elements in both cases).
The NNPDF collaboration has recently released a QED PDF set, compatible with the NNPDF3.1 fit, adopting the LUXqed prescription~\cite{Manohar:2016nzj} (NNPDF31\_nnlo\_as\_0118\_luxqed~\cite{Bertone:2017bme}).
We have checked that the differences in the $A^*_{\rm{FB}}$ predictions obtained between the QED and non-QED sets are small, $|\Delta A^*_{\rm{FB}}| < 2 \times10^{-4}$.
Furthermore, as the LUXqed prescription has been widely accepted, it has been shown that the contribution of photon initiated processes to the Drell-Yan spectrum is negligible~\cite{Accomando:2016ehi}.

\begin{figure}[h]
\begin{center}
\includegraphics[width=0.33\textwidth]{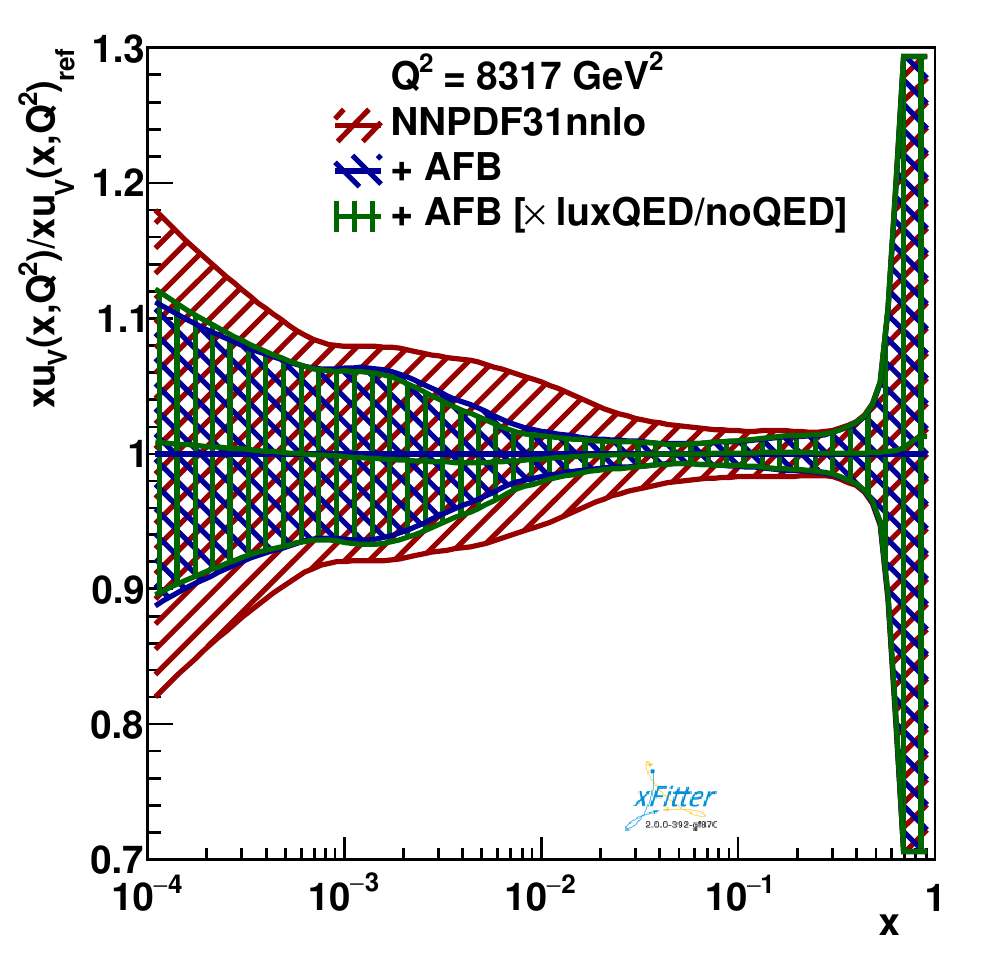}%
\includegraphics[width=0.33\textwidth]{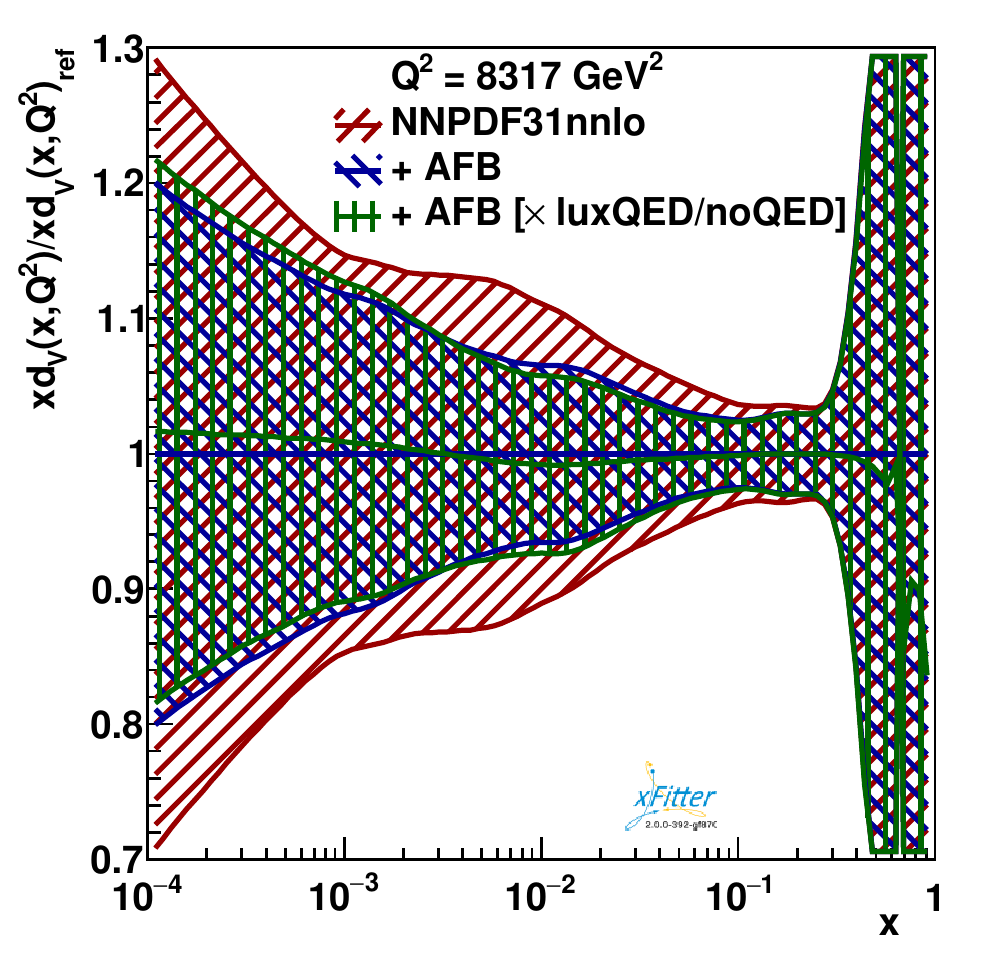}%
\includegraphics[width=0.33\textwidth]{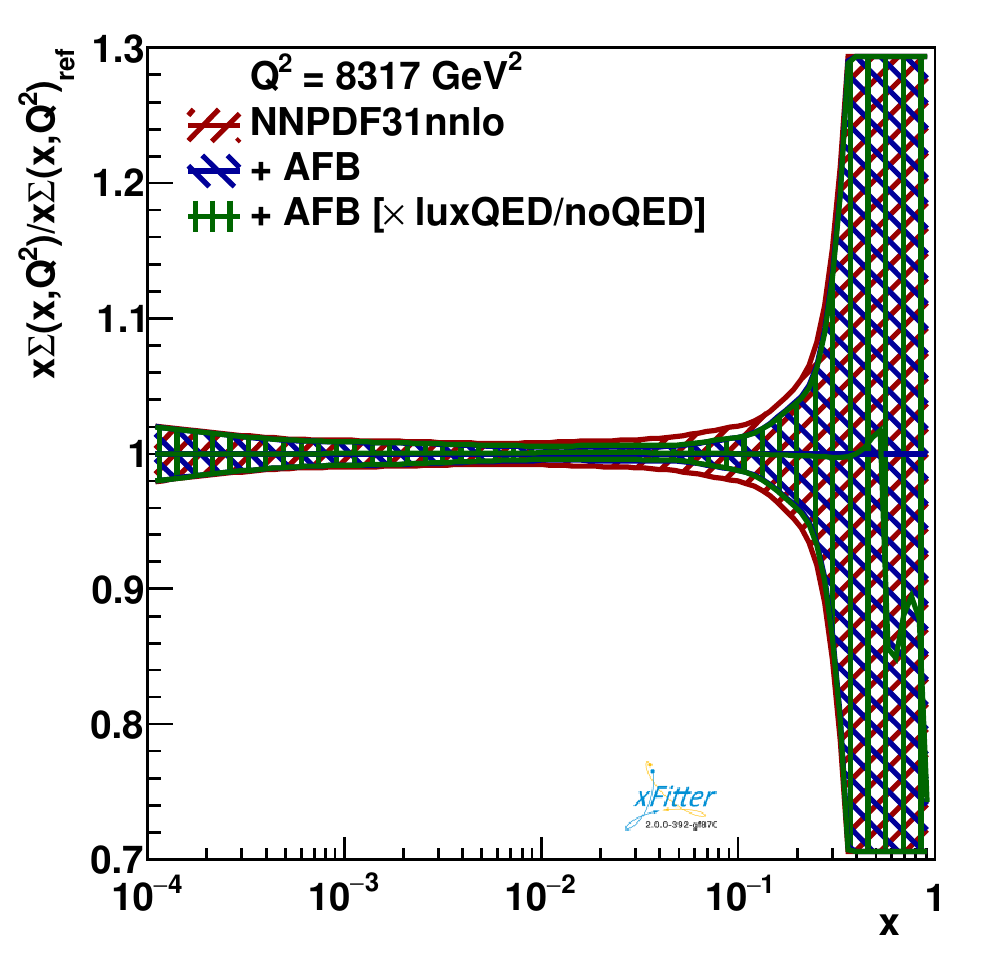}%
\caption{Profiled curves obtained with the NNPDF3.1nnlo and its central value predictions rescaled with a K-factor to match the NNPDF31\_nnlo\_as\_0118\_luxqed predictions.
The pseudodata corresponds to an integrated luminosity of 3000 fb$^{-1}$.}
\label{fig:AFB_NNPDF_QED}
\end{center}
\end{figure}

The profiling of the NNPDF31\_nnlo\_as\_0118\_luxqed set unfortunately cannot be done within the profiling technique implemented in {\tt{xFitter}}, because of the ``replicas'' error method employed in this set whilst no equivalent Hessian PDF set is available.
For this reason we have chosen to study the variations from the QED PDF set in the form of a $k$-factor that was used to rescale the $A^*_{\rm{FB}}$ central value obtained with the NNPDF3.1nnlo set, and we found that the impact on the profiled PDFs is very small. The results of the profiling are visible in Fig.~\ref{fig:AFB_NNPDF_QED}.

\begin{figure}[h]
\begin{center}
\includegraphics[width=0.33\textwidth]{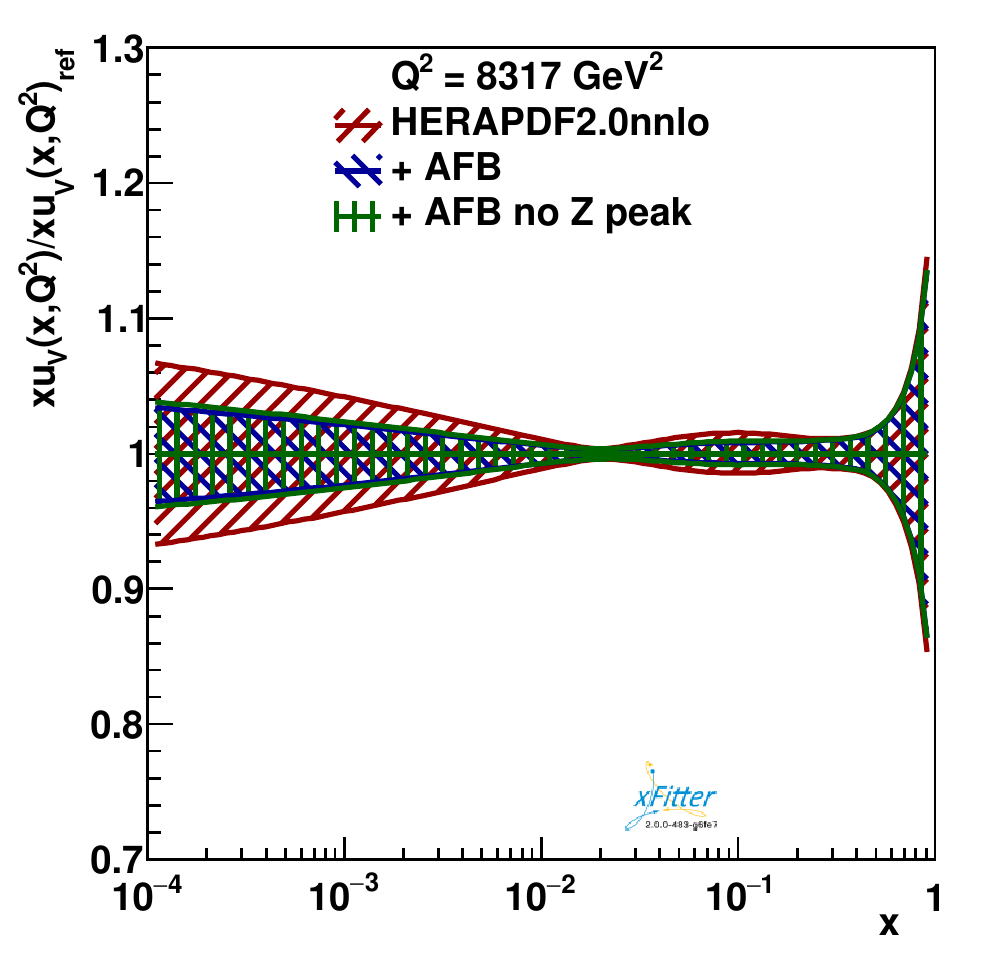}%
\includegraphics[width=0.33\textwidth]{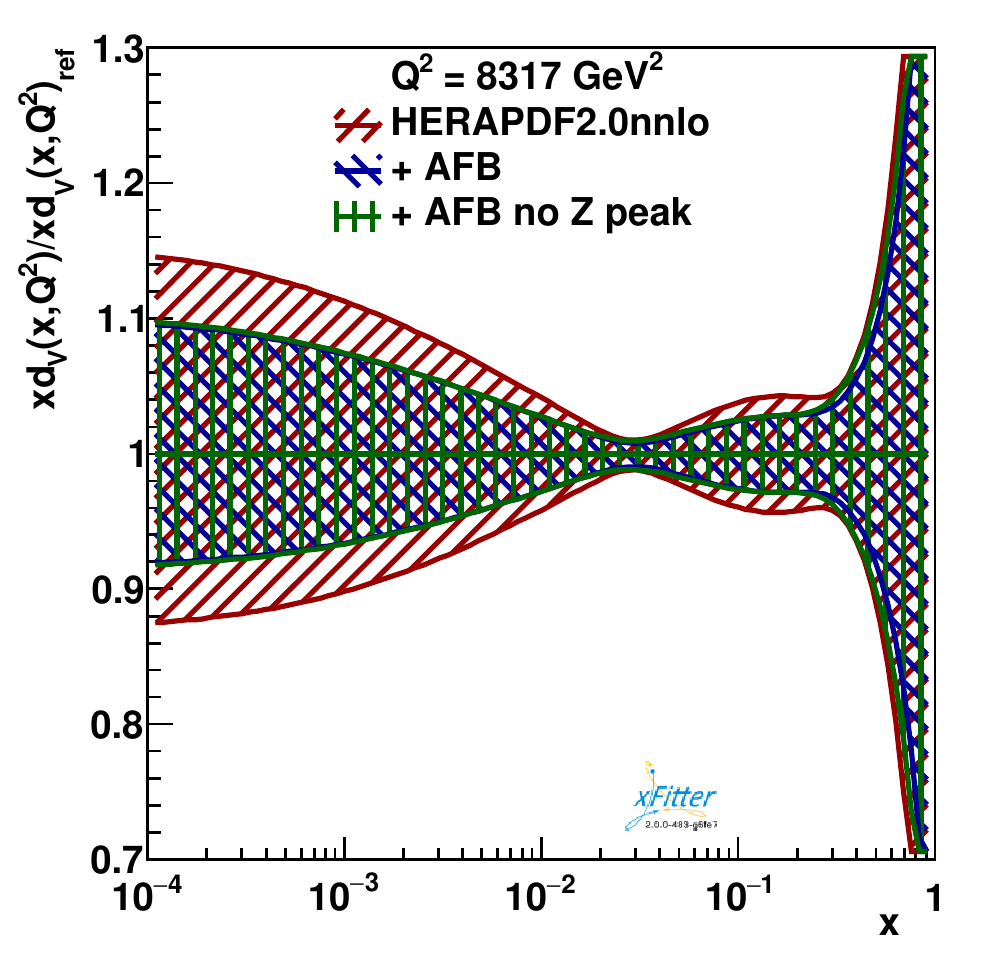}%
\includegraphics[width=0.33\textwidth]{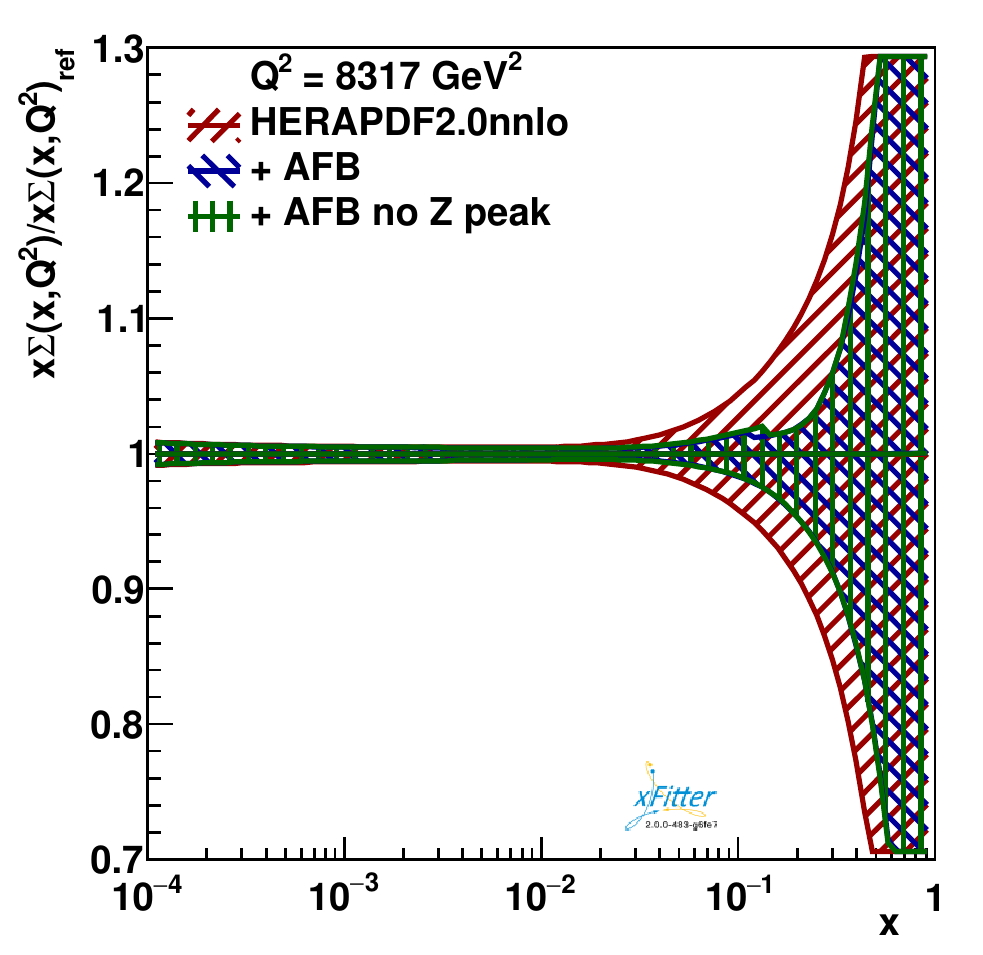}\\%
\includegraphics[width=0.33\textwidth]{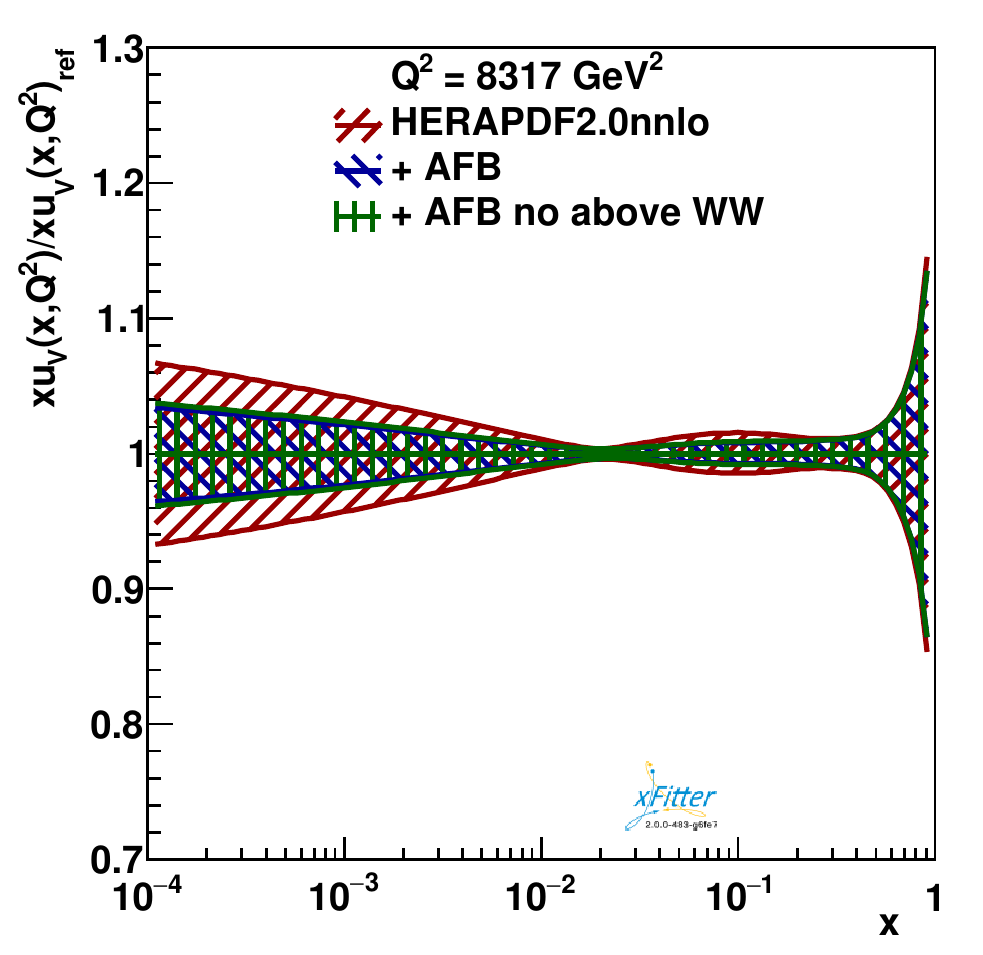}%
\includegraphics[width=0.33\textwidth]{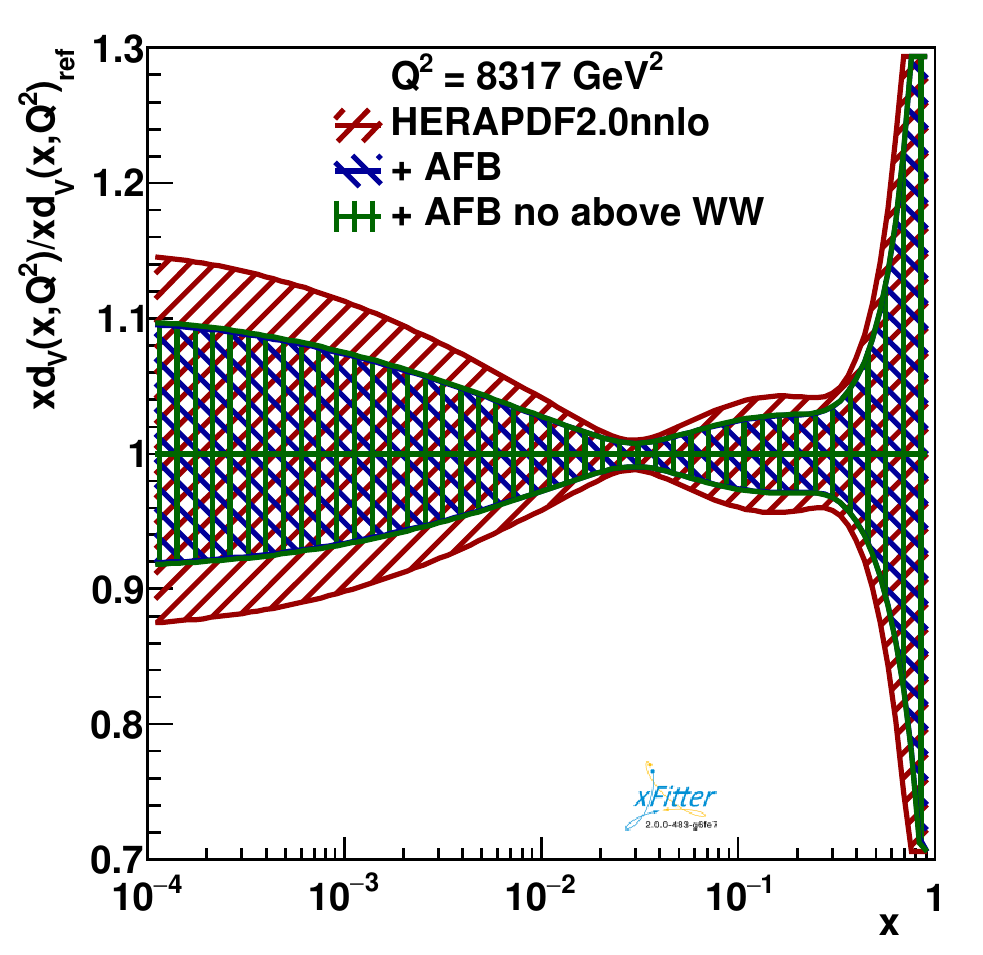}%
\includegraphics[width=0.33\textwidth]{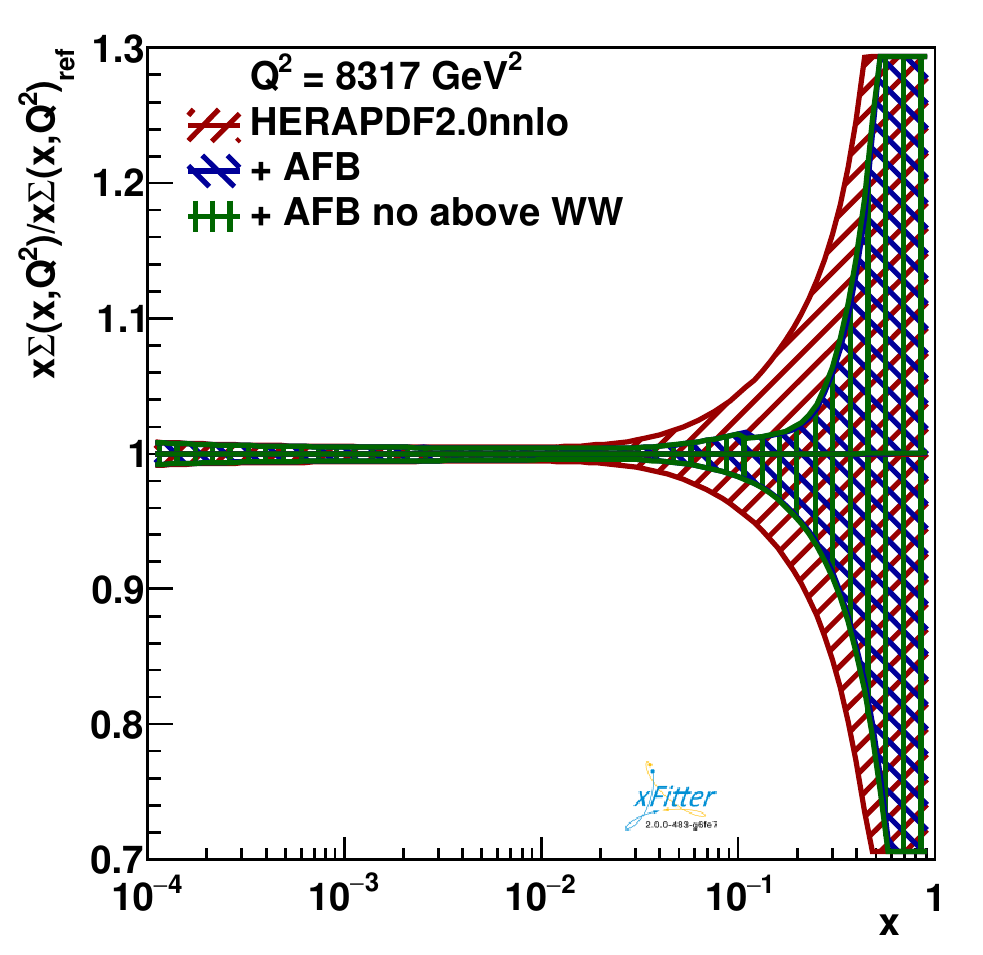}%
\caption{Profiled curves obtained with the HERAPDF2.0nnlo using the full set of data, and when removing the data in the invariant mass region around the $Z$ peak (top row) and when removing the data in the invariant mass region above $WW$ production threshold (bottom row).
The pseudodata corresponds to an integrated luminosity of 3000 fb$^{-1}$.}
\label{fig:AFB_without_points}
\end{center}
\end{figure}

Higher order EW corrections have been shown to be relevant in the TeV region~\cite{Baur:1997wa,Baur:2001ze,Maina:2004rb,Zykunov:2005tc,CarloniCalame:2007cd,Arbuzov:2007db,Dittmaier:2009cr}, however, they could also have an impact in the region around the $Z$ peak, where the high statistics allow for very precise measurements, as well as for $WW$ production~\cite{ATLAS:2018gqq}.
Since they are not included in the current analysis, we want to study the impact of these specific subsets of data in the profiling. For this purpose, 
we employ again the HERAPDF2.0nnlo PDF set.
In the top row of Fig.~\ref{fig:AFB_without_points} we show the profiled curves removing the data in the invariant mass interval $84~\textrm{GeV} < M_{\ell\ell} < 98~\textrm{GeV}$, corresponding to $M_Z \pm 3\Gamma_Z$, while in the bottom row we repeat the same exercise removing the data above the $WW$ production threshold, that is, $M_{\ell\ell} > 161~\textrm{GeV}$.

In the first case there is a small enlargement of the error bands in the $u$-valence and $d$-valence quark distributions, showing some impact of the $Z$ peak data, which is expected because of the large statistic in this invariant mass interval.
In the second case instead only the error band of the $u$-valence quark distribution shows a small increment, meaning that the high invariant mass data has a smaller impact on the profiling, having a worse statistical precision.

\section{Conclusions} 
\label{sec:Conclusions}
High-statistics measurements from the LHC Runs 2, 3 and the HL-LHC stage can be exploited to place constraints on the PDFs. DY processes yielding di-lepton production are a primary channel which may be used to this end. Both cross section and asymmetry distributions can be used for such a purpose.

Concerning the latter, as a counterpart to the lepton charge asymmetry of the CC channel of DY production, in this work we have studied the Forward-Backward Asymmetry $A^*_{\rm{FB}}$, which can be defined in the NC channel of DY production, and we have performed PDF profiling calculations in the {\tt{xFitter}} framework to investigate the impact of $A^*_{\rm{FB}}$ pseudodata on PDF determinations. 
We have found that new PDF sensitivity arises from the di-lepton mass and rapidity spectra of the $A_{\rm{FB}}^*$, which encodes information on the lepton polar angle, or pseudorapidity.

With the partial Run 2 integrated luminosity that we have used in this paper (L = 30 fb$^{-1}$) we observe a significant reduction in PDF uncertainties on the $u$-valence and $d$-valence distributions in the intermediate $x$ region, which can be further improved exploiting the full Run 2 data set (L = 150 fb$^{-1}$).
Adopting the luminosity of Run 3 (L = 300 fb$^{-1}$), we predict the observation of a moderate reduction in PDF uncertainties also on the sea quark distributions.
Above this threshold we observe a saturation effect such that when adopting the projected HL stage luminosity (L = 3000 fb$^{-1}$) we notice a smaller reduction of the uncertainties bands compared to the previous cases.
Furthermore, we have shown that we obtain very different levels of improvement on each PDF, both in magnitude and in range of $x$, depending on the specific PDF set under analysis.

We have also studied the impact of applying cuts on the di-lepton rapidity. 
By increasing the rapidity cut, we obtain enhanced sensitivity to quark distributions in the high $x$ region.
In this case the high statistic collected during the HL stage will be crucial in order to achieve a sufficient precision in the measurement of the $A_{\rm{FB}}^*$.

Performing a rotation of the eigenvectors and sorting them according to their sensitivity to the $A_{\rm{FB}}^*$ data, we noted a strong correlation between $u$-valence and $d$-valence eigenvectors, and that the new data is most sensitive to their charge weighted sum $((2/3)u_V + (1/3)d_V)$, oppositely to the CC lepton asymmetry data, which are instead mostly used to constrain $(u_V - d_V)$.

In summary, $A_{\rm{FB}}^*$ revealed itself a new powerful handle in the quest to contain the systematics associated to PDF determination and exploitation in both SM and BSM studies.

\section*{Acknowledgements}
\noindent
JF and FH thank DESY and the Terascale Physics Helmholtz Alliance for hospitality and support while part of this work was being done.
SM is supported in part through the NExT Institute and acknowledges funding from the STFC CG grant ST/L000296/1.
The work of O.\,Z. has been supported by Bundesministerium f\"ur Bildung und Forschung (contract 05H18GUCC1).

\bibliography{bib}

\end{document}